\documentstyle[aps,preprint,eqsecnum]{revtex}
\tightenlines
\begin{document}
\draft
\title{Gravitational radiation 
from compact binary systems: gravitational waveforms and energy loss
to second post-Newtonian order}
\author{Clifford M. Will$^1$, and Alan G. Wiseman$^2$}
\address{$^1$ McDonnell Center for the Space Sciences,
Department of Physics,
Washington University, St. Louis, Missouri 63130}
\address{$^2$ Theoretical Astrophysics, California Institute of 
Technology, Pasadena, California 91125}
\date{\today}
\maketitle
\begin{abstract}
We derive the gravitational waveform and gravitational-wave energy flux
generated by a binary star system of compact objects (neutron stars or
black holes), accurate through second post-Newtonian order
($O[(v/c)^4] \sim O[(Gm/rc^2)^2]$) beyond the lowest-order quadrupole
approximation.
We cast the Einstein equations into the form of a flat-spacetime wave
equation together with a harmonic gauge condition, and solve it
formally as a retarded integral over the past null cone of the chosen
field point.  The part of this integral that involves the matter
sources and the near-zone gravitational field is evaluated in terms of
multipole moments using standard techniques; 
the remainder of the retarded integral, extending over the
radiation zone, is evaluated in a novel way.  The result is a
manifestly convergent and finite procedure for calculating
gravitational radiation to arbitrary orders in a post-Newtonian
expansion.  Through second post-Newtonian order, the radiation is also
shown to propagate toward the observer along true null rays of the
asymptotically Schwarzschild spacetime, despite having been derived
using flat spacetime wave equations.
The method cures defects that plagued previous ``brute-force'' slow-motion
approaches
to the generation of gravitational radiation, and yields results that
agree perfectly with those recently obtained by a mixed
post-Minkowskian post-Newtonian method.
We display explicit formulae for the gravitational waveform and the
energy flux for two-body systems, both in arbitrary orbits and in
circular orbits.  In an appendix, we extend the formalism to bodies
with finite spatial extent, and derive the  spin corrections to the
waveform and energy loss.
\end{abstract}
\section{INTRODUCTION}
\label{sec:intro}
The generation of gravitational radiation is a long-standing problem
that dates back to the first years following the publication of
general relativity (GR).  In 1916 Einstein calculated the gravitational
radiation emitted by a laboratory-scale object using the linearized
version of GR \cite{AE}.  Some of his assumptions were questionable and
his answer for the energy flux was off by a factor of two (an error
pointed out by Eddington \cite{eddington}).  There followed a lengthy debate
about whether gravitational waves are real or an artifact of general
coordinate invariance, the former interpretation being confirmed by
the rigorous, coordinate free theorems of Bondi and his school
\cite{bondi,bmvdb,sachs} and by the short-wave analysis of Isaacson
\cite{isaacson}.  Shortly after the discovery of the binary pulsar PSR
1913+16 in 1974, questions were raised about the foundations of the
``quadrupole formula'' for gravitational radiation damping \cite{ehlers}
(and in
some quarters, even about its quantitative validity \cite{cranks}).
These questions were answered in part by theoretical 
work designed to shore up the
foundations of the quadrupole approximation
\cite{walkerwill,anderson80,damour84,cs79,iww84}, and in part 
(perhaps mostly) by the
agreement between the predictions of the
quadrupole formula and the {\it observed} 
rate of damping of the pulsar's orbit
\cite{taylor79,taylor94}.

Because it is a slow-motion system ($v/c \sim 10^{-3}$), the binary
pulsar is sensitive only to the lowest-order effects of gravitational
radiation as predicted by the quadrupole formula.  Nevertheless, the
first correction terms of order $v/c$ and $(v/c)^2$ to the quadrupole formula,
were
calculated as early as 1976 \cite{ew,wagwill}.  These are now
conventionally called ``post-Newtonian'' (PN) corrections, with each power
of $v/c$ corresponding to half a post-Newtonian order (1/2PN), in analogy with
post-Newtonian corrections to the Newtonian
equations of motion \cite{convention}.  
In 1976,
the post-Newtonian corrections were of purely academic, rather than
observational interest.  

Recently, however, the issue of higher post-Newtonian corrections in
the theory of gravitational waves has taken on some urgency.  The reason is
the construction of kilometer-scale, laser interferometric
gravitational-wave observatories in the U.S. (LIGO project) and Europe
(VIRGO project), with gravitational-wave searches scheduled to
commence around 2000 (see \cite{snowmass} for a review).  These
broad-band antennae will have the capability of detecting and
measuring the gravitational waveforms from astronomical sources in a
frequency band between about 10 Hz (the seismic noise cutoff) and 
500 Hz (the photon counting noise cutoff), with a maximum
sensitivity to strain at around 100 Hz of $\Delta l/l \sim 10^{-22}$
(rms).  The most promising source for detection and study of
the gravitational-wave signal is the ``inspiralling compact binary''
-- a binary system of neutron stars or black holes (or one of each) in
the final minutes of a death dance leading to a violent merger.
Such is the fate, for example of the Hulse-Taylor binary pulsar PSR
1913+16 in about 300 M years.  Given the expected sensitivity of the
``advanced LIGO'' (around 2001), which could see such sources out to
hundreds of megaparsecs, it has been estimated that from 3 to
100 annual inspiral events could be detectable 
\cite{snowmass,narayan,phinney}.

The urgency derives from the realization \cite{3min} that extremely
accurate theoretical predictions for the orbital evolution, and to a
lesser extent, the gravitational waveform,
will play a central role in the data analysis
from these observatories.  That data analysis is likely to involve
some form of matched filtering of the noisy detector output against an
ensemble of theoretical ``template'' waveforms which depend on the
intrinsic parameters of the inspiralling binary, such as the component
masses, spins, and so on, and on its inspiral evolution.  
How accurate must a template be in order to
``match'' the waveform from a given source (where by a match we mean
maximizing the signal-to-noise ratio)?  In the total accumulated phase
of the wave detected in the sensitive bandwidth, the template must
match the signal to a fraction of a cycle.  For two inspiralling
neutron stars, around 16,000 cycles should be detected; this implies a
phasing accuracy of $10^{-5}$ or better.  Since $v/c \sim 1/10$ during
the late inspiral, this means that correction terms in the phasing
at the level of
$(v/c)^5$ or higher are needed.  More formal analyses confirm this
intuition \cite{finnchern,cutlerflan94,poissonwill,cutlerflan96}.  

The bottom line is that theorists have been challenged to derive the
gravitational waveform and the resulting radiation back-reaction on the orbit
phasing at least to 2PN,
or second post-Newtonian order, $O[(v/c)^4]$, beyond the quadrupole
approximation, and probably to 3PN
order.  Furthermore, because of the extreme complexity of the
calculations at such high PN order, independent calculations are called
for, in order to inspire confidence in the final formulae.  After all,
the formulae will ultimately be compared against real data.

This challenge was recently taken up by two teams of workers, one
composed of Blanchet, Damour and Iyer (BDI), the other composed of the 
present authors.   The goal was to derive the gravitational waveform
and the energy flux for inspiralling compact binaries of arbitrary
masses, through 2PN order.
Each team adopted a different approach to the
calculation, and worked in isolation from the other.  Only at the end
of the calculation were comparisons made for the key formulae for the
waveform and the gravitational energy flux.  
The results agreed precisely \cite{bdiww}.

The BDI approach was based on a mixed post-Newtonian and 
``post-Minkowskian'' framework for
solving Einstein's equations approximately, developed in a long series of
papers by Damour and colleagues \cite{bd86,bd88,bd89,di91,bdtail,luc95}.  
The idea is to solve
the vacuum Einstein equations in the exterior of the
material sources extending out to the radiation zone
in an expansion (``post-Minkowskian'')
in ``nonlinearity'' (effectively an
expansion in powers of Newton's constant $G$), and to express the
asymptotic solutions in terms of a set of formal, time-dependent, 
symmetric and trace-free (STF) multipole moments \cite{thorne80}.  
Then, in a near
zone within one characteristic wavelength of the radiation, the
equations including the material 
source are solved in a slow-motion approximation (expansion in powers
of $1/c$)
that yields a set of STF source multipole moments expressed as integrals over 
the ``effective'' source, including both matter and 
gravitational field contributions.  
The solutions involving the
two sets of moments are then matched in an intermediate zone, resulting
in a connection between the formal radiative moments and the source moments.
The matching also provides a natural way, using analytic continuation,
to regularize integrals involving the non-compact contributions of
gravitational stress-energy, that might otherwise be divergent.   

The approach of this paper is based on a framework developed by
Epstein and Wagoner (EW) \cite{ew}.  Like the BDI approach,
it involves rewriting the Einstein
equations in their ``relaxed'' form, namely as an inhomogeneous,
flat-spacetime wave equation for a field $h^{\alpha\beta}$, whose
source consists of both the material stress-energy, and a
``gravitational stress-energy'' made up of all the terms non-linear in
$h^{\alpha\beta}$.  The wave equation is accompanied by a harmonic or
deDonder gauge condition
on $h^{\alpha\beta}$, which serves to specify a coordinate system, and
also imposes equations of motion on the sources.  Unlike the BDI
approach, a {\it single} formal solution is written down, valid everywhere
in spacetime.  This formal solution,
based on the flat-spacetime retarded Green function, is a retarded integral
equation for $h^{\alpha\beta}$, which is then iterated in a
slow-motion ($v/c<1$), weak-field ($||h^{\alpha\beta}|| <1$ )
approximation, that is very similar to the corresponding procedure in
electromagnetism.  However, because the integrand of this retarded integral   
is not compact by virture of the non-linear field contributions, the
original EW formalism quickly runs up against integrals that are not
well defined, or worse, are divergent.  Although at the lowest
quadrupole and first few PN orders, various arguments can be given to
justify sweeping such problems under the rug \cite{wagwill}, they are not very
rigorous, and provide no guarantee that the divergences do not become
insurmountable at higher orders.  As a consequence, despite 
efforts to cure the problem, the EW formalism fell into
some disfavor as a route to higher orders, although an extension to
3/2PN order was accomplished \cite{magnum}.

One contribution of this paper is a resolution of this problem.
The resolution involves taking
literally the statement that the solution is a {\it retarded} integral,
{\it i.e.} an integral over the {\it entire} past null cone of the field
point.  To be sure, that part of the integral that extends over
the intersection between the past null cone and 
the material source and the near zone 
is still approximated as usual by a slow-motion expansion
involving spatial integrals of moments of the source, including the
non-compact gravitational contributions, just as in the BDI framework.  
But instead of cavalierly extending the
spatial integrals to infinity as was implicit in the original EW
framework, and risking undefined or 
divergent integrals, we terminate the integrals at
the boundary of the near zone, chosen to be at a radius $\cal R$ given
roughly by one wavelength of the gravitational radiation.  
{}For the 
integral over the rest of the past null cone
exterior to the near zone (``radiation zone''), we do not make a slow-motion
expansion, instead we use a coordinate transformation
to convert
the integral into a convenient, easy-to-calculate form, that is
manifestly convergent, subject only to reasonable assumptions about
the past behavior of the source.  
This transformation was
suggested by our earlier work on a non-linear gravitational-wave
phenomenon called the Christodoulou memory \cite{christo}. 
Not only are all integrations now
explicitly finite and convergent, we show explicitly that all
contributions from the near-zone spatial integrals that grow with
$\cal R$ (and that would have diverged had we let ${\cal R} \to
\infty$) are actually {\it cancelled} by corresponding terms from the
radiation-zone integrals.  Thus the procedure, as expected, has no
dependence on the artificially chosen boundary radius $\cal R$
of the near-zone.  In
addition, the method can be carried to higher orders in a
straightforward, albeit very tedious manner.  The result is a 
manifestly finite,
well-defined procedure for calculating gravitational radiation to high,
and we suspect all, PN orders.  

The result of the calculation is 
an explicit formula for the gravitational waveform for a
two-body system, 
the transverse-traceless (TT) part of the radiation-zone field, denoted
$h^{ij}$, and representing the deviation of the metric from
flat spacetime.  In terms of an expansion
beyond the quadrupole formula, it has the schematic form,
\begin{equation}
h^{ij} = {{2G\mu} \over Rc^4} 
\left\{ \tilde Q^{ij} [ 1 + O(\epsilon^{1/2}) + O(\epsilon)
+ O(\epsilon^{3/2}) + O(\epsilon^2) \dots ] \right\} _{TT} \,,
\label{1-1}
\end{equation}
where $\mu$ is the reduced mass, and  $\tilde Q^{ij}$ represents two time
derivatives of
the mass quadrupole moment tensor (the series actually contains
multipole orders beyond quadrupole).
The TT projection operation is described below.
The expansion parameter $\epsilon$ is related to the orbital variables
by $\epsilon \sim Gm/rc^2 \sim (v/c)^2$, where $r$ is
the distance
between the bodies, $v$ is the relative velocity, and $m=m_1 + m_2$ is
the total mass.  The 1/2PN and 1PN terms were derived in
\cite{wagwill}, the 3/2PN terms in \cite{magnum}.  The contribution of
gravitational-wave ``tails'', caused by backscatter of the outgoing
radiation off the background spacetime curvature, at
$O(\epsilon^{3/2})$, were derived and studied in 
\cite{bdtail,poissontail,agwtail}.

This paper derives
the 2PN terms including 2PN tail contributions; 
the results are in complete agreement with BDI
\cite{bdi2pn}.  
We also find that part of the tail terms at 3/2PN and 2PN order serve
to guarantee that the outgoing radiation propagates along true null
directions of the asymptotic curved spacetime, despite the use of flat
spacetime wave equations in the solution.  
The explicit formula for the general two-body waveform 
is given below in Eqs. (\ref{hanswer})
and (\ref{hpieces}).

There are also contributions to the waveform
due to intrinsic spin of the bodies, which occur 
at $O(\epsilon^{3/2})$ (spin-orbit) and
$O(\epsilon^2)$ (spin-spin); these have been calculated elsewhere
\cite{kww,kidder}, and are rederived in the EW framework in Appendix F.  

Equations of motion for the material sources must also be specified to 2PN
order in order to have a consistent solution of 
Einstein's equations.  These have
the schematic form
\begin{equation}
d^2 {\bf x}/dt^2 = -(Gm{\bf x}/r^3)
[1+O(\epsilon)+O({\epsilon}^{3/2})+O({\epsilon}^2)
 + \dots ]\,,
\label{1-2}
\end{equation}
where ${\bf x} ={\bf x}_1 -{\bf x}_2$ is the separation vector.  
The lowest-order contribution is obviously
Newtonian.
The next term $O(\epsilon)$ is the first
post-Newtonian correction, which
gives rise to 
such effects as the advance of the periastron.
The term $O(\epsilon^{3/2})$ comes solely from the spin-orbit
interaction.
The term of $O(\epsilon^2)$
is a {\it second} post-Newtonian correction to the equation of motion
(and also contains spin-spin interactions).
The terms in Eq. (\ref{1-2}) are all non-dissipative, having nothing
to do with gravitational radiation reaction.  Through 2PN order, these
equations are by now standard; see for example 
\cite{deruelle,damour82,damour300} and
Eq. (\ref{eom}) below.

Given the gravitational waveform, we can 
compute the rate energy is carried off by the radiation
(schematically $\int \dot h \dot h d\Omega$,
the gravitational analog of the Poynting
flux).  The result has the schematic form
\begin{eqnarray} 
dE/dt = (dE/dt)_Q 
[1+O(\epsilon)+O({\epsilon}^{3/2})+O({\epsilon}^2)
 + \dots ]\,.
\label{1-3} 
\end{eqnarray} 
Here $(dE/dt)_Q$ denotes the lowest-order quadrupole contribution,
proportional to the square of three time derivatives of the trace-free
mass quadrupole moment tensor of the source.  The explicit formula for
a general two-body system is given below in Eqs. (\ref{Edotanswer})
and (\ref{Edotpieces}).
{}For the special case of 
non-spinning bodies moving on quasi-circular orbits ({\it i.e.}
circular apart from a slow inspiral), the energy flux has the 
form
\begin{eqnarray}
{dE \over dt} = &&{32G \over 5c^5} \eta^2 {\left ({Gm \over rc^2} \right )}^5
\biggl [ 1 - {Gm \over rc^2} \left ( {2927 \over 336} + {5 \over 4} \eta
\right )
\nonumber \\
&&+ 4\pi{\left ( {Gm \over rc^2} \right )}^{3/2} 
+ {\left ( {Gm \over rc^2} \right )}^2 \left (
{293383 \over 9072} + {380 \over 9} \eta \right ) \biggr ] \;,
\label{edot}
\end{eqnarray}
where $\eta= m_1m_2/m^2$.  The first term is the quadrupole
contribution, the second term is the 1PN contributon \cite{wagwill}, the
third term, with the coefficient $4\pi$, is the ``tail'' contribution
\cite{bdtail,poissontail,agwtail,lucschafer}, and the fourth term is the 2PN
contribution derived here.  This new contribution was reported in
\cite{bdiww}, and was also derived using the BDI approach in
\cite{bdi2pn}.  For the contributions of spin-orbit and spin-spin
coupling see \cite{kww,kidder,bdiww} and Appendix F.

Similar expressions can be derived for the loss of angular momentum
and linear momentum.  These losses react
back on the orbit to circularize it and cause it to inspiral.  The
result is that the orbital phase (and consequently the 
gravitational-wave 
phase) evolves non-linearly with time.  It is the sensitivity of the
broad-band LIGO and VIRGO-type detectors to phase that makes the  
higher-order contributions to $dE/dt$ so observationally relevant.
{}For example, for an inspiral of two $1.4M_\odot$ neutron stars, 
the 2PN term  
in Eq. (\ref{edot}) contributes about 9 of the 16,000 cycles
observable in the bandwidth of the advanced LIGO.  More detailed
analyses of the effect of the 2PN terms on the matched filtering can
be found in \cite{poissonwill,krolak,poissonPN}.  A ready-to-use set
of formulae for the 2PN gravitational waveform template, including the
non-linear evolution of the gravitational-wave frequency (not
including spin effects) may be found in \cite{biww}.  Spin corrections
to the waveform templates may be found in Appendix F.

An alternative approach to deriving gravitational waveforms and energy
flux for inspiralling compact binaries, 
in the limit in which one mass is much smaller than the other,
is that of black-hole perturbation theory.  This method provides 
numerical results that are exact in $v/c$, as well as analytical results 
expressed as series in powers of $v/c$, both for
non-rotating and for rotating black holes 
\cite{poissontail,ps95,cfps93,sasaki94,tagoshi94}.  For
non-rotating holes, the analytical expansions have been carried to
{\it fourth} PN order \cite{tagoshi94}.  In all cases of overlap, the
results agree precisely with our post-Newtonian results, in the limit
$\eta \to 0$.

This paper is an attempt to present, in a relatively complete and
self-contained form, the formalism and machinery of our ``improved EW''
approach to higher-order gravitational radiation from binary systems.
Indeed, we begin with the raw Einstein equations, and end with a plot of
the 2PN waveform.
The goal is to provide sufficient detail to allow the reader, using
this paper virtually alone, to verify any of the results reported here (we 
make no statement about the amount of work involved), and to carry the
computations to higher PN orders.  In Section II, we lay out the
foundations of gravitational-wave generation, describing the relaxed
Einstein equations, the matter sources and the near
and radiation zones, and the
formal retarded integral solution of the wave equation, including the
new treatment of integration over the null-cone in the radiation
zone.
We turn in Section III to the weak-field, slow-motion approximation,
and write down the matter and field variables to the accuracy needed
to find the radiation to 2PN order.  The part of the retarded integral
for $h^{\alpha\beta}$ that extends over the near zone can be written
in terms of a set of ``Epstein-Wagoner'' moments; these are evaluated
explicitly in Section IV.  In Section V, we evaluate the contributions
to $h^{\alpha\beta}$ from the radiation-zone integrals, showing both
the explicit cancellation of those terms in the EW moments that grow with
$\cal R$, and the generation of tail terms.  Section VI specializes to
two-body systems, and displays the full formulae for the
gravitational waveform and energy loss.  In Section VII, we further
specialize to circular orbits.  Section VIII makes concluding remarks.
A number of technical details are
relegated to Appendices.

Our conventions and notation generally follow those of
\cite{MTW,thorne80}.  Henceforth
we use units in which $G = c = 1 $.
Greek indices run over four spacetime values 0, 1, 2, 3, while
Latin indices run over three spatial values 1, 2, 3;
commas denote partial derivatives with
respect to a chosen coordinate system, while semicolons denote
covariant derivatives;
repeated indices are summed over;
$\eta^{ \mu \nu } = \eta_{ \mu \nu } = {\rm diag}(-1,1,1,1)$;
$g \equiv \det(  g_{ \mu \nu } )$;
$a^{(ij)} \equiv ( a^{ij} + a^{ji} )/2$;
$a^{[ij]} \equiv ( a^{ij} - a^{ji} )/2$;
$\epsilon^{ijk}$ is the totally antisymmetric Levi-Civita symbol
$( \epsilon^{123} = + 1)$.  We use a multi-index notation for products
of vector components: $x^{ij \dots k} \equiv x^ix^j \dots x^k$, with a
capital letter superscript denoting a product of that dimensionality:
$x^L \equiv x^{ i_1 } x^{ i_2 } ... x^{ i_l } $; angular brackets
around indices denote STF products (see Appendix \ref{appSTF} for
definitions).
Spatial indices are freely raised and lowered with
$\delta^{ij}$ and $\delta_{ij}$.

\section{Foundations of gravitational-wave generation}
\label{sec:foundations}
\subsection{The relaxed Einstein equations}

We begin our development of the gravitational-wave generation problem
with the Einstein Equations
\begin{eqnarray}
R^{\alpha \beta} - {1 \over 2} g^{\alpha \beta} R = 8 \pi 
T^{\alpha \beta} \; .
\label{einstein}
\end{eqnarray}
Here $R^{\alpha \beta}$ is the Ricci curvature tensor,
$g^{\alpha \beta}$ is the spacetime metric and $T^{\alpha \beta}$ 
is the stress-energy tensor of the matter.
Although Eq. (\ref{einstein}) 
is a conceptually powerful statement,
relating the curvature of spacetime on the
left-hand side to the stress-energy of matter on the right-hand side, 
it is not a particularly useful
form of the Einstein equations for practical calculations of 
gravitational-wave 
generation.
{}For that purpose it is conventional first to define the potential
\begin{eqnarray}
h^{\alpha \beta} \equiv \eta^{\alpha \beta} - (-g)^{1/2} g^{\alpha \beta} \; ,
\label{hdefinition}
\end{eqnarray}
(see {\it e.g.} \cite{thorne80})
and to choose a particular coordinate system defined by the deDonder
or harmonic gauge condition
\begin{eqnarray}
h^{\alpha \beta},_{\beta} = 0 \; .
\label{harmonic}
\end{eqnarray}
The spatial components of 
$h^{\alpha \beta}$ evaluated far from the source comprise the
gravitational waveform and are directly related to
the signal which a gravitational-wave detector measures. 
With these definitions 
the Einstein equations (\ref{einstein}) can be recast in the following
form
\begin{eqnarray}
\Box h^{ \alpha \beta } = -16 \pi {\tau}^{ \alpha \beta } \; ,
\label{relaxed}
\end{eqnarray}
where $\Box \equiv  -{\partial}^2 / \partial t^2 + {\nabla}^2 $
is the flat-spacetime wave operator.
The source on the right-hand side is given by the ``effective''
stress-energy pseudotensor
\begin{eqnarray}
\tau^{\alpha\beta} = (-g)T^{\alpha\beta} + (16\pi)^{-1}
\Lambda^{\alpha\beta} \;,
\label{effective}
\end{eqnarray}
where $\Lambda^{\alpha\beta}$ is the non-linear ``field'' contribution
given by
\begin{equation}
\Lambda^{\alpha \beta}
   = 16\pi (-g) t_{LL}^{\alpha \beta } 
   +  ( h^{\alpha \mu},_{\nu} h^{\beta \nu},_{\mu}
                  - h^{\alpha \beta},_{\mu \nu} h^{\mu \nu} ) \; ,
\label{nonlinear}
\end{equation}
and $ t_{LL}^{\alpha \beta }$ is the ``Landau-Lifshitz''
pseudotensor, given by
\begin{eqnarray}
 16 \pi (-g)t_{LL}^{\alpha \beta } &\equiv& \bigl \{ 
 g_{\lambda\mu}g^{\nu\rho}{h^{\alpha\lambda}}_{,\nu}{h^{\beta\mu}}_{,\rho} 
+{1 \over 2} 
 g_{\lambda\mu}g^{\alpha\beta}{h^{\lambda\nu}}_{,\rho}{h^{\rho\mu}}_{,\nu} 
- 2g_{\mu\nu}g^{\lambda (\alpha}{h^{\beta )\nu}}_{,\rho}{h^{\rho\mu}}
_{,\lambda} 
\nonumber \\
&&+ {1 \over 8}
(2g^{\alpha\lambda}g^{\beta\mu}-g^{\alpha\beta}g^{\lambda\mu})
(2g_{\nu\rho}g_{\sigma\tau}-g_{\rho\sigma}g_{\nu\tau})
{h^{\nu\tau}}_{,\lambda}{h^{\rho\sigma}}_{,\mu} \bigr \} \;.
\label{landau}
\end{eqnarray}
By virtue of the gauge condition
(\ref{harmonic}), this source term
satisfies the conservation law
\begin{equation}
{{\tau}^{\alpha \beta}}_{, \beta} = 0 \; ,
\label{conservation}
\end{equation}
which is equivalent to the equation of motion of the matter
${T^{\alpha\beta}}_{;\beta}=0$.

We emphasize that Eq. (\ref{relaxed}) is not an {\it approximate}, 
or {\it weak-field},
form of the Einstein equations;
it is exact, and relies only on the assumption that spacetime can be
covered by harmonic coordinates.

The form of Eq. (\ref{relaxed}) is suggestive of the wave equation
for the vector potential in electromagnetism.  This
analogy with E\&M is at once helpful and deceptive.
It is helpful in that it suggests how to 
proceed to solve the equation, {\it i.e.} use a retarded 
Green function, and an expansion  
in terms of radiative multipole moments.
It further illustrates that, just as the
current density in E\&M is the source for the vector potential,
here the stress-energy of the matter is a
source of the gravitational potential.

However there are several important differences
between Eq. (\ref{relaxed}) and its electromagnetic
counterpart.
{}First, the ``source'' in Eq. (\ref{relaxed}) also contains a
gravitational part that depends explicitly on
$h^{\alpha \beta}$, the very quantity for which we are trying to solve.
Second, unlike the E\&M case where the source (the currents)
has finite spatial extent (compact support), 
we can expect $\tau^{\alpha \beta}$,
which depends on the fields $h^{\alpha \beta}$, to have infinite
spatial extent. Indeed the very outgoing radiation that we
hope to detect, will, at some level of approximation, serve as
a contribution to the source, thus generating
an additional component of the radiation.
However, we have found that, for the physical situations of interest, 
this latter, highly nonlinear effect, 
often referred to as the 
Christodoulou memory, is very weak and can be adequately approximated
by the methods of this paper \cite{christo}.  

Another complication in Eq. (\ref{relaxed}) is that
the second derivative term $h^{\alpha \beta},_{\mu \nu} h^{\mu \nu}$
in the source really ``belongs'' on the left-hand side with the other
second derivative terms in the wave operator.
Such a term in a differential equation modifies the propagation
characteristics of the field from the flat-spacetime characteristics
represented by the d'Alembertian operator.  Physically this is a
manifestation of the fact that the radiation propagates along null
cones of the curved spacetime around the source, which deviate from
the flat null cones of the harmonic coordinates.
Nevertheless, 
the techniques to be presented here do recover the leading
manifestations of this effect, commonly known as ``tails'', including
modification of the phasing of the solutions from their initial
dependance on flat
space retarded time to true retarded time of the asymptotic Schwarzschild
spacetime of the source.

\subsection{Source, near-zone and radiation-zone}

We consider a material source consisting of a collection of fluid
balls (stars) whose size is typically small compared to their
separations.  The material will be modeled as perfect fluid, having
stress-energy tensor
\begin{eqnarray}
T^{\alpha\beta} \equiv (\rho +p)u^\alpha u^\beta +pg^{\alpha\beta}
\;,
\label{fluid}
\end{eqnarray}
where $\rho$ and $p$ are the locally measured energy density and
pressure, respectively, and $u^\alpha$ is the four-velocity of an
element of fluid.  We shall assume that the bodies are sufficiently
compact that we can ignore all intrinsic multipole moments of the
bodies at quadrupole order and beyond.  That is, we treat only the
bodies' monopole (mass) moments (in an Appendix we treat the bodies'
dipole (spin) moments).  For inspiralling
binaries of compact objects, the effects of rotationally induced and
tidally induced quadrupole and higher moments on the orbital evolution or
gravitational radiation have been shown, in the case of binary neutron
stars, to be negligible until the
final coalescence stage, where the post-Newtonian approximation breaks
down anyway \cite{bildcutler}.  
{}For spinning black holes, the effects are small, but
can be non-negligible for sufficiently large spin \cite{japan}.  In
the long run, such finite-size effects should (and can) be
incorporated into our formalism.

To treat the monopole part of the bodies' mass distributions, we
approximate the stress-energy tensor as a distributional tensor
representing ``point'' masses, given by
\begin{eqnarray}
T^{\alpha\beta}_{\rm MONOPOLE} \equiv \sum_A m_A (-g)^{-1/2}
(u^\alpha_A u^\beta_A /u^0_A ) \delta^3 ({\bf x} - {\bf x}_A(t) )
\;,
\label{pointmass}
\end{eqnarray}
where $m_A$ is the gravitational mass of the A-th body, and
$u^\alpha_A$ is the four-velocity of its center of mass, ${\bf x}_A(t)$. 
{}Formally, such a distributional stress-energy tensor is not valid in
general relativity.  On the other hand, it has been shown in a variety
of post-Newtonian contexts to give results that are equivalent to
treating the bodies as almost spherical fluid balls, defining a
suitable approximate center of mass, and carrying out
explicit integrals over the interiors of the balls.  
The resulting self-field and internal energy effects result in a
renormalization of the mass of each body from a ``bare'' 
mass $\int_A \rho d^3x$ to the gravitational mass $m_A$.  Furthermore,
all effects of the internal structure of the bodies are ``effaced'',
so that all aspects of the motion and gravitational radiation are
characterized by a single mass $m_A$ for each body (see \cite{magnum}
for demonstration of this effacement in the waveform at 3/2PN order).  
This is a
manifestation of the Strong Equivalence Principle, which is satisfied
by general relativity.  All these complications, then, can be embodied in
the distributional stress-energy tensor of Eq. (\ref{pointmass}), with
the caveat that all infinite self-field effects that might result from
the use of the delta-function source are to be discarded (self-field
effects having already been renormalized into $m_A$).  An alternative
viewpoint takes the gravitational field in a zone surrounding each
body in a coordinate system that momentarily comoves with the body and
notes that it can be characterized by multipole moments that can be
identified with the body's asymptotially measured mass and
(if desired) higher multipole moments.  The fields surrounding each
body are then matched to an appropriate interbody gravitational field,
with the equations of motion providing consistency conditions for such
matching.  Apart from tidal effects, 
the results depend only on the effective masses of the
bodies, and all self-field effects are automatically accounted for
(see \cite{hartlethorne,soffel} for example, for detailed 
implementations of this
approach in various situations).

The effects of spins can be added to the framework in a
straightforward way; these are reviewed in Appendix \ref{appspin}.  

We consider the bodies to comprise a bound system of characteristic
size ${\cal S} = \max_{\{A,B\}} r_{AB}$, where $r_{AB}=| {\bf x}_A -{\bf
x}_B |$, with a center of mass chosen to be at the origin of
coordinates, ${\bf X}=0$.  The {\it source zone} then consists of
the world tube ${\cal T} =\{ x^\alpha | r<{\cal S}, -\infty <t<\infty \}$.

The bodies are assumed to move with characteristic velocities $v_A
<1$, and for much of their evolution with $v_A \ll 1$.  The
characteristic reduced wavelength  of gravitational radiation,
${\lambda\!\!\!{\scriptscriptstyle{{}^{-}}}} = \lambda /2\pi \sim
{\cal S} /v \equiv {\cal R}$ 
serves to define the boundary of the {\it near zone},
defined to be the world tube  ${\cal D} =\{x^\alpha | r<{\cal R}, -\infty
<t<\infty$ \}.  Within the near zone, the gravitational fields can be
treated as almost instantaneous functions of the source variables, {\it
i.e.}
retardation can be ignored or treated as a small perturbation of
instantaneous solutions.  For most of the evolution, up to the point where 
the
post-Newtonian approximation breaks down, ${\cal R} \gg {\cal S}$.

The region exterior to the near zone is the {\it radiation zone}, $r >
{\cal R}$.  In this zone, we evaluate the fully retarded solutions of
Eq. (\ref{relaxed}), and focus on the parts that fall off as
$r^{-1}$.

The formal solution to Eq. (\ref{relaxed}) can be written down
in terms of the retarded, flat-space Green function:
\begin{eqnarray}
h^{\alpha \beta} (t,{\bf x}) = && 4 \int
{ \tau^{\alpha \beta} (t^\prime, {\bf x^\prime} )
  \delta( t^\prime - t + | {\bf x} - {\bf x^\prime} | )
\over | {\bf x} - {\bf x^\prime} | } d^4x^\prime \;.
\label{bigintegral}
\end{eqnarray}
This represents an integration of $\tau^{\alpha \beta}/|
{\bf x} - {\bf x^\prime} |$ over the past harmonic null cone $\cal C$
emanating from the field point $(t,{\bf x})$
(see Fig. 1).  This past null cone intersects the world tube $\cal D$
enclosing
the near zone at the three-dimensional hypersurface $\cal N$.
Thus the integral of Eq. (\ref{bigintegral}) consists of two pieces,
an integration over the hypersurface $\cal N$, and an integration
over the rest of the past null cone ${\cal C} - {\cal N}$.  Each of
these integrations will be treated differently.  We will also treat
slightly differently the two cases in which (a) the field point is
outside the near zone (Fig. 1), and (b) the field point is within
the near zone (Fig. 2).  The former case will be relevant for
calculating the gravitational-wave signal, while the latter will be
important for calculating field contributions to $\tau^{\alpha\beta}$
that must be integrated over the near zone, as well as for calculating
fields that enter the equations of motion.

\subsection{Radiation-zone field point, near-zone integration }
\label{sec: farnear}

{}For a field point in the  radiation zone, and integration over the
near zone, we first carry out the $t^\prime$ integration in Eq.
(\ref{bigintegral}), to obtain
\begin{eqnarray}
h_{\cal N}^{\alpha \beta} (t,{\bf x}) = && 4 \int_{\cal N} 
{ \tau^{\alpha \beta} (t -| {\bf x} - {\bf x^\prime} |, {\bf x^\prime} )
\over | {\bf x} - {\bf x^\prime} | } d^3x^\prime \;.
\label{nearintegral}
\end{eqnarray}
Within the near zone, the spatial integration variable ${\bf
x^\prime}$ satisfies $|{\bf x^\prime}| \le {\cal R} < r$, where the
distance to the field point $r=
|{\bf x}|$.
We now expand the $x^\prime$-dependence in the integrand
in powers of ${ |{\bf x^\prime }|/ r}$, using the fact that
\begin{eqnarray}
 | {\bf x} - {\bf x^\prime} |^q = \sum_{m=0}^\infty {1 \over {m!}}
(-x^\prime)^{i_1 \dots i_m} (r^q)_{,i_1 \dots i_m} \;.
\label{xprimeexpand}
\end{eqnarray}
We next expand $\tau^{\alpha\beta}$ in a Taylor series about the
retarded time $u \equiv t-r$.
The integration is now over the hypersurface $\cal M$, which is the
intersection of the near-zone world-tube with the constant-time
hypersurface $t_{\cal M}=u=t-r$ (see Fig. 3).
Roughly speaking, each term in the Taylor series is smaller than its
predecessor by a factor of order $v < 1$, thus for any hope of 
convergence of the
series, one must restrict attention to slow-motion sources. 
We now have an infinite series in $x^\prime$  (expansion of $| {\bf x}
- {\bf x^\prime} |^{-1}$) multiplying a double infinite series
(expansion of $| {\bf x} - {\bf x^\prime} |$ inside the Taylor
expansion).  
Grouping terms with the same
powers of $x^\prime$ and carrying out the appropriate combinatorics
(including use of ``Fa\`a di Bruno's formula''
\cite{stegun}), it is straightforward to show that 
\begin{eqnarray}
h_{\cal N}^{\alpha \beta} (t,{\bf x}) = &&4 \sum_{q=0}^\infty
{{(-1)^q} \over {q!}} \left ( {1 \over r} M^{\alpha\beta k_1 \dots
k_q} \right )_{,k_1 \dots k_q} \;,
\label{genexpand}
\end{eqnarray}
where
\begin{eqnarray}
M^{\alpha\beta k_1 \dots
k_q} (u) \equiv \int_{\cal M} \tau^{\alpha\beta} (u,{\bf x^\prime}) 
{x^\prime}^{k_1}
\dots {x^\prime}^{k_q} d^3 x^\prime \;.
\label{genmoment}
\end{eqnarray}
This general expansion, both in powers of $r^{-1}$ and in
retarded-time derivatives of $M^{\alpha\beta k_1 \dots
k_q} (u)$ will prove useful in later integrations of field quantities
over the far zone.  

However, for gravitational-wave detectors, we need only to focus
on the spatial components of $h^{\alpha\beta}$,
and on the leading component in $1/R$, where $R$ is the distance to
the detector.  Using the fact that $u_{,i} =
-\hat N^i$, 
where ${\bf \hat N} \equiv {\bf x}/R$ denotes the detector direction, 
we obtain
\begin{eqnarray}
h_{\cal N}^{ij}(t, {\bf x}) = {4 \over R} \sum_{m=0}^\infty {1 \over {m!}}
{\partial^m \over {\partial t^m}} \int_{\cal M} 
\tau^{ij} (u , {\bf x^\prime} ) ({\bf \hat N \cdot x^\prime}
)^m d^3x^\prime + O(R^{-2})  \;.
\label{series}
\end{eqnarray}

Because of the conservation law Eq. (\ref{conservation}), $\tau^{ij}$
satisfies the identities
\begin{mathletters}
\label{identities}
\begin{eqnarray}
\tau^{ij}&=&{1 \over 2}(\tau^{00}x^ix^j)_{,00}+2(\tau^{l(i}x^{j)})_{,l}
-{1 \over 2}(\tau^{kl}x^ix^j)_{,kl} \;, 
\label{identities1}\\
\tau^{ij}x^k&=&{1 \over 2}(2\tau^{0(i}x^{j)}x^k-\tau^{0k}x^ix^j)_{,0}
+{1 \over 2}(2 \tau^{l(i}x^{j)}x^k-\tau^{kl}x^ix^j)_{,l} \;.
\label{identities2}
\end{eqnarray}
\end{mathletters}
Using these identities in Eq. (\ref{series}) generates the multipole
expansion
\begin{eqnarray}
h_{\cal N}^{ij}(t, {\bf x}) = {2 \over R} {d^2 \over {dt^2}}
\sum_{m=0}^\infty  \hat N_{k_1} \dots \hat N_{k_m} I_{EW}^{ijk_1 \dots k_m}(u)
\;,
\label{EWseries}
\end{eqnarray}
where the ``Epstein-Wagoner'' (EW) moments are given by
\begin{mathletters}
\label{EWmoments}
\begin{eqnarray}
I_{EW}^{ij} &=& \int_{\cal M} \tau^{00}x^ix^j d^3x  + I_{EW({\rm
surf})}^{ij} \;, \label{EWmoment1}\\
I_{EW}^{ijk} &=& \int_{\cal M}(2\tau^{0(i}x^{j)}x^k-\tau^{0k}x^ix^j)d^3x
+ I_{EW({\rm surf})}^{ijk} \;, \label{EWmoment2}\\
I_{EW}^{ijk_1 \dots k_m} &=& {2 \over {m!}} {d^{m-2} \over {dt^{m-2}}}
\int_{\cal M} \tau^{ij}x^{k_1} \dots x^{k_m} d^3x \quad (m \ge 2) \;,
\label{EWmoment3}
\end{eqnarray}
\end{mathletters}
where integrating the spatial derivative terms in Eqs.
(\ref{identities}) by parts generates surface integrals at the two-dimensional
coordinate sphere of radius $\cal R$ bounding the 
hypersurface $\cal M$, denoted
$\partial {\cal M}$,
resulting in surface contributions to the first two EW moments
given by
\begin{mathletters}
\label{surf}
\begin{eqnarray}
(d/dt)^2I_{EW({\rm surf})}^{ij}= \oint_{\partial {\cal M}}
(4\tau^{l(i}x^{j)}-(\tau^{kl}x^ix^j)_{,k}) {\cal R}^2 \hat n^l d^2\Omega \;,
\label{surf1} \\
(d/dt)I_{EW({\rm surf})}^{ijk}= \oint_{\partial {\cal M}}
(2\tau^{l(i}x^{j)}x^k-\tau^{kl}x^ix^j) {\cal R}^2 \hat n^l d^2\Omega \;,
\label{surf2}
\end{eqnarray}
\end{mathletters}
where $\hat n^l$ denotes an outward radial unit vector, and $d^2\Omega$
denotes solid angle.

One advantage of this multipole expansion is that the field and source
variables appearing in the integrand $\tau^{\alpha\beta}$ are
evaluated at the single retarded time $u$; a disadvantage is that
because the field contributions to $\tau^{\alpha\beta}$ fall off as
some power of $r$, one can expect to encounter 
integrals that depend on positive powers of the radius
$\cal R$ of the boundary of integration, especially in some of the
higher-order moments.  If this boundary is
formally taken to $\infty$ (as was previously done), these integrals
would diverge.  However, as we shall see, such $\cal R$-dependent
effects are {\it precisely} cancelled by contributions from the integral
over the rest of the past null cone, to which we now turn.

\subsection{Radiation-zone field point, radiation-zone integration}
\label{sec: farfar}

The integral over the rest of the past null cone ${\cal C} - {\cal N}$
can be written in the form
\begin{eqnarray}
h_{{\cal C}-{\cal N}}^{\alpha\beta} (t,{\bf x}) = 
4 \int_{-\infty}^\infty du^\prime \int_{{\cal C}-{\cal N}} 
{ {\tau^{\alpha\beta} (t^\prime, {\bf x^\prime} ) 
  \delta( t^\prime - t + | {\bf x} - {\bf x^\prime} | )}
\over {| {\bf x} - {\bf x^\prime} | }} \delta (u^\prime -t^\prime
+r^\prime ) d^4x^\prime \;,
\end{eqnarray}
where we have simply inserted $1=\int du^\prime \delta (u^\prime -t^\prime
+r^\prime )$.  We now integrate over $t^\prime$ and $r^\prime$, and
note that
\begin{equation}
\int_{-\infty}^\infty dt^\prime \int_{\cal R}^\infty dr^\prime
\delta (u^\prime-t^\prime+r^\prime) \delta (t^\prime -t+| {\bf x}-{\bf
x^\prime} | ) = 
\left \{ \begin{array}{ll}
 {{| {\bf x} - {\bf x^\prime} |} \over {t-u^\prime-{\bf \hat n^\prime \cdot
x}}} & u^\prime <u \;{\rm and}\; r^\prime > {\cal R} \\
0 & u^\prime > u \; {\rm or} \; r^\prime < {\cal R} \;.
\end{array}
\right.
\label{2deltas}
\end{equation}
The result is
\begin{eqnarray}
h_{{\cal C}-{\cal N}}^{\alpha\beta} (t,{\bf x}) = 
4 \int_{-\infty}^u du^\prime \oint_{{\cal C}-{\cal N}} 
{ {\tau^{\alpha\beta} (u^\prime +r^\prime, {\bf x^\prime} )} 
\over {t-u^\prime- {\bf \hat n^\prime \cdot x} } } 
[r^\prime (u^\prime, \Omega^\prime)]^2  d^2 \Omega^\prime  \;.
\label{hrest}
\end{eqnarray}
Note that $r^\prime$ is a
function of $u^\prime$ and $\Omega^\prime$ via the condition [from the
two delta-functions in Eq. (\ref{2deltas})]: $t-u^\prime =r^\prime + 
| {\bf x} -
{\bf x^\prime} |$, which gives 
\begin{eqnarray}
r^\prime (u^\prime, \Omega^\prime ) = [(t-u^\prime )^2
-r^2]/[2(t-u^\prime- {\bf \hat n^\prime \cdot x})] \;.
\label{rprime}
\end{eqnarray}
The integration over solid angle $d^2 \Omega^\prime$ for a given value
of $u^\prime$, together with the $u^\prime+r^\prime$ ``time''
dependence of $\tau^{\alpha\beta}$, can be seen to represent an
integration over the two-dimensional
intersection of the past null cone $\cal C$ with
the future null cone $t^\prime=u^\prime +r^\prime$ emanating from the
center of mass of the system at $t_{\rm CM}=u^\prime$ (Fig. 4).  The 
integration
over $u^\prime$ then includes all such future-directed cones, starting from the
infinite past, and terminating in
the one emanating from the center of mass at time $u$, which is tangent to
the past null cone of the observation point.

However, for $u \ge u^\prime \ge u-2{\cal
R}$, the two-dimensional intersections meet the boundary of the near
zone, and so the angular integration is not complete.  If we choose the
field point $\bf x$ to be in the z-direction, so that ${\bf \hat n^\prime \cdot
x}=r \cos \theta^\prime$, then the condition $r^\prime \ge {\cal R}$,
together with Eq. (\ref{rprime}) imply that $0 \le \phi^\prime \le 2\pi$,
$1-\alpha \le \cos \theta^\prime \le 1$, where
\begin{eqnarray}
\alpha = (u-u^\prime)(2r-2{\cal R}+u-u^\prime)/2r{\cal R}  \;.
\label{alpha}
\end{eqnarray}
Note that $\alpha$ ranges from 0 ($u^\prime=u$) to 2 ($u^\prime =u-2{\cal
R}$).
{}For $u^\prime < u-2{\cal R}$, the angular integration covers the full
$4\pi$.  Thus we write the radiation-zone integral in the form
\begin{eqnarray}
h_{{\cal C}-{\cal N}}^{\alpha\beta} (t,{\bf x}) &=& 
4 \int_{u-2{\cal R}}^u du^\prime \int_0^{2\pi} d\phi^\prime
\int_{1-\alpha}^1
{ {\tau^{\alpha\beta} (u^\prime +r^\prime, {\bf x^\prime} )} 
\over {t-u^\prime- {\bf \hat n^\prime \cdot x} } } 
[r^\prime (u^\prime, \Omega^\prime)]^2  d \cos \theta^\prime
\nonumber \\
&&+ 4 \int_{-\infty}^{u-2{\cal R}} du^\prime \oint
{ {\tau^{\alpha\beta} (u^\prime +r^\prime, {\bf x^\prime} )} 
\over {t-u^\prime- {\bf \hat n^\prime \cdot x} } } 
[r^\prime (u^\prime, \Omega^\prime)]^2  d^2 \Omega^\prime \;.
\label{houter}
\end{eqnarray}
Note that $\tau^{\alpha\beta}$ contains only field contributions evaluated in
the radiation zone; in determining these we will make use of the general
expansion (\ref{genexpand}).

To obtain the contribution to the gravitational waveform, we evaluate
the spatial components of Eq. (\ref{houter}) at distance $R$ and
direction $\bf \hat N$ and keep
the leading $1/R$ part.

\subsection{Near-zone field point, near-zone integration}
\label{sec: nearnear}

In this case,
in Eq. (\ref{nearintegral}), both $\bf x$ and ${\bf x}^\prime$ are
within the near zone, hence $|{\bf x}-{\bf x}^\prime | \le 2{\cal R}$.
Consequently, the variation in retarded time can be treated as a small
perturbation, since $\tau^{\alpha\beta}$ varies on a time scale $\sim
{\cal R}$.  We therefore expand the retardation in powers of $|{\bf
x}-{\bf x}^\prime |$, to obtain
\begin{eqnarray}
h_{\cal N}^{\alpha \beta} (t,{\bf x}) = && 4 \sum_{m=0}^\infty {1
\over {m!}} {\partial^m \over {\partial t^m}} \int_{\cal M}
\tau^{\alpha \beta} (t,{\bf x}^\prime) 
 | {\bf x} - {\bf x^\prime} |^{m-1} d^3 x^\prime \;,
\label{nearexpand}
\end{eqnarray}
where $\cal M$ here denotes the intersection of the hypersurface $t=$
constant with the near-zone world-tube.

\subsection{Near-zone field point, radiation-zone integration}
\label{sec: nearfar}

The formulae from Section \ref{sec: farfar}, such as (\ref{rprime}) and
(\ref{alpha}), carry over to this case with
only one modification.  The final future null cone that appears in the
integration is the one that intersects the boundary of the near-zone
and the past null cone of the field point simultaneously at $u^\prime
=u-2{\cal R} +2r$, rather than $u^\prime =u$ (Fig. 5) (recall that here,
$r<{\cal R}$).  The result is, for a near-zone field point,
\begin{eqnarray}
h_{{\cal C}-{\cal N}}^{\alpha\beta} (t,{\bf x}) &=&
4 \int_{u-2{\cal R}}^{u-2{\cal R} + 2r} du^\prime \int_0^{2\pi} d\phi^\prime
\int_{1-\alpha}^1
{ {\tau^{\alpha\beta} (u^\prime +r^\prime, {\bf x^\prime} )}
\over {t-u^\prime- {\bf \hat n^\prime \cdot x} } }
[r^\prime (u^\prime, \Omega^\prime)]^2  d \cos \theta^\prime
\nonumber \\
&&+ 4 \int_{-\infty}^{u-2{\cal R}} du^\prime \oint
{ {\tau^{\alpha\beta} (u^\prime +r^\prime, {\bf x^\prime} )}
\over {t-u^\prime- {\bf \hat n^\prime \cdot x} } }
[r^\prime (u^\prime, \Omega^\prime)]^2  d^2 \Omega^\prime \;.
\label{houter2}
\end{eqnarray}

\subsection{Gravitational waveform and energy flux}

To obtain the gravitational 
waveform, we combine the two contributions to $h^{ij}$, 
Eqs. (\ref{EWseries}) and the leading $1/R$ part of
the spatial components of (\ref{houter}), 
and evaluate the transverse-traceless (TT) part, given
by 
\begin{eqnarray}
h_{TT}^{ij} = h^{kl} (P_k^i P_l^j - {1 \over 2} P^{ij} P_{kl}) \;,
\label{TT}
\end{eqnarray}
where $P_k^i = \delta_k^i -\hat N_k \hat N^i$.

Note that the two expressions that contribute to $h^{kl}$ in
Eq. (\ref{TT}) each
depend on the radius
$\cal R$ of the near zone.  Since $\cal R$ was an arbitrarily chosen
radius, the final physical answer should not depend on it.  However,
to check that all terms involving $\cal R$ cancel in the end would be
a  formidable task.  Instead we adopt the following non-rigorous, but
reasonable strategy.  
All
terms in the near-zone EW moments and in the radiation-zone integrals
that are {\it independent} of $\cal R$
are kept.  
All terms that {\it fall off} with $\cal R$ 
will be dropped.  Close examination shows that, despite our formal
choice ${\cal R} \sim {\lambda\!\!\!{\scriptscriptstyle{{}^{-}}}}$, 
nothing in our calculations actually constrains the
value of $\cal R$, apart from the inequality ${\cal R} < R$.  Thus we are
free to
make $\cal R$ sufficiently large, but still less than $R$,
so as to make such terms as small as we wish,
whether or not they ultimately cancel.
In this regard, it is useful to note that, for a LIGO/VIRGO detector 10
Mpc from a source emitting gravitational waves at $f=100$ Hz, $fR
= R/\lambda \sim 10^{16}$, and thus many orders of magnitude of $\cal R$ are
available to achieve this suppression.  Nevertheless, we believe that
all such terms actually cancel.  Finally,
all terms that {\it grow} with powers of $\cal R$ are also kept.  In
this case we
will show explicitly that all terms that vary as positive powers
of $\cal R$ cancel between the near-zone and radiation-zone integrals.  This
procedure thus isolates the finite terms that arise from convergent
integrals, while simultaneously verifying that no truly divergent integrals
arise.  The result is a well-defined, explicitly finite, method for
calculating the gravitational waveform.  It is the explicit inclusion of the
radiation-zone integral in the formulation of Eq. (\ref{houter}) that
cures the apparent divergences that plagued the original EW framework.

The energy flux is given by
\begin{eqnarray}
\dot E = (R^2/32\pi) \oint \dot h_{TT}^{ij}  \dot h_{TT}^{ij}
d^2\Omega \;.
\label{Edotformula}
\end{eqnarray}

\section{Weak field, slow-motion approximation}

\subsection{Iteration of relaxed Einstein equations}

We make the standard assumption that, with respect to the orbital
motion and mutual gravitational interactions, 
\begin{eqnarray}
v_A^2 \sim m_A/{\cal S} \sim \epsilon \ll 1 \;.
\end{eqnarray}
where $\epsilon$ will be used as an expansion parameter.  

Now,
because the field $h^{\alpha\beta}$ appears in the source of the field
equation, the usual method of solution is to iterate:  substitute 
$h^{\alpha\beta}=0$ in the right-hand side of Eq. (\ref{bigintegral})
and solve for the first-iterated $h_1^{\alpha\beta}$; substitute that
into Eq. (\ref{bigintegral}) and solve for 
the second-iterated $h_2^{\alpha\beta}$, and so
on (imposing the gauge condition Eq. (\ref{harmonic}) consistently at
each order).  The first iterated $h_1^{\alpha\beta}$ is $O(\epsilon)$, and
each subsequent iteration improves its accuracy by one order in
$\epsilon$.  Thus, for example,
to obtain a result for the waveform accurate to the
order of the quadrupole formula, $h \sim (m/r) \ddot I^{ij} \sim (m/r)
(v^2 + m/{\cal S}) \sim \epsilon^2$, {\it two} iterations of Eq.
(\ref{bigintegral}) are needed.  To obtain the first post-Newtonian
corrections to the quadrupole approximation, {\it i.e.} $h$ to order
$\epsilon^3$, $h_3^{\alpha\beta}$, or
three iterations, are needed, while to obtain the 2PN contributions
(the goal of this paper), the fourth-iterated field is needed.  This
would be a daunting task, if it weren't for the use of the identities,
Eqs. (\ref{identities}).  Consider for example, the quadrupole
formula.
The source $\tau^{ij}$ of the
second-iterated field $h_2^{ij}$ contains $\rho v^iv^j$ as well as terms
of the form $(\nabla h_1^{00})^2$, both of which are $O(\rho \times
\epsilon)$.  (Note that  $(\nabla h)^2 \sim h \nabla^2 h \sim \rho
\epsilon$).  However, the use of the identity Eq.
(\ref{identities1}) in the near-zone integration 
converts $\tau^{ij}$ into two time derivatives of
$\tau^{00}x^ix^j$ (modulo total divergences); because of the 
slow-motion approximation, two time derivatives increase the order by
$\epsilon$, and thus, to sufficient accuracy, only the dominant
contribution to $\tau^{00}$, namely $\rho$, is needed, {\it without}
explicit recourse to the
first-iterated $h_1^{\alpha\beta}$.
Instead, $h_1^{\alpha\beta}$ is
buried implicitly in the equation of motion (\ref{conservation}) that leads
to the identity 
(\ref{identities1}).  This circumstance is responsible for the
prevalent, but erroneous view that linearized gravity (one iteration)
suffices to derive the quadrupole formula.  The formula so derived 
turns out to
be ``correct'', but its foundation is not (see \cite{walkerwill} for
discussion).

Thus, in practice, in order to evaluate EW
moments required for the N-th
iterated field, we will only need the (N-2)-iterated field contributions
to the sources.  This is not precisely true for the two EW surface
integrals, and formally the full (N-1)-iterated field must be used in
$\tau^{ij}$ there, but with sufficient care, it can be shown without
detailed, explicit
calculations that the contributions of
the (N-1)-iterated fields all fall off sufficiently rapidly with $\cal R$
to have no effect on these surface integrals.  Similarly,  
for the radiation-zone integration, the
full (N-1)-iterated field must be used in $\tau^{ij}$, Eq.
(\ref{houter}).  However, it will also be possible to show that
the contributions of the these fields fall off with $\cal R$.  To
obtain the finite contributions and the contributions needed to
cancel any divergent terms from the EW moments, only the N-2 iterated
fields will be needed in practice.
Thus to 2PN order (fourth iteration), only second-iterated fields will
be
needed explicitly in the source terms.

\subsection{Second-iterated fields in source terms }
\label{sec2iterated}

Because the source contributions are integrated over all space, we
must evaluate the second-iterated fields $h_2^{\alpha\beta}$ in a form
that is valid everywhere (this and the following section follow the
approach and notation of BDI; see \cite{luc95}, for example).
The first iteration of the field equations (\ref{relaxed}) gives the
linearized equations, $\Box h_1^{\alpha\beta} =-16\pi T^{\alpha\beta}$.
Since $T^{\alpha\beta}$ has compact support, the solutions are
standard Lienard-Wiechert-type retarded functions.  
The solutions have the leading order behavior $h^{00} \sim
\epsilon$, $h^{0i}\sim
\epsilon^{3/2}$, $h^{ij}\sim
\epsilon^2$.  Taking these orders into account, together with the
fact that, because of the slow-motion assumption,
$\partial/\partial t \sim \epsilon^{1/2} \partial/\partial x^i$, we can
write the second-iterated field equations in the form (we drop the
subscripts)
\begin{mathletters}
\label{2iterated}
\begin{eqnarray}
\Box h^{00} &=& -16\pi (-g) T^{00} + {7 \over 8} 
h^{00}_{,k} h^{00}_{,k}  + O(\rho \epsilon^2) \;, \\
\Box h^{0i} &=& -16\pi (-g) T^{0i}+O(\rho \epsilon^{3/2}) \;, \\
\Box h^{ij} &=& -16\pi (-g) T^{ij} - {1 \over 4}( 
h^{00}_{,i} h^{00}_{,j} - {1 \over 2} \delta_{ij} 
h^{00}_{,k} h^{00}_{,k}) + O(\rho \epsilon^2) \;,
\end{eqnarray}
\end{mathletters}
where we have kept only contributions required to determine
$h^{00}$, $h^{0i}$, and $h^{ij}$ to the accuracies $\epsilon^2$, 
$\epsilon^{3/2}$,
and $\epsilon^2$, respectively (note that, in identifying orders of
source terms with dimension (length)$^{-2}$, we can use
$\Box^{-1} \rho \sim \epsilon$).
By defining the densities
\begin{mathletters}
\label{sigma}
\begin{eqnarray}
\sigma &\equiv& T^{00} +T^{ii} \;, \\
\sigma_i &\equiv& T^{0i} \;, \\
\sigma_{ij} &\equiv& T^{ij} \;,
\end{eqnarray}
\end{mathletters}
and the retarded potentials
\begin{mathletters}
\label{VW}
\begin{eqnarray}
V(t,{\bf x}) &\equiv& \int_{\cal C} 
{ {d^3x^\prime} \over {|{\bf x}-{\bf x^\prime}|}}
\sigma (t-|{\bf x}-{\bf x^\prime}|,{\bf x^\prime})  \;, \label{VW1}\\
V_i(t,{\bf x}) &\equiv& \int_{\cal C} { {d^3x^\prime} 
\over {|{\bf x}-{\bf x^\prime}|}}
\sigma_i (t-|{\bf x}-{\bf x^\prime}|,{\bf x^\prime})  \;, \label{VW2}\\
W_{ij}(t,{\bf x}) &\equiv& \int_{\cal C} 
{ {d^3x^\prime} \over {|{\bf x}-{\bf x^\prime}|}}
\left [ \sigma_{ij} + {1 \over {4\pi}} ( V_{,i}  V_{,j} -
{1 \over 2} \delta_{ij}  V_{,k}  V_{,k} ) \right ]
 (t-|{\bf x}-{\bf x^\prime}|,{\bf x^\prime})  \;, \label{VW3}
\end{eqnarray}
\end{mathletters}
it is straightforward to solve Eqs. (\ref{2iterated}) to the needed
order, with the result
\begin{mathletters}
\label{h2iterated}
\begin{eqnarray}
h^{00}&=&4V-4(W-2V^2)+O(\epsilon^3) \;, \label{h2iterated00}\\
h^{0i}&=&4V_i +O(\epsilon^{5/2}) \;, \\
h^{ij}&=&4W_{ij} +O(\epsilon^3) \;, 
\end{eqnarray}
\end{mathletters}
where $W=W_{ii}$.  It is useful to note that, although these forms of
$h^{\alpha\beta}$ are of 
sufficient accuracy in practice to be used 
in the effective sources for evaluating
the waveform to 2PN order, they are not sufficiently accurate for use in
the equations of motion that must also be specified consistently to 2PN
order.  The 2PN equations of motion require $h^{00}$ to $O(\epsilon^3)$
and $h^{0i}$ to $O(\epsilon^{5/2})$ ($h^{ij}$ is sufficiently accurate
as it stands).  However, as the 2PN equations of motion are well
known, we shall not undertake their derivation here, and will simply
use the published equations \cite{damour300,linc} when they are needed.

Because the source of $V$ and $V_i$ has compact support, the
integrals (\ref{VW1}) and (\ref{VW2}) can be evaluated simply for
field points within either the near zone or the radiation zone.  But because
the source of $W_{ij}$ contains both compact and non-compact support
pieces, it must be evaluated carefully, with proper attention paid to
contributions from the integration over the radiation-zone part of the null
cone.  The details will depend on the use
to which $W_{ij}$ is being put.  Evaluation of $W_{ij}$ is discussed 
in Appendix \ref{appW}.

When we calculate the EW moments, we shall need the field
contributions to $\tau^{\alpha\beta}$ evaluated at fixed retarded time
$u$ (on the hypersurface $\cal M$), and for field points with $r<{\cal
R}$.  We therefore expand the retardation $t-|{\bf x}-{\bf x^\prime}|$
as a perturbation of the potentials $V$, $V_i$
and $W_{ij}$ about $t=u$, with $|{\bf x}-{\bf x^\prime}|$ acting as
the expansion parameter [see Eq. (\ref{nearexpand})].  The results are
\begin{mathletters}
\label{VWapprox}
\begin{eqnarray}
V &=& U+ {1 \over 2} \partial_t^2 X +O(\epsilon^{5/2}) \;, \label{Vapprox}\\
V_i &=& U_i + O(\epsilon^{5/2}) \;, \\
W_{ij} &=& P_{ij} + (W_{ij})_{{\cal C}-{\cal N}} +O(\epsilon^{5/2}) \;, 
\end{eqnarray}
\end{mathletters}
where the ``instantaneous'' potentials are given by
\begin{mathletters}
\label{UP}
\begin{eqnarray}
U(u,{\bf x}) &\equiv& \int_{\cal M}
 { {d^3x^\prime} \over {|{\bf x}-{\bf x^\prime}|}}
\sigma (u,{\bf x^\prime})  \;, \label{UP1}\\
X(u,{\bf x}) &\equiv& \int_{\cal M}
  d^3x^\prime |{\bf x}-{\bf x^\prime}|
\sigma (u,{\bf x^\prime})  \;, \label{UP2}\\
U_i(u,{\bf x}) &\equiv& \int_{\cal M}
 { {d^3x^\prime} \over {|{\bf x}-{\bf x^\prime}|}}
\sigma_i (u,{\bf x^\prime})  \;, \label{UP3}\\
P_{ij}(u,{\bf x}) &\equiv& \int_{\cal M}
 { {d^3x^\prime} \over {|{\bf x}-{\bf x^\prime}|}}
\left [ \sigma_{ij} + {1 \over {4\pi}} ( U_{,i}  U_{,j} -
{1 \over 2} \delta_{ij}  U_{,k}  U_{,k} ) \right ]
 (u,{\bf x^\prime})  \;. \label{UP4}
\end{eqnarray}
\end{mathletters}
We have used the fact that, by virtue of the conservation of mass and
momentum at lowest order, $\partial_t \int \sigma d^3x \sim
\epsilon^{5/2}$ and $\partial_t \int \sigma_i d^3x \sim
\epsilon^3$.  We will drop the contribution from
the radiation-zone integral 
$(W_{ij})_{{\cal C}-{\cal N}}$, which falls off at least as fast as ${\cal
R}^{-2}$ (see Appendix \ref{appW}).  
Note that these potentials satisfy $U_{i,i}=-\dot U$,
$\nabla^2 X =2U$, $P_{ij,j}=-\dot U_i$.   

\subsection{Near zone metric, matter stress-energy, and effective gravitational
source}

In order to evaluate the components of the stress-energy tensor
$T^{\alpha\beta}$ to the necessary order, we need the components of
the near-zone metric to post-Newtonian order.  These are given from Eqs 
(\ref{hdefinition}) and
(\ref{h2iterated}) by
\begin{mathletters}
\label{g2iterated}
\begin{eqnarray}
g^{00}&=&-(1+2V+2V^2)+O(\epsilon^3) \;, \\
g^{0i}&=&-4V_i +O(\epsilon^{5/2}) \;, \\
g^{ij}&=&(1-2V)\delta^{ij} +O(\epsilon^2) \;, \\ 
(-g) &=& 1 +4V- 8(W-V^2) +O(\epsilon^3) \label{-g}\;.
\end{eqnarray}
\end{mathletters}
These equations, together with the distributional
definition (\ref{pointmass}) of the
stress-energy tensor yield, to the requisite order,
\begin{mathletters}
\label{sigmaPN}
\begin{eqnarray}
\sigma &=& \sum_A m_A \left [ 1-V+ {3 \over 2} v_A^2 \right . \nonumber \\
&&+ \left . {1 \over 2} V^2 + {1
\over 2} Vv_A^2 +4W + {7 \over 8} v_A^4 -4V_i v_A^i  + O(\epsilon^3)
\right ] 
\delta^3({\bf x}-{\bf x}_A ) \;, \\
\sigma_i &=& \sum_A m_A v_A^i \left [1-V+ {1 \over 2} v_A^2 +O(\epsilon^2)
\right ]
\delta^3({\bf x}-{\bf x}_A ) \;, \\
\sigma_{ij} &=& \sum_A m_A v_A^iv_A^j \left [1-V+ {1 \over 2} v_A^2 
+O(\epsilon^2) \right ]
\delta^3({\bf x}-{\bf x}_A ) \;, 
\end{eqnarray}
\end{mathletters}
where the potentials $V$, $V_i$ and $W$ are assumed to be evaluated at
${\bf x_A}$, excluding contributions of the A-th body itself (to avoid
infinite self-field terms).  The components of $T^{\alpha\beta}$ can
be easily constructed from these expressions.  

To the needed order, $\Lambda^{\alpha\beta}$ has the form
\begin{mathletters}
\label{lambda}
\begin{eqnarray}
\Lambda^{00} &=& -14  V_{,k}  V_{,k} + 16 \left [
-V \ddot V +  V_{,k} {\dot V}_k -2V_k  \dot V_{,k} 
+{5 \over 8} {\dot V}^2 \right .\nonumber \\
&&\left . + {1 \over 2}  V_{m,k} ( V_{m,k}
+3  V_{k,m}) +2  W_{,k} V_{,k} 
- W_{kl} V_{,kl} - {7 \over 2} V V_{,k} V_{,k} \right ] 
+O(\rho \epsilon^3)\;, 
\label{lambda00}\\
\Lambda^{0i} &=& 16 \left [ V_{,k} (  V_{k,i} -
 V_{i,k}) + {3 \over 4} \dot V V_{,i} \right ]
+O(\rho \epsilon^{5/2}) \;, 
\label{lambda0i} \\
\Lambda^{ij} &=& 4 \left ( V_{,i} V_{,j} - {1 \over 2}
\delta_{ij} V_{,k} V_{,k} \right ) + 16 \left [
2  V_{,(i} {\dot V}_{j)} - V_{k,i} V_{k,j}
-  V_{i,k} V_{j,k} \right . \nonumber \\
&& \left . + 2   V_{k,(i}  
V_{j),k} - \delta_{ij} ({3 \over 8} {\dot V}^2 + V_{,k}
{\dot V}_k - V_{m,k}  V_{[m,k]} ) \right ] +O(\rho \epsilon^3) \;,
\label{lambdaij}  
\end{eqnarray}
\end{mathletters}
where overdot denotes $\partial/\partial t$.  Notice the presence of the
cubically nonlinear terms in $\Lambda^{00}$, 
involving either $V \times W$ or $V^3$.

\section{Evaluation of Epstein-Wagoner moments}
\label{sec:ewmoments}

\subsection{Basic strategy}
\label{strategy}

The EW moments are integrals over a sphere of harmonic coordinate
radius $\cal R$ about the center of mass of the system, with all
variables entering the integrands to be evaluated at retarded time
$u=t-r$.  We substitute the matter stress-energy tensor
$T^{\alpha\beta}$, and the second-iterated fields evaluated in the
near-zone into Eqs. (\ref{EWmoments}).  
We expand all quantities to the PN order needed to achieve
a 2PN-accurate waveform.  Each volume integral will be split into a
``compact'' (C) piece involving integration of the compact-support
matter source, and a ``field'' (F) piece, involving integration of the
non-linear field contributions.  In $I_{EW}^{ij}$ and $I_{EW}^{ijk}$,
the two surface integrations at the
boundary radius $\cal R$ will involve only the field contributions,
and will require somewhat special treatment.

In integrating the field terms, we will frequently integrate by parts,
but will carefully evaluate and save the surface terms, 
using the identity
\begin{eqnarray}
\label{intbyparts}
\int_{\cal M} \partial_k F^{ij \dots m} d^3x = \oint_{\partial {\cal
M}} F^{ij \dots m}
|_{\cal R} \hat n^k {\cal R}^2 d^2 \Omega \;.
\end{eqnarray}
In order to simplify some of the integrations, 
we will frequently make a change of variables within integrals, in
order to place one of the bodies at the origin of the new variables,
for example ${\bf y} \equiv {\bf x}-{\bf x}_A$.  Even though $d^3y=d^3x$,
this shift has the
consequence that the region of integration ${\cal M}_x= \{ x^i| |{\bf x}|
\le {\cal R} \}$ 
will now appear in the new coordinates to be a region bounded by
$|{\bf y}|=|{\cal R}{\bf
\hat n}-{\bf x}_A|$, {\it i.e.} not centered at ${\bf y}=0$.  
It is much easier in practice
to integrate in $y$-coordinates
over a region ${\cal M}_y= \{ y^i| |{\bf y}|
\le {\cal R} \}$, which is shifted by ${\bf x}_A$ relative to the true
region of integration.  The two integrations can be
related by taking into account the appropriate surface integrals,
using the identity
\begin{eqnarray}
\int_{{\cal M}_x} f({\bf x}) d^3x &=& \int_{{\cal M}_y} g({\bf y}) d^3y
 - \oint_{\partial {\cal M}_y} g({\bf y}) {\bf \hat y \cdot x}_A {\cal R}^2 d^2
\Omega_y \nonumber \\
&&+ {1 \over 2}\oint_{\partial {\cal M}_y} 
{\bf x}_A \cdot {\bf \nabla } g({\bf y}) {\bf \hat y \cdot x}_A {\cal R}^2
d^2 \Omega_y + \dots \;,
\label{surfshift}
\end{eqnarray}
where $ g({\bf y}) \equiv  f({\bf y} +{\bf x}_A)$ and ${\bf \hat y}={\bf
y}/y$.  Again, we evaluate and save the surface terms.

In the end, we will only be interested in the physically measurable,
transverse-traceless (TT) components of the radiation-zone field $h^{ij}$.
We will therefore make frequent use of the identities, which follow
from the definition (\ref{TT}):
\begin{eqnarray}
(\delta^{ij})_{TT} =0 \;, \quad (\hat N^i B^j)_{TT} =0 \;,
\label{TT=0}
\end{eqnarray}
where ${\bf B}$ is arbitrary.  
These identities apply only to indices ``$i$'' and ``$j$'' appearing
in the components of the final waveform; we do not apply them to
fields which ultimately make up source terms.

In the field integrals, we will need explicit forms for the
instantaneous potentials (\ref{UP}) evaluated inside the near zone.  
To the needed order, they are given by
\begin{mathletters}
\label{UPnew}
\begin{eqnarray}
U(u,{\bf x}) &=& \sum_A {m_A^* \over {|{\bf x}-{\bf
x}_A|}} + O(\epsilon^3) \;, \label{UP1new}\\
X(u,{\bf x}) &=& \sum_A m_A |{\bf x}-{\bf x}_A| (1+O(\epsilon))
\;, \label{UP2new}\\
U_i(u,{\bf x}) &=& \sum_A {{m_A v_A^i} \over {|{\bf x}-{\bf
x}_A|}} + O(\epsilon^{5/2} )\;, \label{UP3new}\\
P(u, {\bf x}) &=& \sum_A {{m_A v_A^2} \over {|{\bf
x}-{\bf x}_A|}} +{1 \over 4} U^2  - {1 \over 2} \sum_{A \ne B} {{m_A m_B} 
\over {|{\bf x}-{\bf
x}_A||{\bf x}_A-{\bf x}_B|}} + O(\epsilon^3) \;, \label{UP4new}
\end{eqnarray}
\end{mathletters}
where $P \equiv P_{ii}$, and where
\begin{eqnarray}
m_A^* \equiv m_A \left (1 + {3 \over 2} v_A^2 - \sum_B m_B 
/|{\bf x}_A-{\bf x}_B| +
O(\epsilon^2) \right ) \;.
\label{m*}
\end{eqnarray}
Equation (\ref{UP4new}) can be easily obtained from Eq. (\ref{UP4})
(after contraction on $ij$) by integrating by parts, carefully
checking the vanishing of all surface terms.  Although the full
potential $P_{ij}$ appears (via $W_{ij}$) in $\Lambda^{00}$, we will
not need its explicit form, as the integration of that particular term
will be handled by a ``trick'' (see Appendix \ref{appcubic}).  Note 
that the so-called
``superpotential'' $X(u, {\bf x})$ is needed only to lowest order
because it always appears twice time-differentiated, {\it e.g.} in Eq.
(\ref{Vapprox}), and so its contribution is already
$O(\epsilon)$ relative to that of
$U$.

\subsection{The two-index moment $I_{EW}^{ij}$}

We write Eq. (\ref{EWmoment1}) in the form
\begin{eqnarray}
I_{EW}^{ij} = I_C^{ij} + I_F^{ij} +I_S^{ij} \;, 
\end{eqnarray}
where the three terms represent the compact (C), field (F) and surface
(S) contributions.
Substituting Eqs. (\ref{sigma}), (\ref{VWapprox}), (\ref{-g}),  
(\ref{sigmaPN}), and (\ref{UPnew}) into
$(-g)T^{00}$ and expanding through $O(\rho \epsilon^2)$, we obtain
\begin{eqnarray}
I_C^{ij} &=& \sum_A m_A x_A^{ij} \left ( 1+ {1 \over 2} v_A^2 +3 \sum_B
{m_B \over r_{AB}} \right ) +{3 \over 8} \sum_A m_A x_A^{ij} v_A^4 \nonumber \\
&&+ \sum_{AB} m_A m_B {x_A^{ij} \over r_{AB}} \left ( 2v_B^2 + {7
\over 2} v_A^2 - 4{\bf v}_A \cdot {\bf v}_B - {3 \over 2} ({\bf v}_B
\cdot {\bf \hat n}_{AB})^2 - \sum_C {m_C \over r_{BC}} \right .\nonumber \\
&&\left . \qquad + {7 \over 2} \sum_C {m_C
\over r_{AC}} - {3 \over 2} {\bf a}_B \cdot {\bf x}_{AB} \right ) 
+O(\epsilon^3) \times mx_A^2 \;,
\label{IijC}
\end{eqnarray}
where ${\bf x}_{AB} \equiv {\bf x}_A - {\bf x}_B$, $r_{AB} \equiv
|{\bf x}_{AB}|$, ${\bf \hat n}_{AB}
\equiv {\bf x}_{AB}/r_{AB}$, and ${\bf a}_A \equiv d^2 {\bf
x}_A/dt^2$.  All sums are assumed to exclude cases where a
denominator ({\it e.g.} $r_{BC}$) might vanish. 

To the required order for calculating $I_F^{ij}$, $\Lambda^{00}$ can
be written in terms of the instantaneous potentials,
\begin{eqnarray}
\Lambda^{00} &=&
 -14 U_{,k}U_{,k} + 16 \left ( -{7 \over 8} U_{,k} {\ddot X}_{,k}  -U
\ddot U +U_{,k} \dot U_k -2U_k \dot U_{,k} + {5 \over 8} \dot U^2
\right .
\nonumber \\
&& \left .+ {1 \over 2} U_{m,k}(U_{m,k} +3U_{k,m}) +2P_{,k} U_{,k}
- P_{km}U_{,km} - {7 \over 2} UU_{,k}U_{,k} \right ) 
+O(\rho \epsilon^3) 
\;.
\label{lambda00inst}
\end{eqnarray}
{}For the first term, the integral $-(14/16\pi)\int_{\cal M} U_{,k}U_{,k} 
x^{ij} d^3x$ is
straighforward: integrating twice by parts and showing that the
surface terms are proportional to ${\cal R} \delta^{ij}$, which has no
TT part, we are left with the integral $(14/16\pi) \int_{\cal M} U \nabla^2 U
x^{ij} d^3x= -(7/2)\sum_{AB} m_A^* m_B^* x_A^{ij}/r_{AB}$.  This
term is of 1PN and 2PN order via the PN contributions to $m^*$.  The next
term, $-(14/16\pi) \int_{\cal M} U_{,k} {\ddot X}_{,k}x^{ij} d^3x$ is already
of 2PN order.  We integrate once by parts to remove the derivative from
$U$. 
Using the fact
that $\nabla^2 X=2U$, we find a surface 
integral $ -(14/16\pi) \oint_{\partial {\cal M}} U
\ddot X_{,k} x^{ij} \hat n^k {\cal R}^2 d^2\Omega$, and  
the new integrals $(28/16\pi) \int_{\cal M} U
\ddot U x^{ij} d^3x + (28/16\pi)\int_{\cal M} U \ddot X_{,k} \delta^{k(i}
x^{j)} d^3x$.  The first of these volume integrals can be combined
with that arising from the third term in Eq. (\ref{lambda00inst}).
We substitute Eqs. (\ref{UP1new}) and (\ref{UP2new}), including a
delta-function term that arises in $\ddot U$ (see Appendix 
\ref{appderivatives}).
In the surface term, we expand the integrand in powers of
$r^{-1}$, and obtain $-(7/15) \sum_{AB} m_Am_B{\cal R}
(v_A^{ij} + x_A^{(i} a_A^{j)}) + O({\cal R}^{-1}) $.  We drop all terms
that fall off with increasing $\cal R$.  
In the volume integrals, for each term in the sums $\ddot U= \sum_A
\ddot U_A$ and $\ddot X_{,k} = \sum_A \ddot X_{A,k}$, we
change integration variables from ${\bf x}$ to ${\bf y} = {\bf x}-{\bf
x}_A$ so that, for a given $A$, the potentials $\ddot U_A$ 
and $\ddot X_{A,k}$ are centered
at the origin of the new ${\bf y}$ coordinate, while $U$ now takes the
form $\sum_B m_B/|{\bf y}+{\bf }x_{AB}|$.     
We calculate the surface contributions that result from this change of
variables using Eq. (\ref{surfshift}).  For example, 
the first integral then becomes
\begin{eqnarray}
\int_{\cal M} U
\ddot U x^{ij} d^3x &=& \sum_{AB}  m_Am_B 
\int_{{\cal M}_y} {1 \over {|{\bf y}+{\bf x}_{AB}|}}  \left [
y {\bf a}_A \cdot {\bf \hat y} - v_A^2 +3({\bf v}_A \cdot {\bf \hat y})^2 -
{{4\pi} \over 3} v_A^2 y^3 \delta^3 ({\bf y}) \right ] \nonumber \\
&&  \qquad \times \left (y^2
\hat y^{ij} +2y \hat y^{(i}x_A^{j)}+ x_A^{ij} \right ) y^{-3} d^3y \;.
\label{firstintegral}
\end{eqnarray}
We use the spherical harmonic expansion
\begin{eqnarray}
{1 \over {|{\bf y}+{\bf x}_{AB}|}} \equiv \sum_{l,m} {{4\pi} \over
{2l+1}} {{(-r_<)^l} \over {r_>^{l+1}}} Y_{lm}^* ({\bf \hat n}_{AB})
Y_{lm} ({\bf \hat y}) \;,
\end{eqnarray}
where $r_{<(>)}$ denotes the lesser (greater) of $r_{AB}$ and $y$,
express all products of unit vectors $\hat y^k$ in terms of
symmetric, trace-free (STF) products using Eqs. (\ref{STFformulae}), and
integrate over directions ${\bf \hat y}$, using the identity
\begin{eqnarray}
\sum_m \int Y_{lm}^* ({\bf \hat n}) Y_{lm} ({\bf \hat y}) \hat 
y^{<L^\prime>}
d^2\Omega_y \equiv \hat n^{<L>} \delta_{ll^\prime} \;,
\label{angleident}
\end{eqnarray}
(see Appendix \ref{appSTF})
where the superscript $<L>$ over a unit vector denotes an
$l$-dimensional STF product.  We then integrate over $y$, using the formula
\begin{eqnarray}
\int_0^{\cal R} {{r_<^l} \over {r_>^{l+1}}} y^q dy = {{(2l+1)r_{AB}^q
} \over {(l+q+1)(l-q)}} \left [ 1- {{l+q+1} \over {2l+1}} \left (
{{\cal R} \over {r_{AB}}} \right )^{q-l} \right ] \;. \quad (q-l \ne 0)
\label{radialintegral}
\end{eqnarray}
The result is a series of terms of three types: those
with non-vanishing TT part 
that are independent of $\cal R$ and linear in $\cal R$,
which we keep; terms with vanishing TT part which we discard
(regardless of their dependence on $\cal R$); and
terms that fall off with increasing $\cal R$, which we also discard.
An example of the second type of term would be a contribution to
$I^{ij}_{EW}$ proportional to ${\cal R} \delta^{ij}$.  The
contribution of such a term to $h^{ij}$ has
no TT part; equivalently, it can be eliminated to the necessary order
by a finite gauge
or coordinate transformation.

Many of the field integrations that we encounter
in evaluating the EW moments are amenable to this general method: 
(i) integrate by parts to leave one potential undifferentiated, (ii)
change variables to put the center of the differentiated potentials at
the origin, (iii) expand the undifferentiated potential in spherical
harmonics, (iv) express all unit vector products in STF terms, (v)
integrate over $d^3y$ using the identites (\ref{angleident}) and 
(\ref{radialintegral}), (vi)
retain all relevant contributions from surface integrals that arise in
steps (i) and (ii).  

Terms 2 through 8 contributed by $\Lambda^{00}$ [Eq (\ref{lambda00inst})]
can be handled using
this method, as can the compact contributions to $P_{ij}$ and $P$
(proportional to velocities) in terms 9 and 10.  However the
non-linear field contributions to $P_{ij}$ and $P$ lead to additional
complications, although the basic method still applies.  These terms
are discussed in Appendix \ref{appcubic}.  Finally, the cubically non-linear
term 11 can be calculated
easily by integrating by parts.  Computation 
of these terms is straightforward but
tedious.  In evaluating 2PN terms, we make
repeated use of the fact, valid to Newtonian order, that $\sum_A m_A
{\bf x}_A=0$.  

We now turn to the surface term $I_S^{ij}$, given by Eq.
(\ref{surf1}).  Because the surface lies outside the matter source,
only the field contribution, $\Lambda^{ij}$ is needed.  The term can
be rewritten in the form
\begin{eqnarray}
(d/dt)^2I_S^{ij}=(1/16\pi) \oint_{\partial {\cal M}}
\left ( 2\Lambda^{k(i} \hat n^{j)k} {\cal R}^3 - {\Lambda^{kl}}_{,l}
\hat n^{ijk}
{\cal R}^4 \right ) d^2 \Omega \;.  
\label{surfint}
\end{eqnarray}

However,
because $I_S^{ij}$ is essentially two anti-time-derivatives of
the surface integral, reducing its order by $\epsilon$, we need to
know $\Lambda^{ij}$ to $O(\rho \epsilon^3)$, {\it i.e.} to $O(\epsilon^2)$
beyond its leading order terms, at least in principle.
This is in contrast to having to know $\Lambda^{00}$ in the spatial
integral $I_F^{ij}$ only to $O(\epsilon)$ beyond its leading order.  
This would present considerable complications, except for the fact that we only
need to calculate a surface integral, and retain terms that are either
independent of or grow with $\cal R$.  Consequently we only need to
retain contributions to $\Lambda^{ij}$ that vary as ${\cal R}^{-2}$ or
${\cal R}^{-3}$.  To see what terms must be
retained, we return to the definition of $\Lambda^{ij}$, Eq. 
(\ref{nonlinear}). 
{}Far from the source, the fields $h^{\alpha\beta}$ have the
leading $\epsilon$ and $r$ dependences $h^{00} \sim \epsilon/r 
$, $h^{0i} \sim
\epsilon^{3/2}/r^2$ ($r^{-2}$ here
because the net momentum of the system vanishes),
and $h^{ij} \sim \epsilon^2/r$; $\Lambda^{ij}$
has the schematic form $(h_{,\lambda})^2 + h
(h_{,\lambda})^2 + h^2 (h_{,\lambda})^2 +\dots$.  By
combining the leading forms of $h^{\alpha\beta}$ with the knowledge
that time-derivatives increase the order by $\epsilon^{1/2}$, while
spatial derivatives either increase the rate of fall-off by one power of 
$r^{-1}$ or increase the order by $\epsilon^{1/2}$ via the retarded
time dependence, it can be
shown by inspection that terms of order $h(h_{,\lambda})^2$ and
higher are either of higher than 2PN order, or fall off faster than
${\cal R}^{-3}$, or generate angular dependence that leads to no TT
parts.
However, the purely quadratic terms proportional to $(h_{,\lambda})^2$ do
contribute; their explicit contribution is given by the non-linear terms of Eq.
(\ref{nonlinear}) with $g_{\mu\nu}$
replaced by $\eta_{\mu\nu}$.  Again, inspection shows that, to the
required order, we can write
\begin{eqnarray}
\Lambda^{ij} = -h^{00} \ddot h^{ij} + {1 \over 4} {h^{00}}_{,i }
{h^{00}}_{,j} +2 h^{00,(i} \dot h^{j)0} 
 - \delta^{ij} ({1 \over 8} {h^{00}}_{,k }{h^{00}}_{,k } +h^{00,k}
\dot h^{k0} ) \;.
\label{lambdaforsurf}
\end{eqnarray}
{}Further inspection shows that knowing $h^{\alpha\beta}$ to the
accuracy shown in Eq. (\ref{h2iterated}) suffices;  the
higher-order terms not explicitly shown in those expressions
contribute terms either at higher-than-2PN order, or at 
faster-than-${\cal R}^{-3}$ fall-off.  We do need to evaluate $V$, $V_i$ and
$W_{ij}$ carefully, however.  Expanding these functions in powers of 
$|{\bf x}-{\bf x^\prime}|$
about $t=u$, but to higher orders than that shown in Eq.
(\ref{VWapprox}), using Eqs. (\ref{sigmaPN}) for $\sigma$, $\sigma_i$
and $\sigma_{ij}$, and displaying only terms that lead to the appropriate
contributions in $\Lambda^{ij}$, we find in the vicinity of $r={\cal R}$
\begin{mathletters}
\label{VWforsurf}
\begin{eqnarray}
V &=&  {{\tilde m} \over r}  + {1 \over {4r}} \ddot Q^{kl}
(3 \delta^{kl}-n^kn^l) +O(\epsilon^3/r) 
- {2 \over 3}  \stackrel{(3)}{Q} + O(\epsilon^{3}r^0)
\nonumber \\
&&+ {r \over 16} \stackrel{(4)}{Q^{kl}}
(\delta^{kl}+n^kn^l) + O(\epsilon^4 r) + O(\epsilon^2/r^2) +
O(\epsilon^2/r^3) \;, \\
V_i &=& -{1 \over {2r^2}} (\epsilon^{ijk} J^k - \dot Q^{ij})n^j
+O(\epsilon^{5/2})/r^2 
- {1 \over 4} \stackrel{(3)}{Q^{ij}}n^j + O(\epsilon^3
r^0) + O(\epsilon^{3/2}/r^3)\;,  \\
W^{ij} &=& {1 \over {2r}} \ddot Q^{ij} +O(\epsilon^2/r^2) +
O(\epsilon^3/r) \;, 
\end{eqnarray}
\end{mathletters}
where we define here and for future use
\begin{mathletters}
\label{mQJ}
\begin{eqnarray}
\tilde m &\equiv&  m + E ;, \label{mQJ-m}\\
E &\equiv&  {1 \over 2} \sum_A (m_A v_A^2 - \sum_B
m_Am_B/r_{AB}) \;, \label{mQJ-E}\\
{\bf X} &\equiv& \tilde m^{-1} \sum_A m_A {\bf x}_A \left ( 1 + {1 \over 2}
v_A^2 - {1 \over 2} \sum_B m_B/r_{AB} \right )=0 \;, \label{mQJ-X} \\
Q^{ij} &\equiv& \sum_A m_A x_A^{ij} \;, \label{mQJ-Q}\\
Q^{ijk} &\equiv& \sum_A m_A x_A^{ijk} \;, \label{mQJ-Q2}\\
J^{i}&\equiv& \sum_A m_A \epsilon^{ilm} x_A^{l}v_A^{m} \label{mQJ-J} \;, \\
J^{ij}&\equiv& \sum_A m_A \epsilon^{ilm} x_A^{l}v_A^{m} x_A^j \label{mQJ-J2} 
\;, 
\end{eqnarray}
\end{mathletters}
where $m=\sum_A m_A$, and $Q=Q^{ii}$.  In Eq. (\ref{VWforsurf}) we
show schematically the $\epsilon$ order and the $r$ dependence of the
terms neglected.  Note that, by virtue of the Newtonian
equations of motion, $E$ and $J^{i}$ are constant to leading order.
Here $\tilde m$, $Q^{ij}$ and $J^{i}$ are to be evaluated at $u=t-R$.  
Combining Eqs. (\ref{VWforsurf}), (\ref{h2iterated}),
(\ref{lambdaforsurf}) and (\ref{surfint}), we find, to the required
order that $I_S^{ij} =
-(7/6)m{\cal R} \ddot Q^{ij}$.  

Combining $I_C^{ij}$, $I_F^{ij}$ and $I_S^{ij}$, we obtain finally
\begin{eqnarray}
I_{EW}^{ij} &=& \sum_A m_A x_A^{ij} \left ( 1 + {1 \over 2} v_A^2
-{1 \over 2} \sum_B {m_B \over r_{AB}} \right ) +  
{3 \over 8} \sum_A m_A x_A^{ij}
v_A^4 \nonumber \\
&&+ {1 \over 12} \sum_{AB} m_A x_A^{ij} {m_B \over r_{AB}} \biggl \{ 
28v_A^2
-11 v_B^2 -22{\bf v}_A \cdot {\bf v}_B  \nonumber \\
&& - ({\bf v}_A
\cdot {\bf \hat n}_{AB})^2 
+2 ({\bf v}_B \cdot {\bf \hat n}_{AB})^2 
-2{\bf v}_A \cdot {\bf \hat n}_{AB}{\bf
v}_B \cdot {\bf \hat n}_{AB}  \nonumber \\
&& +2 ({\bf a}_A+ {\bf a}_B)
 \cdot {\bf x}_{AB} +6 \sum_C {m_C \over r_{BC}} \biggr \}
\nonumber \\
&&- {1 \over 12} \sum_{AB} {{m_Am_B} \over r_{AB}} \biggl \{ 
{1 \over 2}[({\bf v}_A + {\bf v}_B )^2 - (({\bf v}_A + {\bf v}_B )
\cdot {\bf \hat n}_{AB} )^2 ] x_A^{(i}x_B^{j)} \nonumber \\
&& +2 ({\bf v}_A + {\bf v}_B ) \cdot {\bf x}_{AB}
(10 v_A^{(i}x_A^{j)} + 11v_A^{(i} x_B^{j)} ) -(26v_A^{ij}
-49v_A^{(i}v_B^{j)} )r_{AB}^2 \biggr \} \nonumber  \\
&&- {1 \over 12} \sum_{AB} m_Am_B r_{AB}
 \left \{ {\bf a}_A \cdot {\bf \hat n}_{AB} x_A^{(i}{\hat n}_{AB}^{j)} 
- a_A^{(i}x_A^{j)} +23 
(a_A+a_B)^{(i}x_A^{j)} 
\right \} \nonumber \\
&&-3 \sum_{AB} m_A^2 m_B \hat n_{AB}^{ij}
+{\cal G}_{(3)}^{ij} - {14 \over 5}m{\cal R} \ddot Q^{ij} +
O(\epsilon^3) \times Q^{ij} \;,
\label{Iijfinal}
\end{eqnarray}
where ${\cal G}_{(3)}^{ij}$ is a complicated 3-body term arising from
the $P_{km}U_{,km}$ term in Eq. (\ref{lambda00inst}), that vanishes
identically for two-body systems.  It is evaluated in Appendix \ref{appcubic}.

\subsection{The three-index moment $I_{EW}^{ijk}$} 

Since $I_{EW}^{ijk}$ is dominantly of 1/2PN order, we need to
calculate only the first post-Newtonian corrections to it, {\it i.e.} terms
of 3/2PN order. 
We first note that $I_{EW}^{ijk}$  [Eqs. (\ref{EWmoment2}) and
(\ref{surf2})] can be written
\begin{eqnarray}
I_{EW}^{ijk}=\tilde I_{EW}^{ijk} + \tilde I_{EW}^{jik} - \tilde
I_{EW}^{kij} \;, 
\label{tildeIijk}
\end{eqnarray}
where we separate $\tilde I_{EW}^{ijk}$ into compact, field and surface
contributions, given by
\begin{mathletters}
\label{Itilde}
\begin{eqnarray}
\tilde I_{C}^{ijk} + \tilde I_{F}^{ijk} &=& \int_{\cal M}
\tau^{0i}x^jx^kd^3x \;, \nonumber \\
(d/dt) \tilde I_{S}^{ijk} &=& (1/16\pi) \oint_{\partial {\cal M}}
\Lambda^{li} \hat n^{jkl} {\cal R}^4 d^2\Omega \;.
\end{eqnarray}
\end{mathletters}
Substituting Eqs. (\ref{sigma}), (\ref{Vapprox}), (\ref{-g}),
(\ref{sigmaPN}) and (\ref{UPnew}) into $(-g)T^{0i}$ and expanding
through $O(\rho \epsilon^{3/2})$, we obtain
\begin{eqnarray}
\tilde I_C^{ijk} =
\sum_A m_A v_A^i x_A^{jk} \left ( 1+ {1 \over 2} v_A^2 +3
\sum_B {m_B \over r_{AB}} \right ) + O(\epsilon^{5/2}) \times Q^{ij} \;.
\label{IijkC}
\end{eqnarray}
To the required order, 
\begin{eqnarray}
\Lambda^{0i}=16 \left [ U_{,k} (U_{k,i}-U_{i,k} ) + {3 \over 4} \dot U
U_{,i} \right ] \;.
\label{lambda0iinst}
\end{eqnarray}
We then calculate ${\tilde I}_F^{ijk}$ 
following the method laid out in
Sec. \ref{strategy}.
In the course of this calculation we find no TT terms dependent on
positive powers of $\cal R$.  Finally we evaluate the surface
contribution using Eqs. (\ref{lambdaforsurf}) and (\ref{VWforsurf}) 
evaluated to lowest order,
and find no contributions.  The final result is 
\begin{eqnarray}
\tilde I_{EW}^{ijk} &=& \sum_A m_A v_A^i x_A^{jk} \left ( 1+ {1 \over
2} v_A^2 - {1 \over 2} \sum_B {m_B \over r_{AB}} \right ) \nonumber
\\
&&-{1 \over 2}\sum_{AB} {{m_Am_B} \over r_{AB}} {\bf v}_A \cdot {\bf
\hat n}_{AB} \hat n_{AB}^i x_A^{jk} \nonumber \\
&&-{1 \over 12} \sum_{AB} m_Am_B r_{AB} \left [ 2{\bf v}_A \cdot {\bf
\hat n}_{AB} \hat n_{AB}^{ijk} + 11(2 v_A^i \hat n_{AB}^{jk} -v_A^j
\hat n_{AB}^{ik}-v_A^k \hat n_{AB}^{ij}) \right ] \nonumber \\
&&+{1 \over 2}  \sum_{AB} m_Am_B \left [ {\bf v}_A \cdot {\bf
\hat n}_{AB} \hat n_{AB}^{i(j} x_A^{k)} +7v_A^ix_A^{(j} \hat
n_{AB}^{k)} -7v_A^{(j}x_A^{k)} \hat
n_{AB}^{i} \right ] + O(\epsilon^{5/2}) \times Q^{ij} \;.
\label{Iijkfinal}
\end{eqnarray}
The result agrees with Eq. (A52) of \cite{magnum}.
The full moment $I_{EW}^{ijk}$ can be constructed from this using Eq.
(\ref{tildeIijk}).  

\subsection{The four-index moment $I_{EW}^{ijkl}$}

Since this moment contributes to the waveform already at PN order, we only
need to evaluate the integrands through their first PN corrections.
We write Eq. (\ref{EWmoment3}) in the form
\begin{eqnarray}
I_{EW}^{ijkl} = I_C^{ijkl} + I_F^{ijkl}  \;, 
\end{eqnarray}
(there is no surface contribution).
Expanding $(-g)T^{ij}$ through $O(\rho\epsilon^2)$, we obtain
\begin{eqnarray}
I_C^{ijkl} = \sum_A m_A v_A^{ij}x_A^{kl} \left ( 1+ {1 \over 2} v_A^2 +3 \sum_B
{m_B \over r_{AB}} \right ) + O(\epsilon^3) \times Q^{ij} \;.
\label{IijklC}
\end{eqnarray}
To the required order, $\Lambda^{ij}$ can be written in terms of the
instantaneous potentials
\begin{eqnarray}
\Lambda^{ij} &=& 4 (  U_{,i}U_{,j} - {1 \over 2}
\delta_{ij}  U_{,k}U_{,k} ) + 16 \left [ {1 \over 4} U_{,(i} \ddot
X_{,j)}  +
2 U_{,(i} \dot U_{j)}- U_{k,i}U_{k,j}
- U_{i,k}U_{j,k}
+2 U_{k,(i} U_{j),k} \right .\nonumber \\
&& \left . 
-\delta_{ij} ({1 \over 8} U_{,k} \ddot X_{,k}+ {3 \over 8} {\dot U}^2 
+ U_{,k} {\dot U}_k -U_{m,k}  U_{[m,k]} ) \right
] +O(\rho \epsilon^3) \;.
\label{lambdaijinst}
\end{eqnarray}
The term proportional to $\delta^{ij}$ produces no TT contributions to
the waveform, so we drop it.

The method proceeds as in the previous cases, without the
complications of cubic non-linearities.  The result is 
\begin{eqnarray}
I_{EW}^{ijkl} &=& \sum_A m_A \left ( v_A^{ij} - {1 \over 2} \sum_B
m_B \hat n_{AB}^{ij}/r_{AB} \right ) x_A^{kl} 
+{1 \over 12} \sum_{AB} m_Am_Br_{AB} \hat n_{AB}^{ij} (\hat n_{AB}^{kl}
-\delta^{kl} ) \nonumber \\
&& + {1 \over 2} \sum_A m_A v_A^2 v_A^{ij} x_A^{kl} \nonumber \\
&& -{1 \over 4} \sum_{AB} m_Am_B x_A^{kl}/r_{AB} \left ( 2 v_A^{ij}
+ (4v_{AB}^2 - v_B^2) \hat n_{AB}^{ij} 
\right .\nonumber
\\
&&\left .  - 3 ({\bf v}_A \cdot {\bf \hat n}_{AB})^2 \hat n_{AB}^{ij}
-4 (3{\bf v}_A \cdot {\bf \hat n}_{AB}-4{\bf v}_B \cdot {\bf \hat n}_{AB})
v_A^{(i} \hat n_{AB}^{j)} \right . \nonumber \\
&& \left .  -16a_B^{(i}x_{AB}^{j)}
- 2 a_A^{(i}x_{AB}^{j)}+  {\bf a}_A \cdot {\bf
x}_{AB} \hat n_{AB}^{ij} 
\right . \nonumber \\
&& \left . -2 \sum_C m_C (1/r_{AC} +1/r_{BC})  \hat n_{AB}^{ij} 
\right ) \nonumber \\
&& -{1 \over 24} \sum_{AB} m_Am_B r_{AB} \delta^{kl} \left ( 
(4v_{AB}^2-v_A^2)\hat n_{AB}^{ij} + 2 (8v_{AB}^{ij} -23
v_B^{ij}) \right . \nonumber \\
&& \left . - ({\bf v}_A \cdot {\bf \hat n}_{AB})^2 \hat n_{AB}^{ij}
-8 {\bf v}_{AB} \cdot {\bf \hat n}_{AB} v_{AB}^{(i}\hat n_{AB}^{j)}
+4 {\bf v}_{A} \cdot {\bf \hat n}_{AB} v_{A}^{(i}\hat n_{AB}^{j)}
\right . \nonumber \\
&& \left . -14  a_A^{(i} x_{AB}^{j)}
+ {\bf a}_A \cdot {\bf x}_{AB}\hat n_{AB}^{ij} 
- 4 \sum_C
(m_C/r_{AC})\hat n_{AB}^{ij} \right )
\nonumber \\
&& +{1 \over 24} \sum_{AB} m_Am_B r_{AB} \hat n_{AB}^{ijkl} \left ( 
4v_{AB}^2-5v_A^2 +9 ({\bf v}_A \cdot {\bf \hat n}_{AB})^2
-4 \sum_C m_C/r_{AC} - 3 {\bf a}_A \cdot {\bf
x}_{AB} \right ) \nonumber \\
&&+{1 \over 3} \sum_{AB} m_Am_B \hat n_{AB}^{(k}x_A^{l)} \left (
v_A^2\hat n_{AB}^{ij}
+ 10 v_A^{ij} + 4 ({\bf v}_A \cdot {\bf
\hat n}_{AB} -3{\bf v}_{AB} \cdot {\bf
\hat n}_{AB} ) v_A^{(i} \hat n_{AB}^{j)} \right . \nonumber \\
&&  \left . - 3({\bf v}_A \cdot {\bf
\hat n}_{AB})^2 \hat n_{AB}^{ij} 
-14 a_A^{(i}x_{AB}^{j)} + {\bf
a}_A \cdot {\bf x}_{AB} \hat n_{AB}^{ij}  \right ) \nonumber \\
&& +{1 \over 12} \sum_{AB} m_Am_B r_{AB}\hat n_{AB}^{ij}  \left ( v_A^{kl}
+2{\bf v}_{A} \cdot {\bf \hat n}_{AB} v_{A}^{(k}\hat n_{AB}^{l)}
- a_A^{(k}x_{AB}^{l)} +2 a_A^{(k}x_A^{l)} \right
) \nonumber \\
&& + {1 \over 12}\sum_{AB} m_Am_B r_{AB}\hat n_{AB}^{kl} \left ( 
 4v_{AB}^{ij}-21v_A^{ij}
+8 {\bf v}_{AB} \cdot {\bf \hat n}_{AB} v_{AB}^{(i}\hat n_{AB}^{j)}
\right . \nonumber \\
&& \left .
-6{\bf v}_{A} \cdot {\bf \hat n}_{AB} v_{A}^{(i}\hat n_{AB}^{j)}
 +35 a_A^{(i}x_{AB}^{j)} \right ) \nonumber \\
&&+{1 \over 3}\sum_{AB} m_Am_B r_{AB} \left (
  2v_{AB}^{(i} \hat n_{AB}^{j)} v_{AB}^{(k} \hat n_{AB}^{l)}
- v_A^{(i} \hat n_{AB}^{j)} v_A^{(k} \hat n_{AB}^{l)}
\right ) \nonumber \\
&& + {1 \over 3} \sum_{AB} m_Am_B \left ( 2 
v_{A}^{(i}\hat n_{AB}^{j)}v_{A}^{(k}x_A^{l)} - {\bf v}_{A} \cdot
{\bf \hat n}_{AB}\hat n_{AB}^{ij}v_{A}^{(k}x_A^{l)}
-12 v_{A}^{(i}\hat n_{AB}^{j)}v_{AB}^{(k}x_A^{l)}  \right ) \nonumber \\
&& -{8 \over 35} m{\cal R} \ddot Q^{ij} \delta^{kl} +O(\epsilon^3)
\times Q^{ij} \;.
\label{I4final}
\end{eqnarray}

\subsection{The five- and six-index moments $I_{EW}^{ijklm}$ and
$I_{EW}^{ijklmn}$}

These moments contribute to the waveform at 3/2PN and 2PN order,
respectively, thus we only
need to evaluate the dominant, Newtonian contributions to the
integrands.  Splitting the moments into a compact and a field piece,
substituting the lowest order contributions to $\tau^{ij}$ at
$O(\rho\epsilon)$, into Eq. ({\ref{EWmoment3}),
namely $(-g)T^{ij}= \sum_A m_A v_A^{ij} \delta^3
({\bf x} -{\bf x}_A)$, and $\Lambda^{ij}=
 4 (  U_{,i}U_{,j} - {1 \over 2}
\delta_{ij}  U_{,k}U_{,k} )$, and carrying out the integration
procedures as above, we obtain
\begin{mathletters}
\label{I56final}
\begin{eqnarray}
I_{EW}^{ijklm} &=& {1 \over 3} {d \over {dt}} \left \{ 
\sum_A m_A x_A^{klm} \left ( v_A^{ij}- {1 \over 2} \sum_B {m_B \over r_{AB}}
\hat n_{AB}^{ij} \right )  \right . \nonumber \\
&& \left . +{1 \over 4} \sum_{AB} m_Am_B r_{AB} \hat n_{AB}^{ij} x_A^{(k}
(\hat n_{AB}^{lm)} -\delta^{lm)} ) \right \}+O(\epsilon^{5/2}) \times
Q^{ij}  \;, \label{I5final} \\
I_{EW}^{ijklmn} &=& {1 \over 12} {d^2 \over {dt^2}} \biggl \{
\sum_A m_A x_A^{klmn} \left ( v_A^{ij}- {1 \over 2} \sum_B {m_B \over
r_{AB}}
\hat n_{AB}^{ij} \right )  \nonumber \\
&& 
+{1 \over 2} \sum_{AB} m_Am_B r_{AB} \biggl [ \hat n_{AB}^{ij} x_A^{(kl}
(\hat n_{AB}^{mn)} -\delta^{mn)} ) \nonumber \\
&& -{1 \over 10}x_{AB}^{ij} \left ( 2\hat n_{AB}^{klmn}-
2\hat n_{AB}^{(kl}\delta^{mn)}-\delta^{(kl}\delta^{mn)}
\right ) \biggr ] 
\nonumber \\
&& -{8 \over 105} m{\cal R} Q^{ij} \delta^{(kl}\delta^{mn)} \biggr \}
+O(\epsilon^3) \times Q^{ij} \;. \label{I6final}
\end{eqnarray}
\end{mathletters}
Equation (\ref{I5final}) is equivalent to Eq. (A53d) of \cite{magnum}.

\section{Evaluation of radiation-zone contributions}

We now turn to the evaluation of the contribution $h_{{\cal
C}-{\cal N}}^{ij}(t, {\bf x})$ given by the integral over the remainder of
the past light cone of the observer, Eq. (\ref{hrest}).  There is no
material source now, so $\tau^{ij}=\Lambda^{ij}/16\pi$.  
On the other hand, the time dependence in the integrand of Eq.
(\ref{hrest}) is not the simple fixed retarded time $u=t-R$ of the EW
moments. The $(u^\prime +r^\prime)$-dependence of $\tau^{ij}$ in Eq.
(\ref{hrest}) reflects the variation in retarded time along each
two-dimensional intersection of the past light cone of the event
$(t,{\bf x})$ with the future light cone of the event $(u^\prime,0)$.
However, $\tau^{ij}$ is a functional of retarded potentials, such as
$V$.  When evaluated at $u^\prime +r^\prime$, $V$ has the form
\begin{eqnarray}
V(u^\prime +r^\prime,{\bf x^\prime}) = \int {{d^3x^{\prime\prime}} \over
{|{\bf x^\prime}-{\bf x^{\prime\prime}}|}} \sigma(u^\prime +r^\prime
-|{\bf x^\prime}-{\bf x^{\prime\prime}}|,{\bf x^{\prime\prime}}) \;,
\label{Vu+r}
\end{eqnarray}
Notice that, because $|{\bf x^{\prime\prime}}| \ll {\cal R}$, while
$|{\bf x^{\prime}}| > {\cal R}$, we can approximate
\begin{eqnarray}
u^\prime +r^\prime
-|{\bf x^\prime}-{\bf x^{\prime\prime}}| &\approx& u^\prime + {\bf
\hat n^\prime} \cdot {\bf  x^{\prime\prime}} + 
\nonumber \\
&& + (2r^\prime)^{-1} [ ({\bf
\hat n^\prime} \cdot {\bf  x^{\prime\prime}})^2 - 
{r^{\prime\prime}}^2 ] +\dots \;,
\label{uprimeexpand}
\end{eqnarray}
where ${\bf \hat n^\prime}={\bf x^\prime}/r^\prime$, and then expand such
retarded functions about $u^\prime$ in powers of the small quantity
$r^{\prime\prime}/r^\prime$.  For a given $u^\prime$, the
retarded fields that contribute to 
$\Lambda^{ij}$ along the intersection between the two
light cones in Fig. 5 all have their source in the near zone, on
slices of the near zone world tube that pass through the center of
mass at time $u^\prime$.  The expansion (\ref{uprimeexpand}) simply
reflects the fact that, as one moves around the source in angle
(integration over $d^2\Omega$ in Eq. (\ref{hrest})), the orientation
of the slice of the near-zone world tube that generates the fields 
precesses (see Fig. 6).  

Since the ingredients of $\Lambda^{ij}$ are all fields evaluated in
the radiation zone, we can use expansions in powers of $1/r^\prime$, such as
those of Eq. (\ref{genexpand}).  The angular dependence of such expansions
can always be expressed in terms of STF products of radial unit vectors 
${\bf \hat n}^\prime$ (analogues of spherical harmonics).  
Thus 
$\Lambda^{ij}$ can be written, in the regime $r^\prime \gg {\cal R}$,
as a sequence of terms of the generic form $f_{N,l}^{ij} (u^\prime)
\hat n{^\prime}^{<L>} {r^\prime}^{-N}$. 
Then a change of variables
\begin{eqnarray}
\zeta \equiv (t-u^\prime)/r=1+(u-u^\prime)/r
\label{change}
\end{eqnarray}
puts Eq. (\ref{houter}) into
the form, for each $(N,l)$ term,
\begin{eqnarray}
h_{{\cal C}-{\cal N}}^{ij}(N,l) &=& 
\left ( {2 \over r} \right )^{N-2}\int_1^{1+2{\cal R}/r} {{d\zeta}
\over {(\zeta^2-1)^{N-2}}} f_{N,l}(u-r(\zeta-1)) \nonumber \\
&& \quad \times (4\pi)^{-1} \int_0^{2\pi}
d\phi^\prime \int_{1-\alpha}^1 \hat n{^\prime}^{<L>} (\zeta - {\bf
\hat n}^\prime \cdot {\bf \hat n} )^{N-3} d \cos \theta^\prime \nonumber \\
&&+\left ( {2 \over r} \right )^{N-2} \int_{1+2{\cal R}/r}^\infty {{d\zeta}
\over {(\zeta^2-1)^{N-2}}} f_{N,l}(u-r(\zeta-1)) \nonumber \\
&& \quad \times (4\pi)^{-1}  \int_0^{2\pi}
d\phi^\prime \int_{-1}^1 \hat n{^\prime}^{<L>} 
 (\zeta-{\bf \hat n}^\prime \cdot {\bf
\hat n} )^{N-3} d\cos \theta^\prime \;,
\label{houterzeta}
\end{eqnarray}
where ({\it cf.} Eq. (\ref{alpha})) $\alpha=
(\zeta -1)(\zeta+1-2{\cal R}/r )(r/2{\cal R})$.
We first carry out the
angular integrals, which yield $\hat n^{<L>}A_{N,l}(\zeta,\alpha)$,
where $A_{N,l}$ can be computed from Legendre polynomials $P_l(z)$ by
$A_{N,l}(\zeta,\alpha)={1 \over 2} \int_{1-\alpha}^1 P_l(z)
(\zeta-z)^{N-3}dz$  [see Appendix \ref{appSTF}, Eq. (\ref{ANl});
$\alpha=2$ corresponds to the full $4\pi$ angular integration].
Then, in the
$\zeta$-integration from 1 to $1+2{\cal R}/r$, we expand the retarded
time dependence of the $f_{N,l}$ about $u$, then integrate; this is
valid since ${\cal R} < r$.  In the integrals from $1+2{\cal R}/r$
to $\infty$, we integrate by parts numerous times, each time
increasing the number of time derivatives of $f_{N,l}$, stopping when
the result exceeds the PN order required.  The boundary terms that
arise are evaluated at
$\zeta=1+2{\cal R}/r$ and $\zeta=\infty$, corresponding to retarded
time $u-2{\cal R}$ and $-\infty$ respectively.  At the former boundary, we
again expand the functions about $u$; at the latter boundary the
contributions are assumed to be zero, which is equivalent to making
the usual and reasonable assumption that the source is not
extraordinarily dynamical in the infinite past.  

{}For the cases where the field point is inside the near-zone, Eq.
(\ref{houterzeta}) still applies, except that now $r<R$, and the
first $\zeta$ integral runs from $-1+2{\cal R}/r$ to  $1+2{\cal R}/r$
(Fig. 5).

In working to 2PN order, just
as in the case of the EW surface integrals, Eqs. (\ref{surf}), here we
must also evaluate the integrand $\Lambda^{ij}$ to
$O(\rho\epsilon^3)$.  Here, as before, it can be shown that only the twice
iterated fields are needed in practice.  This can be seen as follows.  
We are only interested in the $1/r$ part of the waveform.  Thus a
contribution to $\Lambda^{ij}$ that is already $O(\rho\epsilon^3)$ but
that falls off faster than ${r^\prime}^{-3}$ ($N>3$) can be dropped.  
This would apply to all terms that are
quartically non-linear and higher, such as terms of the form
$h^2(h_{,\lambda})^2$, which fall off as ${r^\prime}^{-6}$.  
Cubically non-linear terms of the form $h(h_{,\lambda})^2$ can also be
dropped;  at leading order, they are $O(\rho\epsilon^2)$, but fall off
as ${r^\prime}^{-5}$.  One might worry that by expanding $f_{N,l}^{ij}$ in
Eq. (\ref{houterzeta}) about $u$ (the value of retarded time at which all
contributions to the waveform are to be evaluated in the end), one
could reduce the rate of fall off by one power of $r$ for each
retarded time-derivative.  But each time-derivative either raises the
order of the term by $\epsilon^{1/2}$ or kills it outright via a
conservation law, such as
for the Newtonian potential $h \sim m/r$.  Thus the leading cubically
non-linear contribution turns out to be  of order
$\rho\epsilon^3/{r^\prime}^5$, which can be dropped.  Thus only
quadratically non-linear terms of the form $(h_{,\lambda})^2$ need to
be considered.  As before, a knowledge of 
$h^{\alpha\beta}$ to the
accuracy shown in Eq. (\ref{h2iterated}) suffices; higher-order terms
contribute terms of order $\rho\epsilon^3/{r^\prime}^4$.  However, we
must now be careful in evaluating the terms which {\it do} contribute.
{}For example a term of $O(\rho\epsilon)$ that falls off as
${r^\prime}^{-6}$, can, after three terms in the Taylor expansion of its
retarded time dependence about $u$ in powers of $r(\zeta-1)$, lead to
a $1/r$ contribution to the waveform at $O(\Box^{-1}\rho
\epsilon^{5/2})$, which is 3/2PN order beyond quadrupole order.  A term
of this form would arise from the cross term between the gradient of
the Newtonian potential $m/r$ and the Newtonian quadrupole potential
$\sim Q^{ij}/r^3$.  Similarly a $(\rho \epsilon) {r^\prime}^{-7}$ term would
contribute a 2PN contribution to the waveform.  Such a term would
arise from a cross term between the Newtonian potential and the
Newtonian octupole potential $\sim Q^{ijk}/r^4$.
A consequence of these considerations is that, in
expanding the second-iterated fields $h^{\alpha\beta}$, we must use
the general multipole expansions of Eq. (\ref{genexpand}), expanded through
octupole order.  This amounts to expanding $h^{00}$ through $q=3$,
$h^{0i}$ through $q=2$, and $h^{00}$ through $q=1$.  Evaluating the
integrals $M^{\alpha\beta k_1 \dots k_q}$ to the needed order, using
the general method for integrating over the near-zone hypersurface
$\cal M$ described in Sec. \ref{sec: farfar}, and adding any contributions to
$h^{\alpha\beta}$ from
the radiation-zone integrations (primarily from $W^{ij}$; see Appendix 
\ref{appW}),  
we obtain
\begin{mathletters}
\label{hforouter}
\begin{eqnarray}
h^{00} &=& 4m/r^\prime + 7(m/r^\prime)^2 
+ 2 \left \{ {r^\prime}^{-1}Q^{ij}(u^\prime) \right \}_{,ij} 
-  {2 \over 3}  \left \{ {r^\prime}^{-1}Q^{ijk}(u^\prime) \right \}_{,ijk} \;,
\nonumber \\ 
h^{0i} &=&  -2 \left \{ {r^\prime}^{-1} [ \dot Q^{ij}(u^\prime) 
-\epsilon^{ija}J^a(u^\prime) ] \right \}_{,j} +
{2 \over 3} \left \{ {r^\prime}^{-1} [ \dot Q^{ijk}(u^\prime) 
-2\epsilon^{ika}J^{aj}(u^\prime) ] \right \}_{,jk} \;, \nonumber \\
h^{ij} &=& (m/r^\prime)^2 \hat n{^\prime}^{ij} +
2 \ddot Q^{ij}(u^\prime)/r^\prime - {2 \over 3} \left \{
{r^\prime}^{-1} [ \ddot Q^{ijk}(u^\prime) 
-4\epsilon^{(i|ka} \dot J^{a|j)}(u^\prime) ] \right \}_{,k}\;,
\end{eqnarray}
\end{mathletters}
where $Q^{ij}$, $Q^{ijk}$, $J^a$ and $J^{aj}$ are defined in 
Eqs. (\ref{mQJ-Q}) -- (\ref{mQJ-J2}),
and where
the superscript notation $^{(i|a \dots k|j)}$ denotes symmetrization
only on $i$ and $j$.

To the required order, we then have
\begin{eqnarray}
\Lambda^{ij} = -h^{00} \ddot h^{ij} + {1 \over 4} {h^{00}}_{,i }
{h^{00}}_{,j} + {1 \over 2} {h^{00}}_{,(i}
{h^{kk}}_{,j)} +2 h^{00,(i} \dot h^{j)0} \;,
\label{lambdaforouter1}
\end{eqnarray}
with the result
\begin{eqnarray}
\Lambda^{ij} = {4m \over {r^\prime}^2}  &&\left [  \hat n{^\prime}^{<ijkl>}
\left ( {{15Q^{<kl>}} \over {r^\prime}^4} + {{15\dot Q^{<kl>}} \over
{r^\prime}^3} +  {{6 \ddot Q^{<kl>}} \over {r^\prime}^2} +
 {{\stackrel{(3)\qquad}{Q^{<kl>}}} \over {r^\prime}} \right )
\right . \nonumber \\
&&\left .
+ \hat n{^\prime}^{<k(i>}
\left ( {{18Q^{<j)k>}} \over {7{r^\prime}^4}} + {{18 \dot Q^{<j)k>}} \over
{7{r^\prime}^3}} -  {{18 \ddot Q^{<j)k>}} \over {7{r^\prime}^2}} -
 {{24\stackrel{(3)\qquad}{Q^{<j)k>}}} \over {7{r^\prime}}} \right )
\right . \nonumber \\
&&\left .
- \left ( {{6 \ddot Q^{<ij>}} \over {5{r^\prime}^2}} +
 {{6\stackrel{(3)\qquad}{Q^{<ij>}}} \over {5{r^\prime}}}
 +2\stackrel{(4)\qquad}{Q^{<ij>}} 
\right ) \right . \nonumber \\
&& \left .
 \hat n{^\prime}^{<ijklm>}
\left ( {{35Q^{<klm>}} \over {r^\prime}^5} + {{35\dot Q^{<klm>}} \over
{r^\prime}^4} +  {{15 \ddot Q^{<klm>}} \over {r^\prime}^3} +
 {10{\stackrel{(3)\quad\quad}{Q^{<klm>}}} \over {3r^\prime}^2} + 
{{\stackrel{(4)\quad\quad}{Q^{<klm>}}} \over {3r^\prime}} \right )
\right . \nonumber \\
&& \left .
+ \hat n{^\prime}^{<kl(i>}
\left ( {{25Q^{<j)kl>}} \over {3{r^\prime}^5}} + {{25 \dot Q^{<j)kl>}}
\over {3{r^\prime}^4}} -  
{{25 \stackrel{(3)\quad\quad}{Q^{<j)kl>}}} \over {9{r^\prime}^2}} -
 {{10\stackrel{(4)\quad\quad}{Q^{<j)kl>}}} \over {9{r^\prime}}} \right )
\right . \nonumber \\
&& \left . 
+ \hat n{^\prime}^{<kl(i>} \epsilon^{j)ka}
\left ( {{8 \dot J^{ak}} \over  {r^\prime}^3} +{{8 \ddot J^{ak}} \over
{r^\prime}^2} + {{8 \stackrel{(3)\quad}{J^{ak}}} \over  {3r^\prime}}
\right )
\right . \nonumber \\
&& \left .
- \hat n{^\prime}^{<ijl>} \epsilon^{kla}
\left ( {{4 \dot J^{ak}} \over  {r^\prime}^3} +{{4 \ddot J^{ak}} \over
{r^\prime}^2} + {{4 \stackrel{(3)\quad}{J^{ak}}} \over  {3r^\prime}}
\right )
\right . \nonumber \\
&& \left .
- \hat n{^\prime}^k
\left ( {{10 \ddot Q^{<ijk>}} \over {7r^\prime}^3} +
 {{10\stackrel{(3)\quad\quad}{Q^{<ijk>}}} \over {7r^\prime}^2} - 
{{4\stackrel{(4)\quad\quad}{Q^{<ijk>}}} \over {21r^\prime}} 
+ {{4\stackrel{(4)\quad}{Q^{ijk}}} \over {3r^\prime}} +
{2 \over 3}  {\stackrel{(5)\quad}{Q^{ijk}}} \right )
\right . \nonumber \\
&& \left .
+ \hat n{^\prime}^k \epsilon^{(i|ka}
\left ( {{8 \dot J^{a|j)}} \over  {5r^\prime}^3} +{{8 \ddot J^{a|j)}}
\over 
{5r^\prime}^2} + {{16 \stackrel{(3)\quad}{J^{a|j)}}} \over  {5r^\prime}}
+{8 \over 3} \stackrel{(4)\quad}{J^{a|j)}}
\right )
\right . \nonumber \\
&& \left .
- \hat n{^\prime}^{(i} \epsilon^{j)ka}
\left (  {{4 \dot J^{ak}} \over  {5r^\prime}^3} +{{4 \ddot J^{ak}}
\over
{5r^\prime}^2} - {{16 \stackrel{(3)\quad}{J^{ak}}} \over  {15r^\prime}}
\right )
\right ] \;.
\label{lambdaforouter2}
\end{eqnarray}
The terms in 
Eq. (\ref{lambdaforouter2}) are of the generic form $f_{N,l}(u^\prime)
\hat n{^\prime}^{<L>} {r^\prime}^{-N}$.  We substitute each such term
into Eq. (\ref{houterzeta}), integrate using the procedure outlined
above,
and keep only terms through 2PN order 
that fall-off as $1/r$.  Evaluating at the detector distance $R$,
we obtain, finally 
\begin{eqnarray}
h_{{\cal C}-{\cal N}}^{ij}(t,{\bf x}) &=& {{4m} \over R} \int_0^\infty ds
\stackrel{(4)}{Q^{ij}}(u-s) \left [ \ln \left ( {s \over
{2R+s}} \right ) + {11 \over 12} \right ] \nonumber \\
&&+  {{4m} \over 3R} \hat N^k \int_0^\infty ds
\stackrel{(5)\quad}{Q^{ijk}}(u-s) \left [ \ln \left ( {s \over
{2R+s}} \right ) + {97 \over 60} \right ] \nonumber \\
&&-  {{16m} \over 3R} \epsilon^{(i|ka} \hat N^k \int_0^\infty ds
\stackrel{(4)\quad}{J^{a|j)}}(u-s) \left [ \ln \left ( {s \over
{2R+s}} \right ) + {7 \over 6} \right ] \nonumber \\
&&+ {1912 \over 315} {m \over R} \stackrel{(4)}{Q^{ij}}(u) {\cal R} \;.
\label{houterfinal}
\end{eqnarray}
As with the calculation of EW moments, we discard terms that fall off
with increasing $\cal R$.  

The integrals involving the logarithm
of retarded time are the tail terms, and are in complete agreement with
\cite{luc95}, including the constants (11/12, 97/60, 7/6) added to the
logarithms.  Their origin is the backscatter of the outgoing
gravitational waves off the lowest-order, Schwarzschild-like, static
background curvature of the spacetime surrounding the source.  More
precisely, the logarithmic integrals can be seen to arise directly
from the term $- h^{00} \ddot h^{ij}$ in Eq. (\ref{lambdaforouter1}), which
represents a modification of the flat spacetime characteristics by the
potential $h^{00} \sim m/r$.
The first tail term, arising from the $2 d^4 Q^{<ij>}/du^4$ term in Eq.
(\ref{lambdaforouter2}), is actually of 3/2PN order, while the second
and third terms, arising from the ${2 \over 3} d^5 Q^{ijk}/du^5$ and 
$ {8 \over 3}d^4 J^{aj}/du^4$ terms in Eq. (\ref{lambdaforouter2}), 
are of 2PN order.  On the other hand, only the 3/2PN
tail term contributes to the energy flux at 2PN order, resulting in
the ``$4\pi$'' term for circular orbits in Eq. (\ref{edot}).
Notice that the tail terms show no dependence on the near-zone
boundary radius ${\cal R}$.  In 
the BDI
framework, the tail terms contain a scale $b$ which is associated with
a gauge transformation from harmonic coordinates to a set of radiative
coordinates used in that framework; physical results in the end do not
depend on $b$, and the tail effects in the two frameworks
are in complete agreement.

It is easy to see that the final term in Eq. (\ref{houterfinal}),
which depends linearly on $\cal
R$ {\it exactly cancels} the sum of the 
corresponding terms arising from the two-
four- and six-index EW moments [Eqs. (\ref{Iijfinal}), (\ref{I4final})
and (\ref{I6final})].  

Thus combining the contributions of the six EW moments to Eq.
(\ref{EWseries}) with these contributions 
gives the gravitational waveform, valid to 2PN order, for a general
N-body system.  {\it The waveform is explicitly finite, with no divergent
integrals or undefined terms.}   Henceforth, we shall not display any $\cal
R$-dependent terms.

\section{Reduction to two-body systems}

\subsection{Center of mass and equations of motion to 2PN order}

We now specialize to the case of two bodies.  Through 2PN order the
dynamics of two-body systems are well known.  The motion is governed
by a Lagrangian that admits a conserved total energy and angular
momentum, as well as a ``conserved'' center-of-mass definition.  We
define the system's center of mass $\bf X$ and the relative position
$\bf x$ by
\begin{mathletters}
\begin{eqnarray}
{\bf X} &\equiv& m^{-1} (m_1 {\bf x}_1 +m_2 {\bf x}_2 ) + {\bf f}^{(1)}({\bf
x}_1, {\bf x}_2) + {\bf f}^{(2)}({\bf x}_1, {\bf x}_2) +O(\epsilon^3)
\times {\bf X} \;, \\
{\bf x} &\equiv& {\bf x}_1 - {\bf x}_2 \;,
\end{eqnarray}
\end{mathletters}
where $m=m_1+m_2$, and ${\bf f}^{(1)}$ and  ${\bf f}^{(2)}$ denote 1PN
and 2PN corrections to the center-of-mass definition.  Inverting these
expressions and setting ${\bf X}=0$, we obtain
\begin{mathletters}
\label{x1x2}
\begin{eqnarray}
{\bf x}_1 &=& (m_2/m) {\bf x} - {\bf f}^{(1)} -{\bf f}^{(2)} +O(\epsilon^3)
\times {\bf x}_1 \;, \\
{\bf x}_2 &=& -(m_1/m) {\bf x} - {\bf f}^{(1)} -{\bf f}^{(2)}+O(\epsilon^3)
\times {\bf x}_2  \;.
\end{eqnarray}
\end{mathletters}
The only place the 2PN correction ${\bf f}^{(2)}$ could conceivably
be needed is in the
lowest-order quadrupole moment, but in this case it is straightforward
to show that it is not, since
\begin{eqnarray}
Q^{ij} =\sum_A m_A x_A^ix_A^j = \mu x^ix^j + m{f^{(1)}}^i{f^{(1)}}^j +
O(\epsilon^3) \times Q^{ij} \;,
\label{relquad}
\end{eqnarray}
where ${f^{(1)}}^i=-{1 \over 2} \eta (\delta m/m)(v^2-m/r)x^i$
(see, {\it e.g.} \cite{linc}), and where we define the two-body variables
$\mu =m_1m_2/m$ (reduced mass), $\eta=\mu/m$, $\delta m =m_1-m_2$, 
${\bf v}={\bf
v}_1-{\bf v}_2$, and $r=|{\bf x}|$.
The two-body equations of motion then take the effective one-body
relative form, through 2PN order:
\begin{eqnarray}
{\bf a} = {\bf a}_N + {\bf a}_{PN}^{(1)} + 
 {\bf a}_{SO}^{(3/2)}+ 
{\bf a}_{2PN}^{(2)}
+ {\bf a}_{SS}^{(2)} 
 + O({\bf a}^{(5/2)}) \;,
\label{motion}
\end{eqnarray}
where the subscripts denote the nature of the term, post-Newtonian
(PN), spin-orbit (SO), post-post-Newtonian (2PN), and spin-spin (SS); and
the superscripts denote the order in $\epsilon$.  The individual terms
(excluding spins) are given by
\begin{mathletters}
\label{eom}
\begin{eqnarray}
{\bf a}_N = && - {m \over r^2} {\bf \hat n} \,, \label{eomN}\\
{\bf a}_{PN}^{(1)} = && - {m \over r^2} \biggl\{  {\bf \hat n} \left[
-2(2+\eta)
{m \over r} + (1+3\eta)v^2 - {3 \over 2} \eta \dot r^2 \right]
  -2(2-\eta) \dot r {\bf v} \biggr\} \,, \label{eomPN}\\
{\bf a}_{2PN}^{(2)} = && - {m \over r^2} \biggl\{ {\bf \hat n} \biggl[ {3
\over 4}
(12+29\eta) ( {m \over r} )^2 + \eta(3-4\eta)v^4 + {15 \over 8}
\eta(1-3\eta)
\dot r^4 \nonumber \\
 && - {3 \over 2} \eta(3-4\eta)v^2 \dot r^2 
- {1 \over 2} \eta(13-4\eta) {m \over r} v^2 - (2+25\eta+2\eta^2)
{m \over
r} \dot r^2 \biggr] \nonumber \\
 && - {1 \over 2} \dot r {\bf v} \left[ \eta(15+4\eta)v^2 -
(4+41\eta+8\eta^2)
{m \over r} -3\eta(3+2\eta) \dot r^2 \right] \biggr\} \,. \label{eom2PN}
\end{eqnarray}
\end{mathletters}

\subsection{Two-body Epstein-Wagoner Moments}

Restricting the summations in the EW moments to two bodies and
substituting Eqs. (\ref{x1x2}),  we obtain, through 2PN order,
\begin{mathletters}
\label{2bodyEW}
\begin{eqnarray}
I_{EW}^{ij} &=& \mu x^{ij} \left [ 1 + {1 \over 2} (1-3\eta)v^2
-{1 \over 2}(1-2\eta)m/r \right ] 
\nonumber \\
&&  + \mu x^{ij} \left [ {3 \over 8} (1-7\eta+13\eta^2)v^4 + {1 \over
12}(28-79\eta-54\eta^2)v^2(m/r) 
\right. \nonumber \\
&& \left . -{1 \over 4}(5+27\eta-4\eta^2)(m/r)^2
-{1 \over 12}(1-13\eta+30\eta^2)\dot r^2(m/r) \right ]
\nonumber \\
&& + \mu mr \left [{1 \over 6} (13+23\eta) v^{ij} - {5 \over 3}(1 -4\eta)
\dot r v^{(i} \hat n^{j)}\right ] \;,  \label{2bodyEW2}\\
I_{EW}^{ijk} &=& \mu (\delta m/m) \left \{ x^{ij}v^k -2 v^{(i}x^{j)k}
-v^{(i}x^{j)k} \left [ (1-5\eta)v^2+{1 \over 3}(7+12\eta)(m/r) \right
] \right . \nonumber \\
&& \left . + {1 \over 2}x^{ij}v^k \left [ (1-5\eta)v^2+{1 \over
3}(17+12\eta)(m/r) \right ] + {1 \over 6}(1-6\eta) (m \dot r /r^2)
x^{ijk} \right \} \;,  \label{2bodyEW3}\\
I_{EW}^{ijkl} &=& \mu x^{kl}(1-3\eta)(v^{ij}- {1 \over 3}{\hat
n}^{ij}m/r)  - {1 \over 6}\mu mr {\hat n}^{ij} \delta^{kl} 
\nonumber \\
&& + \mu x^{kl} \left [ {1 \over 2}(1-9\eta+21\eta^2)v^2v^{ij}- {1 \over
24}(13-46\eta+36\eta^2)v^2{\hat n}^{ij}m/r 
\right . \nonumber \\
&& \left . +{1 \over
4}(7-10\eta-36\eta^2)v^{ij}m/r 
+{1 \over 6}(7-12\eta-36\eta^2) \dot r v^{(i}{\hat
n}^{j)}m/r \right . \nonumber \\
&& \left.  +{1 \over 8}(1-6\eta+12\eta^2) \dot r^2 {\hat n}^{ij}m/r
+{1 \over 24}(37-122\eta+48\eta^2){\hat n}^{ij}(m/r)^2 \right ]  
\nonumber \\
&&+ \mu mr \delta^{kl}\left [ {1 \over 12}(7-46\eta) v^{ij} -{1 \over
24}(7+2\eta)v^2 {\hat n}^{ij} +{1 \over 6}(3+2\eta) \dot r v^{(i}{\hat
n}^{j)} \right . \nonumber \\
&& \left .  +{1 \over 24}(1-2\eta) \dot r^2 {\hat n}^{ij} - {3 \over 8}
{\hat n}^{ij}m/r  \right ]  \nonumber \\
&& +\mu mr \left [ {1 \over 12}(1-2\eta) {\hat n}^{ij} v^{kl} -{1
\over 6} (1-4\eta) \dot r {\hat n}^{ij} v^{(k} {\hat n}^{l)} -{1 \over
3}(7-20\eta) v^{(i} {\hat n}^{j)}v^{(k} {\hat n}^{l)} \right ] \;, 
\label{2bodyEW4}\\
I_{EW}^{ijklm} &=& -{1 \over 3} (d/dt) \mu (\delta m/m) \left [
(1-2\eta)(v^{ij} -{1 \over 4}{\hat n}^{ij}m/r )x^{klm} -{ 1 \over 4}mr
{\hat n}^{ij} x^{(k} \delta^{lm)} \right ] \;,  \label{2bodyEW5}\\
I_{EW}^{ijklmn} &=& {1 \over 12} \mu (d/dt)^2 \left [
(1-5\eta+5\eta^2) (v^{ij}-{1 \over 5}{\hat n}^{ij}m/r )x^{klmn} \right . 
\nonumber \\
&&  \left . - {1 \over 10}(3-10\eta) mr{\hat n}^{ij} x^{(kl} \delta^{mn)} +{1
\over 10} mr x^{ij} \delta^{(kl}\delta^{mn)} \right ] 
\label{2bodyEW6}\;.
\end{eqnarray}
\end{mathletters}
In addition, the moments that appear in the tail terms, Eq.
(\ref{houterfinal}), reduce, to the required order, to 
\begin{eqnarray}
Q^{ij} &=& \mu x^{ij} \;, \\
Q^{ijk} &=& -\mu (\delta m/m) x^{ijk} \;, \\
J^{aj} &=& -\mu (\delta m/m) ({\bf x} \times {\bf v})^a x^j \;.
\label{tailmoments}
\end{eqnarray}

\subsection{Two-body gravitational waveform and energy loss }
\label{sec:waveform}

Substituting the EW 2-body moments (\ref{2bodyEW}) into Eq.
(\ref{EWseries}), calculating the time derivatives using the 2PN
equations of motion (\ref{eom}) to the accuracy needed, and adding
the tail terms from the radiation zone integral (\ref{houterfinal}), 
we obtain the
gravitational waveform.  An alternative method is first to calculate
the so-called ``symmetric trace-free'' (STF) moments defined by Thorne
\cite{thorne80} and used by BDI, and then to calculate the waveform.  The
procedures and formulae needed to do this are given in Appendix 
\ref{appSTFdecomp}.
The result for the waveform is
\begin{eqnarray}
h^{ij} = && {{2\mu} \over R} \biggl[ \tilde Q^{ij} + P^{1/2}Q^{ij} 
+ PQ^{ij} + PQ_{SO}^{ij} + P^{3/2}Q^{ij}
 + P^{3/2}Q_{TAIL}^{ij} \nonumber \\
&& + P^{3/2}Q_{SO}^{ij} + P^{2}Q^{ij} + P^{2}Q_{TAIL}^{ij}
+ P^{2}Q_{SO}^{ij}+ P^{2}Q_{SS}^{ij}
+ O(\epsilon^{5/2}) \biggr]_{TT} \;,
\label{hanswer}
\end{eqnarray}
where, as before, the superscripts denote the effective PN order, 
and subscripts label the nature of the term, and where the
individual non-spin pieces are given by
\begin{mathletters}
\begin{eqnarray}
\tilde Q^{ij} = &&  2 \left ( v^iv^j - {m \over r}{\hat n}^i{\hat n}^j \right )
\,, \label{hpieces0} \\
P^{1/2}Q^{ij} = && {{\delta m} \over m} \left \{ 3({\bf \hat N \cdot \hat n})
{m \over
r} \left[ 2{\hat n}^{(i}v^{j)} - \dot r {\hat n}^i{\hat n}^j \right] + 
({\bf \hat N \cdot
v})
\left[ {m \over r} {\hat n}^i{\hat n}^j - 2v^iv^j \right] \right \} \,, \\
PQ^{ij} = && {1 \over 3} \biggl\{ (1-3\eta) \biggl[
({\bf \hat N \cdot \hat n}
)^2 {m \over r} \left[ (3v^2 - 15\dot r^2 + 7{m \over r}){\hat n}^i{\hat
n}^j +
30\dot r {\hat n}^{(i}v^{j)} - 14v^iv^j \right] \nonumber \\ &&
+ ({\bf \hat N \cdot \hat n})({\bf \hat N \cdot v}){m \over r} \left[
12\dot r {\hat n}^i{\hat n}^j - 32{\hat n}^{(i}v^{j)} \right] + 
({\bf \hat N \cdot
v})^2
\left[ 6v^iv^j - 2{m \over r} {\hat n}^i{\hat n}^j \right] \biggr] \nonumber \\
&& + \left[ 3(1-3\eta)v^2 - 2(2-3\eta){m \over r} \right] v^iv^j +
4 {m \over r} \dot r (5+3\eta){\hat n}^{(i}v^{j)} \nonumber \\
&& + {m \over r} \left[ 3(1-3\eta)\dot r^2 - (10+3\eta)v^2 + 29 {m
\over r}
\right] {\hat n}^i{\hat n}^j \biggr\} \,, \\
P^{3/2}Q^{ij} = && {{\delta m} \over m} (1-2\eta) \biggl\{ ({\bf \hat N
\cdot \hat n})^3 {m \over r} \biggl[ {5 \over 4}(3v^2-7\dot r^2 +6{m
\over r})
\dot r {\hat n}^i{\hat n}^j - {17 \over 2} \dot r v^iv^j  \nonumber \\ &&
- {1 \over 6}(21v^2-105\dot r^2 +44{m \over r}){\hat n}^{(i}v^{j)}
\biggr] \nonumber \\
&& +{1 \over 4} ({\bf
\hat N \cdot \hat n})^2({\bf \hat N \cdot v}) {m \over r} \biggl[
58v^iv^j + (45\dot r^2
-9v^2 -28{m \over r}){\hat n}^i{\hat n}^j -108 \dot r {\hat n}^{(i}v^{j)}
 \biggr] \nonumber \\ &&
+ {3 \over 2}({\bf \hat N \cdot \hat n})({\bf \hat N \cdot
v})^2 {m \over r} \biggl[ 10{\hat n}^{(i}v^{j)}
- 3 \dot r {\hat n}^i{\hat n}^j \biggr]
+ {1 \over 2} ({\bf \hat N \cdot v})^3 \left[ {m \over r}{\hat n}^i{\hat
n}^j -
4v^iv^j \right] \biggr\}  \nonumber \\ && +
 {1 \over 12} {{\delta m} \over m} ({\bf \hat N \cdot \hat n}){m
\over r}
\biggl[ 2{\hat n}^{(i}v^{j)} \left( \dot r^2 (63+54\eta) - {m \over
r}(128-36\eta)
+ v^2(33-18\eta) \right) \nonumber \\ && + {\hat n}^i{\hat n}^j\dot r 
\left( \dot
r^2(15
-90\eta)-v^2(63-54\eta)+{m \over r}(242-24\eta) \right)
-\dot r v^iv^j(186+24\eta) \biggr] \nonumber \\ &&
+ {{\delta m} \over m} ({\bf \hat N \cdot v}) \biggl[
{1 \over 2}v^iv^j \left( {m \over r}(3-8\eta)-2v^2(1-5\eta) \right)
-{\hat n}^{(i}v^{j)}{m \over r}\dot r (7+4\eta) 
\nonumber \\
&&
- {\hat n}^i{\hat n}^j{m \over r}\biggl( {3 \over 4}(1-2\eta)\dot r^2  
+ {1 \over 3}(26-3\eta){m \over r} - {1 \over 4}(7-2\eta)v^2 \biggr)
\biggr] \,, \\
P^{3/2}Q_{TAIL}^{ij}= && 4m \int_0^\infty \biggl\{ {m \over r^3} \left [
(3v^2+{m \over r}-15 \dot r^2 ){\hat n}^i{\hat n}^j+18 \dot r 
{\hat n}^{(i}v^{j)}-4v^iv^j
\right ] \biggr\}_{u-s} \nonumber \\
&& \times \left [ \ln \left ( {s \over {2R+s}}
\right ) + {11 \over 12} \right ] ds  \;, \\
P^{2}Q^{ij} = && {1 \over 60 } (1-5\eta+5\eta^2)
\biggl\{ 
24({\bf \hat N \cdot v})^4 
\biggl[ 5 v^i v^j - {m \over r} {\hat n}^i {\hat n}^j \biggr] 
\nonumber \\
&& 
+{m \over r} ({\bf \hat N \cdot \hat n})^4
\biggl[ 2 \left( 175 {m \over r} - 465 \dot r^2 + 93 v^2 \right) v^i v^j
\nonumber \\
&&
+ 30 \dot r \left( 63 \dot r^2 - 50{m \over r} - 27 v^2 \right) {\hat
n}^{(i}v^{j)}
\nonumber \\
&& 
+ \left(1155 {m \over r} \dot r^2 - 172 \left({m \over r}\right)^2 
- 945 \dot r^4 - 159 {m \over r} v^2 
+ 630 \dot r^2 v^2 - 45 v^4 \right) {\hat n}^i {\hat n}^j
\biggr]
\nonumber \\
&& 
+24 {m \over r} ({\bf \hat N \cdot \hat n})^3 ({\bf \hat N \cdot v}) 
\biggl[ 87 \dot r v^i v^j 
+ 5 \dot r \left( 14 \dot r^2 - 15 {m \over r} - 6v^2 \right) {\hat
n}^i {\hat n}^j
\nonumber \\
&& 
+ 16 \left( 5 {m \over r} - 10 \dot r^2 + 2v^2 \right) 
{\hat n}^{(i} v^{j)} \biggr]
+288 {m \over r} ({\bf \hat N \cdot \hat n}) ({\bf \hat N \cdot v})^3
\biggl[ \dot r {\hat n}^i {\hat n}^j - 4 {\hat n}^{(i} v^{j)} \biggr]
\nonumber \\
&& 
+24 {m \over r} ({\bf \hat N \cdot \hat n})^2 ({\bf \hat N \cdot v})^2
\biggl[ \left( 35 {m \over r} - 45 \dot r^2 + 9 v^2 \right) {\hat n}^i {\hat
n}^j
               - 76 v^i v^j + 126 \dot r {\hat n}^{(i} v^{j)}
\biggr]  
\biggr\}
\nonumber \\
&& 
+ {1 \over 15} ({\bf \hat N \cdot v})^2
\biggl\{
            \biggl[ 5 ( 25-78\eta+12\eta^2 ) {m \over r}
                    - (18 - 65 \eta + 45 \eta^2 ) v^2
\nonumber \\
&& 
                    + 9 ( 1 - 5 \eta + 5 \eta^2 ) \dot r^2
            \biggr] {m \over r} {\hat n}^i {\hat n}^j
\nonumber \\
&& 
          +3\biggl[ 5 ( 1 - 9\eta + 21\eta^2 ) v^2
                   -2 ( 4 - 25 \eta + 45 \eta^2 ) {m \over r}
            \biggr] v^i v^j
\nonumber \\
&& 
          + 18 ( 6 - 15 \eta - 10 \eta^2 ) {m \over r} \dot r {\hat
n}^{(i} v^{j)}
\biggr\}
\nonumber \\
&& 
+{1 \over 15}({\bf \hat N \cdot \hat n})({\bf \hat N \cdot v}){m \over r}
\biggl\{
           \biggl[ 3 ( 36-145\eta+150\eta^2 ) v^2
                  -5 ( 127 - 392 \eta + 36 \eta^2 ) {m \over r}
\nonumber \\
&& 
                  -15( 2 - 15 \eta + 30 \eta^2 ) \dot r^2 
           \biggr] \dot r {\hat n}^i {\hat n}^j
           + 6 (98 - 295 \eta - 30 \eta^2 ) \dot r v^i v^j
\nonumber \\
&& 
         +2\biggl[ 5 ( 66 - 221\eta + 96 \eta^2 ) {m \over r}
                  -9 ( 18 -  45\eta - 40 \eta^2 ) \dot r^2
\nonumber \\
&& 
                  -  ( 66 - 265\eta +360 \eta^2 ) v^2
           \biggr] {\hat n}^{(i} v^{j)}
\biggr\}
\nonumber \\
&& 
+{1 \over 60}({\bf \hat N \cdot \hat n})^2 {m \over r}
\biggl\{ \biggl[ 3   (33- 130\eta + 150\eta^2) v^4
                + 105( 1 - 10 \eta + 30 \eta^2 ) \dot r^4
\nonumber \\
&& 
                + 15 (181-572 \eta + 84 \eta^2) {m \over r} \dot r^2
                -    (131-770 \eta + 930\eta^2) {m \over r} v^2
\nonumber \\
&& 
                - 60 (  9- 40 \eta +  60\eta^2) v^2 \dot r^2
                -  8 (131-390 \eta +  30\eta^2) \left( {m \over r} \right)^2
           \biggr] {\hat n}^i {\hat n}^j
\nonumber \\
&& 
        + 4 \biggl[     (12+   5\eta - 315\eta^2) v^2
                   -9   (39- 115\eta -  35\eta^2) \dot r^2
\nonumber \\
&& 
                   +5   (29- 104\eta +  84\eta^2) {m \over r} 
            \biggr] v^i v^j 
\nonumber \\
&& 
        + 4 \biggl[15   ( 18-  40\eta -  75\eta^2) \dot r^2
                   -5   (197- 640\eta + 180\eta^2) {m \over r }
\nonumber \\
&& 
                   +3   (21- 130\eta + 375\eta^2) v^2
            \biggr] \dot r {\hat n}^{(i} v^{j)}
\biggr\}
\nonumber \\
&& 
+ {1 \over 60} 
\biggl\{ \biggl[    (467+780\eta-120\eta^2) {m \over r} v^2
                - 15( 61- 96\eta+ 48\eta^2) {m \over r} \dot r^2
\nonumber \\
&& 
                -   (144-265\eta-135\eta^2) v^4
                +  6( 24- 95\eta+ 75\eta^2) v^2 \dot r^2
\nonumber \\
&& 
                -  2(642+545\eta          ) \left( {m \over r} \right)^2
                - 45(  1-  5\eta+  5\eta^2) \dot r^4 
         \biggr] {m \over r} {\hat n}^i {\hat n}^j
\nonumber \\
&& 
       + \biggl[  4 ( 69+ 10\eta-135\eta^2) {m \over r} v^2
                - 12(  3+ 60\eta+ 25\eta^2) {m \over r} \dot r^2
\nonumber \\
&& 
                + 45(  1-  7\eta+ 13\eta^2) v^4
                - 10( 56+165\eta- 12\eta^2) \left( {m \over r} \right)^2
         \biggr] v^i v^j
\nonumber \\
&& 
       +4\biggl[  2 ( 36+  5\eta- 75\eta^2) v^2
                - 6 (  7- 15\eta- 15\eta^2) \dot r^2
\nonumber \\
&& 
                + 5 ( 35+ 45\eta+ 36\eta^2) {m \over r}
         \biggr] {m \over r} \dot r {\hat n}^{(i} v^{j)}
\biggr\} \,, \\
P^{2}Q_{TAIL}^{ij} = &&  2\delta m \int_0^\infty \biggl\{ {m \over r^3} 
\left [
15(3v^2+2{m \over r}-7 \dot r^2 )\dot r {\hat n}^i{\hat n}^j {\bf \hat
n \cdot \hat N} \right . \nonumber \\
&& \left . - (13v^2+{22 \over 3}{m \over r} -65 \dot r^2)(\hat
n^i \hat n^j {\bf v \cdot \hat N} +2 \hat n^{(i} v^{j)} {\bf \hat n
\cdot \hat N} ) \right .\nonumber \\
&& \left . -40 \dot r (v^iv^j {\bf \hat n \cdot \hat N}+2 \hat
n^{(i}v^{j)}{\bf v \cdot \hat N} ) +20v^iv^j {\bf v \cdot \hat N}
\right ] \biggr\}_{u-s} 
\left [ \ln \left ( {s \over {2R+s}}
\right ) + {97 \over 60} \right ] ds  \nonumber \\
&& + 8 \delta m  \int_0^\infty \biggl\{ {m \over r^3} \left [
(v^2-{2 \over 3}{m \over r}-5 \dot r^2 )({\hat n}^i{\hat n}^j {\bf v
\cdot \hat N} - n^{(i}v^{j)} {\bf \hat n \cdot \hat N} ) \right .
\nonumber \\
&& \left . -2 \dot r (v^iv^j
{\bf \hat n \cdot \hat N} - {\hat n}^{(i}v^{j)} {\bf v \cdot \hat N})
\right ] \biggr\}_{u-s} \left [ \ln \left ( {s \over {2R+s}}
\right ) + {7 \over 6} \right ] ds  \;. 
\end{eqnarray}
\label{hpieces}
\end{mathletters}
The leading PN and 3/2PN spin-orbit and the 2PN spin-spin contributions to the
waveform can be found in Eqs. (3.22) of \cite{kidder} and in Appendix
{}F.  There will
also be in principle 2PN spin-orbit terms; these have not been
calculated to date.  

Although we have differentiated the
moments appearing in the tail terms explicitly using the equations of
motion in order to display the waveform contributions in a consistent
manner, this is not the best form of the tail terms 
for explicit numerical evaluation in the case of general orbits.
The reason is the slow fall-off of the logarithmic term with
increasing $s$.  Instead, it is preferable to revert to the forms
of the tail terms given in Eq. (\ref{houterfinal}), split each
integral over $s$ into a finite part from $0$ to $s_0$, where $s_0$
corresponds to several dynamical timescales
of the source, and a remaining integral from $s_0$
to $\infty$.  The first integral can be done using the expressions
given in Eqs. (\ref{hpieces}).  The remaining integral is integrated
by parts twice.  One can then show \cite{agwtail} that the
latter integral falls off as $1/s_0$ generally, and for nearly
periodic orbits, as $1/s_0^2$.  By choosing $s_0$ sufficiently large
(generally a few dynamical timescales or orbital periods),
one then can obtain accurate numerical representations of the tail terms,
without having to integrate over the entire past history of the
source.

Differentiating $h^{ij}$ with respect to time, using the 2PN
equation of motion (\ref{eom}) where required, and substituting into 
Eq. (\ref{Edotformula}); or
equivalently,
taking the appropriate time derivatives of the STF moments 
(Appendix \ref{appSTFdecomp}), 
and substituting into Eq. (\ref{EdotSTF}), one finds for the energy
flux, 
\begin{equation}
{dE \over dt} = {\dot E}_N + {\dot E}_{PN} + {\dot E}_{SO} + {\dot
E}_{TAIL}+ {\dot E}_{2PN} +  {\dot E}_{SS} + O(\epsilon^{5/2}){\dot E}_N  \,,
\label{Edotanswer}
\end{equation}
where the non-spin contributions are
\begin{mathletters}
\begin{eqnarray}
{\dot E}_N = && {8 \over 15} {m^2 \mu^2 \over r^4} \left\{ 12v^2 -
11 \dot
r^2 \right\} \,, \\
{\dot E}_{PN} = &&  {8 \over 15} {m^2 \mu^2 \over r^4} \biggl\{ {1
\over 28}
\biggl[ (785-852\eta)v^4 - 2(1487-1392\eta)v^2 \dot r^2 +
3(687-620\eta) \dot
r^4 \nonumber \\ && - 160(17-\eta) {m \over r}v^2
+ 8(367-15\eta){m \over r} \dot r^2 +
16(1-4\eta)({m \over r})^2 \biggr] \biggr\} \,, \\
{\dot E}_{TAIL}= && -{4m \over 5} \stackrel{(4)\qquad}{Q^{<ij>}}(u)
\int_0^\infty \stackrel{(4)\qquad}{Q^{<ij>}}(u-s)) \ln [s/(2R+s)] ds 
\,, \label{Edottail} \\
{\dot E}_{2PN} = && {8 \over 15} {m^2 \mu^2 \over r^4} 
\biggl\{ {1 \over 756} 
\biggl[   18 (  1692 - 5497\eta + 4430 \eta^2 ) v^6
\nonumber \\
&& 
       -  54 (  1719 -10278\eta + 6292 \eta^2 ) v^4 \dot r^2
\nonumber \\
&& 
       +  54 (  2018 -15207\eta + 7572 \eta^2 ) v^2 \dot r^4
\nonumber \\
&& 
       -  18 (  2501 -20234\eta + 8404 \eta^2 ) \dot r^6
\nonumber \\
&& 
       -  12 (  33510-60971\eta +14290 \eta^2 ) {m \over r} \dot r^4
\nonumber \\
&& 
       -  36 (   4446- 5237\eta + 1393 \eta^2 ) {m \over r} v^4
\nonumber \\
&& 
       + 108 (   4987- 8513\eta + 2165 \eta^2 ) {m \over r} \dot r^2 v^2
\nonumber \\
&& 
       -   3 ( 106319+ 9798\eta + 5376 \eta^2 ) \left({m \over r}\right)^2 
                                                \dot r^2
\nonumber \\
&& 
       +     ( 281473+81828\eta + 4368 \eta^2 ) \left({m \over r}\right)^2 v^2
\nonumber \\
&& 
       -  24 (    253- 1026\eta +   56 \eta^2 ) \left({m \over r}\right)^3
\biggr]
\biggr\} \,.
\end{eqnarray}
\label{Edotpieces}
\end{mathletters}
The 3/2PN spin-orbit and 2PN spin-spin contributions can be found in
Eqs. (3.25) of \cite{kidder} and Appendix F.  
The tail contribution is formally of
3/2PN order, arising from a cross term involving $P^{3/2}Q_{TAIL}^{ij}$
and $\tilde Q^{ij}$; for simplicity, we do not write it out
explicitly (for circular orbits we evaluate it below).   
The ``11/12'' term in Eq.
(\ref{houterfinal}) contributes a term of the schematic form 
$(d^4 Q/du^4)( d^3 Q/du^3)$,
which can be written as a total time derivative and absorbed into a
redefinition of the energy $E$ at an order above that at which it is
well defined as a conserved quantity (see {\it e.g.} 
\cite{iyerwill,iyerwill2}
for a discussion of this point).
In the same way, the form of the tail term shown in Eq. (\ref{Edottail}) has
been achieved by integrating
the tail contribution once by parts
and moving the total time derivative
over to the left-hand side.
The 2PN tail terms in the
waveform make no contribution to the energy flux to 2PN order 
because their cross product with the quadrupole piece contains an odd
number of unit vectors ${\bf \hat N}$, and thus vanishes on integration
over solid angle.  They will, however, produce 5/2PN contributions to $\dot E$
via cross terms with the 1/2PN waveform terms $P^{1/2}Q^{ij}$.

Through first PN order, Eqs. (\ref{Edotpieces}) agree with
\cite{wagwill,lucgerhard}.

\section{Quasi-circular orbits}
\label{sec:circular}

\subsection{Orbit equations and gravitational waveforms}

Because gravitational radiation reaction circularizes orbits, the late
stage of inspiral of a compact binary, such as that of the binary pulsar
PSR 1913+16, will be characterized by a quasi-circular orbit, that is,
an orbit which is circular apart from the slow inspiral caused by
radiation damping.  We define the Newtonian angular momentum ${\bf L}_N
\equiv \mu {\bf x} \times {\bf v}$, the unit vector 
$ \mbox{\boldmath$\lambda$} 
\equiv {\bf \hat L}_N \times {\bf \hat n}$, and the angular velocity
$\omega \equiv |{\bf L}_N|/\mu r^2$.  
A circular orbit is given by the conditions
$\ddot r = \dot r =0$.  Solving the 2PN two-body equations of motion
(\ref{eom}) under these conditions gives 
\begin{eqnarray}
\omega^2 &=& {m \over r^3} \biggl[ 1
- {m \over r} ( 3- \eta ) 
+ \left( {m \over r} \right)^2 \left (6 + {41 \over 4} \eta + \eta^2
\right )
\biggr ] \,.
\label{omegacirc}
\end{eqnarray}
Then the orbital velocity is ${\bf v}=r\omega 
\mbox{\boldmath$\lambda$}$  and
the orbital energy through 2PN order is 
\begin{eqnarray}
 E  &=&-\eta {m^2 \over 2r} \biggl [ 1
- {1 \over 4} {m \over r} ( 7- \eta ) 
- {1 \over 8} \left({m \over r} \right)^2 
  ( 7 - 49 \eta - \eta^2 ) 
\biggr ] \ .
\label{energycirc}
\end{eqnarray}
In order to calculate waveforms as observed by an Earth-bound
detector, we must choose conventions for the direction and
orientation of the orbit.  The standard convention is to choose a
triad of vectors composed of ${\bf \hat N}$, the radial direction to
the observer, $\bf \hat p$, lying
along the intersection of the orbital plane with the plane of the sky
(line of nodes),
and ${\bf \hat q} = {\bf \hat N} \times {\bf \hat p}$ (see Fig. 7).
The normal to the orbit 
${\bf \hat L}_N$ is inclined an angle $i$ relative to $\bf
\hat N$ ($0 \le i \le \pi$).  The orbital phase $\phi =\omega u + {\rm const}$ 
of body 1 is measured from the line of nodes in a positive
(out of the plane) sense (orbits seen to be moving clockwise
correspond to $i \ge \pi/2$).  The two basic waveform 
polarizations $h_+$ and $h_\times$ are
given by
\begin{mathletters}
\begin{eqnarray}
h_+ = {1 \over 2} (\hat p_i \hat p_j -\hat q_i \hat q_j)h^{ij} \;, \\
h_\times = {1 \over 2} (\hat p_i \hat q_j +\hat q_i \hat p_j)h^{ij} \;.
\end{eqnarray}
\label{h+-}
\end{mathletters}
(There is no need to apply the TT projection in Eq. (\ref{hanswer})
before contracting on $\bf \hat p$ and $\bf \hat q$.)  From 
our conventions, we have that ${\bf \hat n}= {\bf \hat p} \cos
\phi + ({\bf \hat q} \cos i + {\bf \hat N} \sin i ) \sin \phi$ and 
$\mbox{\boldmath$\lambda$}= -{\bf \hat p} \sin
\phi + ({\bf \hat q} \cos i + {\bf \hat N} \sin i ) \cos \phi$.  Since
$h^{ij}$ consists of terms of the form $\hat n^i \hat n^j$, $\hat
\lambda^i \hat \lambda^j$ or $\hat n^{(i} \hat \lambda^{j)}$, we find
the following formulae to be useful in evaluating the
polarizations:
\begin{mathletters}
\begin{eqnarray}
(\hat n^i \hat n^j)_+ &=& {1 \over 4} \sin^2 i + {1 \over 4} (1+\cos^2
i) \cos 2\phi \;, \\
(\hat \lambda^i \hat \lambda^j)_+ 
&=& {1 \over 4} \sin^2 i - {1 \over 4} (1+\cos^2
i) \cos 2\phi \;, \\
(\hat n^{(i} \hat \lambda^{j)})_+ &=&  - {1 \over 4} (1+\cos^2
i) \sin 2\phi \;, \\
(\hat n^i \hat n^j)_\times &=&  {1 \over 2} \cos i \sin 2\phi \;, \\ 
(\hat \lambda^i \hat \lambda^j)_\times &=&  - {1 \over 2} \cos i \sin 2\phi 
\;, \\ 
(\hat n^{(i} \hat \lambda^{j)})_\times &=&  {1 \over 2} \cos i \cos 2\phi 
\;, \\ 
{\bf \hat N \cdot \hat n} &=& \sin i \sin \phi \;, \\
{\bf \hat N \cdot} \mbox{\boldmath$\lambda$} &=& \sin i \cos \phi \;.
\end{eqnarray}
\label{polarization}
\end{mathletters}
Substituting $\dot r=0$ and Eq. (\ref{omegacirc}) into Eqs.
(\ref{hpieces}) (keeping PN and 2PN corrections in Eq.
(\ref{omegacirc}) as needed), and using Eqs. (\ref{polarization}), we 
can evaluate
$h_+$ and $h_\times$ explicitly as functions of orbital phase and
orbital orientation.  The waveforms can be expressed in terms
of powers of $m/r$, but it is observationally more useful to express
them in terms of $m\omega \approx (m/r)^{3/2}$, since $\omega$ is
directly related to the observed gravitational-wave frequency.  Instead
of showing the result here, we refer the reader to 
\cite{biww} where the complete,
``ready-to-use'' pair of 2PN waveform polarizations are displayed and
discussed.  Similar substitution into Eqs. (\ref{Edotpieces}) results
in Eq. (\ref{edot}).  

\subsection{Tail Terms}
\label{sec:tailterms}

Because they involve integration over the past history of the source,
the tail contributions to the waveform and energy flux require
additional discussion.  For circular orbits, the $+$ and $\times$ 
polarizations of the
quantity $P^{3/2}Q_{TAIL}^{ij}$ are given by
\begin{mathletters}
\begin{eqnarray}
(P^{3/2}Q_{TAIL})_+ &=& 8m(1+\cos^2 i) \int_0^\infty \left ( {m^2 \over r^4}
\cos 2 \phi \right )_{u-s} \left [ \ln \left ( {s \over {2R+s}} \right
) + {11 \over 12} \right ] ds \;, \\
(P^{3/2}Q_{TAIL})_\times &=& 16m \cos i \int_0^\infty \left ( {m^2
\over r^4}
\sin 2 \phi \right )_{u-s} \left [ \ln \left ( {s \over {2R+s}} \right
) + {11 \over 12} \right ] ds \;.
\end{eqnarray}
\end{mathletters}
Because $r$ and $\omega$ evolve on a radiation-reaction timescale
$\tau_{RR}$ which is long compared to an orbital period, we can
approximate them to be constant in the above integrals; the results
will be valid up to corrections of order $(\omega \tau_{RR} )^{-1} \ln
(\omega \tau_{RR} )
\ll 1$ \cite{lucschafer}.  
Notice that the integrals converge as $s \to \infty$, even if
we approximate $m^2/r^4 \approx {\rm constant}$ (in fact, $r \to
\infty$ in the infinite past \cite{walkerwill2}, so the integrals
truly converge).  
Thus we can substitute $\omega (u-s)$ for $\phi$ with $\omega ={\rm
const}$ in the tail
integrals, pull out the $m^2/r^4$ factor, 
and use the fact that, for any integer $n$, 
\begin{mathletters}
\begin{eqnarray}
{\cal P}_S^{(n)} \equiv \int_0^\infty \sin (n\omega s) \ln \left ( {s
\over {2R+s}} \right ) ds &=& -  {1 \over {n\omega}} 
( \gamma + \ln (2n\omega R) + O[(2n\omega R)^{-2}] ) \;, \label{calPs}\\
{\cal P}_C^{(n)} \equiv \int_0^\infty \cos (n\omega s) \ln \left ( {s
\over {2R+s}} \right ) ds &=&- {1 \over {n\omega}} 
( {\pi \over 2} +O[(2n\omega R)^{-1}] )\;,\label{calPc} 
\end{eqnarray}
\label{calP} 
\end{mathletters}
where $\gamma$ is Euler's constant.
The result is
\begin{mathletters}
\begin{eqnarray}
(P^{3/2}Q_{TAIL})_+ &=& -4(1+\cos^2 i) \left ( {m \over r} \right
)^{5/2} \left \{ {\pi \over 2} \cos 2\phi + [\gamma + \ln (4\omega R)
- {11 \over 12} ] \sin 2\phi \right \} \;, \\
(P^{3/2}Q_{TAIL})_\times &=& -8 \cos i \left ( {m \over r} \right
)^{5/2} \left \{ {\pi \over 2} \sin 2\phi - [\gamma + \ln (4\omega R)
- {11 \over 12} ] \cos 2\phi \right \} \;, 
\end{eqnarray}
\end{mathletters}
It is useful to combine these tail terms with the lowest-order
quadrupole terms, given from Eq. (\ref{hpieces0}) by $\tilde Q_+ = -
(m/r)(1+\cos^2 i) \cos 2\phi$ and $\tilde Q_\times =-2(m/r) \cos i \sin
2\phi$, into the forms
\begin{mathletters}
\begin{eqnarray}
\tilde Q_+ &\approx& -{m \over r}(1+\cos^2 i) \left [ 1+ 2\pi \left ( {m \over
r} \right )^{3/2} \right ] \cos 2\psi \;, \\
\tilde Q_\times &\approx& -2 {m \over r}\cos i \left [ 1+ 2\pi \left ( {m \over
r} \right )^{3/2} \right ] \sin 2\psi \;,
\end{eqnarray}
\end{mathletters}
where 
\begin{eqnarray}
\psi &=& \phi - 2(m/r)^{3/2}[\gamma + \ln(4\omega R e^{-11/12})]
\nonumber \\
&=& \omega \{ u-2m \ln R - 2m[\gamma + \ln(4\omega e^{-11/12})] \}
\;.
\label{phaseshift}
\end{eqnarray}
We first note that one effect of the tail term is to shift the phase
of the quadrupole piece by an irrelevant constant, and by a term which
varies logarithmically with $\omega$ as the inspiral proceeds.  
This slowly varying phase shift was studied in \cite{agwtail}.

We also recognize that $u-2m\ln R =t-R-2m\ln R$ is 
retarded time with respect to
the ``true'' null cone that intersects the observation point at
$(t,R)$.  This can be seen by noting that, in the asymptotic,
Schwarzschild-like spacetime of the source, in harmonic coordinates,
outgoing radial null geodesics obey $t-r-2m \ln r + O(1/r) = {\rm
const}$.   An identical $R$-dependence in the phase 
shows up at the next 1/2PN order,
when one combines the two
polarization states of $P^{1/2} Q^{ij}$ with those of $P^2 Q_{TAIL}^{ij}$.
We thus conclude that, at least through the 2PN order considered,
our procedure for calculating the tail terms
yields gravitational waves that asymptotically propagate along the
true harmonic null cones, toward true future null infinity, despite
the use of a flat-spacetime wave equation for $h^{\alpha\beta}$.  This
avoids the need for further matching or other devices to connect
our solutions to true null infinity, and answers another
long-standing criticism of the EW framework \cite{ehlers}.  It is
useful to note also that, in the BDI approach, a similar logarithmic term
appears in the phase shift (\ref{phaseshift}), but there the term
depends on the parameter $b$ used in the transformation from harmonic
to radiative coordinates.  The appearance of
such
a parameter can be shown to have no physical consequences, as
expected \cite{agwtail,lucsathya}.  Our method is explicitly free of
such arbitrary parameters, all effects of $\cal R$ having 
cancelled.  The only external radius which appears is that of the
observer.

The tail contribution to the energy flux, given by Eq. (\ref{Edottail})
can also be calculated in closed form using the above assumptions
together with Eq. (\ref{calPc}).
The result is the ``$4\pi$'' term in Eq. (\ref{edot}).

\subsection{Display of the waveforms}
\label{sec:display}

We now display our results explicitly by plotting the waveform
for an inspiralling binary as a function of time. 
We will assume that the binary is in a quasi-circular orbit
in its last few moments before the final plunge to coalescence.
The time evolution of the orbital phase-velocity in this 
regime can be obtained
by integrating the equation
\begin{equation}
{ d \omega \over dt } = { \dot E \over dE/d\omega } \;,
\end{equation}
where $\dot E$ is given by Eqs. (\ref{edot}) and 
$dE/d\omega$ can be obtained from 
Eqs. (\ref{omegacirc}) and (\ref{energycirc}).
The orbital phase angle $\phi$ can, in turn,  be obtained by integrating
the orbital phase velocity.
The results are
\begin{mathletters}
\label{freqphase}
\begin{eqnarray}
\omega(t) = && {1 \over 8 m} (T_c-T)^{-3/8} 
\biggl\{ 1 + 
\left[{ 743  \over 2688} + {11 \over 32 } \eta \right] (T_c -T)^{-1/4}
- {3 \pi \over 10 } (T_c -T)^{-3/8}
\nonumber \\
&& \; \; \; +
\left[{ 1855099 \over 14450688} + {56975  \over 258048 } \eta
 + {371 \over 2048 } \eta^2 \right ] (T_c -T)^{-1/2}
\biggr\} \;,\\
\phi(t) = && \phi_c - {1 \over \eta } (T_c-T)^{5/8} 
\biggl\{ 1 + 
\left[{ 3715 \over 8064} + {55 \over 96} \eta \right] (T_c -T)^{-1/4}
- {3 \pi \over 4 } (T_c -T)^{-3/8}
\nonumber \\
 && \; \; \;  +
\left[{ 9275495\over 14450688} + {284875\over 258048 } \eta
 + {1855\over 2048 } \eta^2 \right] (T_c -T)^{-1/2}
\biggr\} \; ,
\end{eqnarray}
\end{mathletters}
where $T$ is a dimensionless time variable  related to the
coordinate retarded time $u$ by
$T = \eta (u/5m)$,  and
$\phi_c$ and $T_c$ are constants of integration.
The constant $T_c$ is the dimensionless retarded time at coalescence
(the time at which the frequency in Eq. (\ref{freqphase})
formally becomes infinite),
and $\phi_c$ is the orbital phase at coalescence.

We can now use the orbital phase evolution along with
Eqs. (\ref{h+-}), (\ref{polarization}) and 
(\ref{omegacirc}) to write $h_+$ and $h_\times$
as explicit functions of time.
We will not display the result here (there are enough large
equations in this paper already), but rather refer the reader
to Eqs. (2) to (4) in \cite{biww} for ``ready-to-use'' waveforms.
The ``ready-to-use'' waveforms are essentially Eqs. (\ref{hpieces})
boiled down to the circular orbit case.

 For the case of a $1.4M_{\odot}$ neutron star spiralling into a
$10M_{\odot}$ black hole 
the resulting  frequency sweep and waveform  
are shown in Fig 8.
The observer is viewing the orbital motion
edge on, so that $i=\pi/2$ in Eqs. (\ref{polarization}).
In this case the gravitational radiation is
linearly polarized (only $h_+$ is present). 
The upper cut-off frequency in Fig. 8 is chosen to be 
180 Hz; this is approximately the orbital frequency
at the innermost stable circular orbit \cite{isco1,isco2}
for this type of system.  For the initial 
LIGO detector, Finn \cite{finngrqc} has
shown that a substantial fraction of the signal-to-noise
ratio available is accumulated when integrating a matched
filter against the signal in the frequency range we have displayed.
In other words, the segment of the waveform shown in Fig. 8b, 
sweeping from 75 Hz to 180 Hz, 
is the portion of the waveform which is actually most
{\it detectable} for the initial LIGO detector.

As energy is extracted from the system by the radiation,
the orbital radius shrinks  and the orbital frequency 
increases.
This gives rise to the dominant ``chirp'' feature of the waveform
in Fig. 8b: 
the growing amplitude and the bunching of peaks at late
time. 
However, 
because the coordinate velocity rises to  about $0.5c$, this system is
quite relativistic,
and thus the inclusion of higher multipoles of the
radiation causes the waveform to differ 
considerably from the simple {\it cosine} chirp 
that one would compute just using quadrupole radiation.
The pairing of wave crests (alternately closer together
and farther apart) signifies the onset of 
the gravitational analogue of synchrotron spikes.
Just as in electricity and magnetism
this feature comes from the inclusion of many harmonics of the 
radiation. 
In our analysis we have included multipoles through the six-index
multipole $I_{EW}^{ijklmn}$. This allows us consistently to
include components of the radiation in our waveform
at multiples of the orbital phase
$n \phi_{\rm orbital}$ where $n$ ranges from $1$ to $6$,
$n=2$ being the dominant quadrupole contribution.

Another interesting feature of Fig. 8b
is that adjacent troughs are not
the same depth, but adjacent crests are essentially the same
height.
This effect also has a discernable physical origin.
The deeper troughs arise when 
the lighter mass is coming toward the observer;
thus the observer is in the forward synchrotron beam pattern of the lighter,
faster-moving mass.
The shallower troughs arise when the lighter mass is receding
from the observer.
[At the left-hand-side of the figure,
the phase is arbitrarily set to zero, {\it i.e.} the heavier
mass (chosen to be $m_1$) is passing through the ascending node
coming toward
the observer and the lighter mass is receding.
The waveform is clearly in the not-so-deep trough
at this left-most point.]
The crests are essentially the same height because the
radiation is virtually the same when the masses
are moving transverse to the line of sight of the observer
regardless of which mass is closer to the observer
(see \cite{magnum} for further discussion of the asymmetric 
radiation emission).
The extent to which the harmonic structure might be measurable 
by a gravitational-wave detector is currently under
investigation \cite{bradywiseman}.
Preliminary analysis shows that neglecting the harmonic
structure ({\it i.e.} just using the quadrupole amplitude
to describe the wave) results in approximately a $4\%$ loss
in signal-to-noise ratio.
In Appendix F we show how the effects of spin modify
the waveform and frequency evolution.
\section{Discussion}

We have extended the Epstein-Wagoner framework for calculating gravitational
radiation from slow-motion systems to produce a method that is free of
divergences or undefined integrals.  The extension involved adding to
the original framework the
integral of the effective source over that part of the 
past null cone of the field
point that is {\it exterior} to the near zone.  When expressed in
appropriate variables, that integral can be shown to be convergent,
and can be evaluated in a straightforward way, to any chosen PN order.
The exterior integral yielded (a) terms that explicitly cancel terms from
the EW framework previously thought to be divergent (b) tail terms, in
agreement with other methods based on matching, and (c) phasing terms
that verify that the radiation asymptotically propagates along true
null cones of the curved spacetime. 

This new, well-defined framework, provides a basis for extending the
calculation of gravitational radiation to higher PN orders.  An
extension to 5/2PN order in the BDI framework has been achieved by
Blanchet \cite{luc5/2}; such an extension in the improved EW
framework is in progress.  Extension to 3PN order will be a bigger
challenge, simply because of the complexity of the terms, including
quadratically non-linear integrals, and the rapidly increasing number
of computations.  However, we foresee no obstacle in principle to such
an extension in the
improved framework.  

This improved framework will also allow derivation of near-zone
gravitational fields in a form that will yield
equations of motion for the sources to high PN orders.  It should
be possible to derive radiation-reaction terms in the two-body
equations of motion, at order
$\epsilon^{5/2}$ and $\epsilon^{7/2}$ beyond Newtonian gravity
\cite{iyerwill,iyerwill2,lucreact}, without
the presence of ill-defined or divergent terms, and without the need
for matching between zones.  One goal would be to derive the
non-dissipative, 3PN terms in the equations of motion.  This would
improve the accuracy of estimates, using a hybrid Schwarzschild-PN
equation of motion, of the transition point between inspiral and
unstable plunge in the late stage of compact binary inspiral
\cite{isco1,isco2}.   Calculation of the near-zone fields will also be
important in developing interfaces between the post-Newtonian approach,
which works well for most of the inspiral, 
and numerical relativity methods which must be used for the final few
orbits and the coalescence.  Work on this latter subject is in
progress.

\acknowledgments

We are grateful to Luc Blanchet, Thibault Damour, Bala Iyer, Eric
Poisson, and Wai-Mo Suen for useful discussions.
This work is supported in part by the National Science Foundation
under Grants No. PHY 92-22902 and PHY 96-00049 (Washington University), 
and AST 94-17371 and PHY 94-24337 (Caltech),
and NASA under Grant No. NAGW 3874 (Washington University)

\appendix

\section{STF Tensors and their Properties}
\label{appSTF}

In calculating field integrals we make frequent use of the properties
of symmetric, trace-free (STF) products of unit vectors.  
The general
formula for such STF products is
\begin{eqnarray}
\hat n^{<L>} \equiv \sum_{p=0}^{[l/2]} (-1)^p {{(2l-l-2p)!!} \over
{(2l-1)!!}} \left [ \hat n^{L-2P} \delta^P + {\rm sym(q)} \right ] \;,
\label{STFgen}
\end{eqnarray}
where $[l/2]$ denotes the integer just less than or equal to
$l/2$, the capitalized superscripts denote the dimensionality, $l-2p$
or $p$, of
products of $\hat n^i$ or $\delta^{ij}$ respectively, 
and ``sym(q)'' denotes all
distinct terms arising from permutations of 
indices, where $q=l!/[(2^p p!(l-2p)!]$ is the total number of such
terms (see \cite{thorne80,bd86} for compendia of formulae).  For 
convenience, we display the
first several examples explicitly
\begin{mathletters}
\label{STFformulae}
\begin{eqnarray}
\hat n^{<ij>} &=& \hat n^{ij} - {1 \over 3} \delta^{ij} \;, \\
\hat n^{<ijk>} &=& \hat n^{ijk} - {1 \over 5} (\hat n^i \delta^{jk} +\hat
n^j \delta^{ik} + \hat n^k \delta^{ij}) \;, \\
\hat n^{<ijkl>} &=& \hat n^{ijkl} - {1 \over 7}(\hat n^{ij}\delta^{kl}
+ {\rm sym(6)}) +{1 \over
35}(\delta^{ij}\delta^{kl}+\delta^{ik}\delta^{jl}+\delta^{il}\delta^{jk})
\;, \\
\hat n^{<ijklm>} &=& \hat n^{ijklm} - {1 \over 9}(\hat n^{ijk}\delta^{kl}
+ {\rm sym(10)}) +{1 \over
63}(\hat n^i \delta^{jk}\delta^{lm} +{\rm sym(15)})
\;, \\
\hat n^{<ijklmn>} &=& \hat n^{ijklmn} - {1 \over 11}(\hat n^{ijkl}\delta^{mn}
+ {\rm sym(15)}) +{1 \over
99}(\hat n^{ij}\delta^{kl}\delta^{mn}+ {\rm sym(45)}) \nonumber \\
&&-{1 \over 693} (\delta^{ij}\delta^{kl}\delta^{mn}+ {\rm sym(15)} )
\;. 
\end{eqnarray}
\end{mathletters}
There is a close connection between these STF tensors and spherical
harmonics.  For example, it is straightforward to show that, for any
unit vector $\bf \hat N$, the contraction of $\hat N^L$ with $\hat
n^{<L>}$ is given by
\begin{eqnarray}
\hat N^L \hat n^{<L>} = {{l!} \over {(2l-1)!!}} P_l ({\bf \hat N \cdot
\hat n}) \;,
\label{legendre}
\end{eqnarray}
where $P_l$ is a Legendre polynomial.
This latter property can be used to establish the identity
(\ref{angleident})
\begin{eqnarray}
\sum_m \int Y_{lm}^* ({\bf \hat n}) Y_{lm} ({\bf \hat y}) \hat y^{<L^\prime>}
d^2\Omega_y \equiv \hat n^{<L>} \delta_{ll^\prime} \;.
\label{angleident2}
\end{eqnarray}
Since the left-hand-side is STF, and depends only on the unit vector
${\bf \hat n}$, then it must be proportional to the STF combination
$\hat n^{<L^\prime>}$.  To
establish the normalization, contract both sides with the 
$L^\prime$-dimensional
non-STF product
$\hat N^{L^\prime}$, where $\bf \hat N$ represents the $z$-direction.  Using 
Eq. (\ref{legendre}), and 
recalling that $P_{l^\prime} = {\cal N}_{l^\prime} Y_{{l^\prime}0}$, where
${\cal N}_{l^\prime}$ is a normalization coefficient, we find that the integral
yields $[l! / (2l-1)!!]
P_{l^\prime}({\bf \hat N \cdot \hat n}) \delta_{l{l^\prime}}$, 
establishing the
unit coefficient in Eqs. (\ref{angleident}) and  (\ref{angleident2}).  

In calculating the radiation-zone contributions to $h^{ij}$, we must
also evaluate the integrals
$(4\pi)^{-1} \int_0^{2\pi}
d\phi^\prime \int_{1-\alpha}^1 \hat n{^\prime}^{<L>} (\zeta - {\bf
\hat n}^\prime \cdot {\bf \hat n} )^{N-3} d \cos \theta^\prime$, 
where $\alpha=
(\zeta-1)(\zeta+1-2{\cal R}/r)(r/2{\cal R})$.  The result must be an
$l$-dimensional STF tensor, dependent on the only vector in the
problem, $\bf \hat n$, and thus must be proportional to $\hat
n^{<L>}$.  To determine the proportionality factor, which will be a
function of $\zeta$ and $\alpha$, we contract with $\hat n^{L}$,
chose $\bf \hat n$ to be in the
$z$-direction. and substitute Eq. (\ref{legendre}).  The result is
\begin{mathletters}
\begin{eqnarray}
(4\pi)^{-1} \int_0^{2\pi} 
d\phi^\prime \int_{1-\alpha}^1 \hat n{^\prime}^{<L>} (\zeta - {\bf
\hat n}^\prime \cdot {\bf \hat n} )^{N-3} d \cos \theta^\prime &=&
A_{N,l}(\zeta,\alpha) \hat n^{<L>} \;, \\
A_{N,l}(\zeta,\alpha) &=& {1 \over 2}  \int_{1-\alpha}^1 P_l(z)
(\zeta-z)^{N-3}dz
\;.
\end{eqnarray}
\label{ANl}
\end{mathletters}

\section{Derivatives of Gravitational Potentials}
\label{appderivatives}

In evaluating the ``field'' parts of EW moments, we have repeated
occasion to integrate expressions involving two derivatives, spatial,
time, and mixed, of the
potential $U$ and two spatial derivatives of $\ddot X$.  For a field 
point external to the bodies,
such derivatives can be calculated easily from the expressions
(\ref{UP1new}) and (\ref{UP2new}).  However, because the integrations
run over the locations ${\bf x}_A$ of the bodies themselves, we must
carefully evaluate the singular behavior of such double derivatives at
${\bf x}={\bf x}_A$.  Consider, for example, the expression for $\ddot
U$, written in terms of a smooth density distribution: 
\begin{eqnarray}
\ddot U = \int \rho^\prime \left [ {{{\bf a^\prime} \cdot ({\bf x}-{\bf
x^\prime})} \over {|{\bf x}-{\bf x^\prime}|^3}} + {{3{v^\prime}^{ij}
(x-x^\prime)^{<ij>}} \over {|{\bf x}-{\bf x^\prime}|^5}} \right ]
d^3x^\prime \;,
\label{deriv1}
\end{eqnarray}
where ${\bf a^\prime}=d{\bf v^\prime}/dt$.  For a field point outside
the bodies, shrinking the density distribution to a point yields a
result equivalent to that obtained by differentiating Eq.
(\ref{UP1new}).  For a point inside, say, body A, we find that the
integral $\int_{\rm body\; A} \ddot U d^3x \to  -(4\pi/3)m_Av_A^2$ as
the size of body A shrinks to a point.
Consequently we must add a delta-function term to all double derivatives
of $U$ and $X$ found using Eqs. (\ref{UP1new}) and (\ref{UP2new}).   
The results are
\begin{mathletters}
\begin{eqnarray}
U_{,ij} &=&  U_{,ij}^\dagger - (4\pi/3)\sum_A m_A
\delta^{ij}\delta^3({\bf x}-{\bf x}_A) \;, \\
\dot U_{,i} &=& \dot U_{,i}^\dagger + (4\pi/3)\sum_A m_A
v_A^i \delta^3({\bf x}-{\bf x}_A) \;, \\
\ddot U &=& \ddot U^\dagger - (4\pi/3)\sum_A m_A
v_A^2 \delta^3({\bf x}-{\bf x}_A) \;, \\
\ddot X_{,ij} &=&  \ddot X_{,ij}^\dagger - (8\pi/15)\sum_A m_A
(v_A^2 \delta^{ij} +2v_A^{ij}) \delta^3({\bf x}-{\bf x}_A) \;, 
\end{eqnarray}
\label{deriv2}
\end{mathletters}
where $\dagger$ denotes derivatives computed from Eqs. (\ref{UP1new}) and
(\ref{UP2new}).

\section{The second-iterated fields}
\label{appW}

In Section \ref{sec2iterated}, we wrote down the second-iterated
solution for $h^{\alpha\beta}$ in terms of the potentials $V$, $V_i$
and $W_{ij}$.  Here we discuss the solutions for these potentials,
Eqs. (\ref{VW}), in more detail, especially the potential $W_{ij}$,
whose source is non-compact.  

We first consider field points in the radiation zone.   Since their
sources have compact support, the potentials $V$ and $V_i$ do not have
to be divided into contributions from integrals over 
the near zone and over the radiation
zone.  They can be expanded using the analogue of Eq.
(\ref{genexpand}), and
written to the needed order in the form
\begin{mathletters}
\begin{eqnarray}
V(t,{\bf x}) &=&  \tilde m/r 
+ {1 \over 2} \left \{ {r}^{-1}Q^{ij}(u) \right \}_{,ij} 
-  {1 \over 6}  \left \{ {r}^{-1}Q^{ijk}(u) \right \}_{,ijk} \nonumber
\\
&& + {1 \over 2} \ddot Q /r -\left \{ r^{-1} F^i (u) \right \}_{,i}  
+ O(\epsilon^3)\;, \label{appV}\\
V_i(t,{\bf x}) &=& -
{1 \over 2} \left \{ {r}^{-1} [ \dot Q^{ij}(u) 
-\epsilon^{ija}J^a(u) ] \right \}_{,j} 
\nonumber \\
&&+ {1 \over 6} \left \{ {r}^{-1} [ \dot Q^{ijk}(u) 
-2\epsilon^{ika}J^{aj}(u) ] \right \}_{,jk} + O(\epsilon^{5/2}) 
\;, \label{appVi}
\end{eqnarray}
\label{appVVi}
\end{mathletters}
where $\ddot Q$ and $F^i \equiv \sum_A m_A x_A^i (v_A^2 - 
\sum_B m_B/2r_{AB} )$ 
respectively represent
the difference between the monopole and dipole moments 
of the potential $V$, and the 1PN accurate, constant total mass $\tilde m$, Eq.
(\ref{mQJ-m}), and the 
vanishing center of mass ${\bf X}$, Eq. (\ref{mQJ-X}).  In
constructing $h^{00}$ using Eq. (\ref{h2iterated00}) these two terms 
are cancelled by terms from $W=W^{ii}$.

The potential $W_{ij}$ must first be divided into near-zone and
radiation-zone contributions, $W_{ij} = (W_{ij})_{\cal N}
+(W_{ij})_{{\cal C}-{\cal N}}$.  To the $O(\epsilon^2)$ needed for the
use of $W_{ij}$ in source terms for higher iterations
[see Eqs. (\ref{h2iterated})], we can approximate the integrand in Eq.
(\ref{VW3}) by
$\sigma_{ij} + (4\pi)^{-1} (U_{,i}U_{,j} - {1 \over 2} \delta_{ij}
U_{,k}U_{,k})
\equiv \tau_{(1)}^{ij}/4$, with $\sigma_{ij}=\sum_A m_A v_A^i v_A^j
\delta^3 ({\bf x}-{\bf x}_A)$.  Here $\tau_{(1)}^{ij}$  denotes the
first-iterated effective 
stress-energy.  Then $ (W_{ij})_{\cal N}$ can be expanded
using the analogue of Eq. (\ref{genexpand}), with the result
\begin{eqnarray}
 (W_{ij})_{\cal N} = \sum_{q=0}^\infty {{(-1)^q} \over {q!}} \left (
{1 \over r} M^{ijk_1 \dots k_q} \right )_{, k_1 \dots k_q} \;,
\label{appWexpand}
\end{eqnarray}
where 
\begin{eqnarray}
M^{ijk_1 \dots  k_q} (u) &=& \int_{\cal M} \tau_{(1)}^{ij}(u,{\bf x})
x^{k_1 \dots k_q} d^3x \;.
\label{genmomentapp}
\end{eqnarray}
Using the expression above for $ \tau_{(1)}^{ij}$ in each of the
moments in Eq. (\ref{genmomentapp}), and using the strategy for
evaluating field integrals described in Sections IV.A and B, we find,
to the needed accuracy,
\begin{mathletters}
\begin{eqnarray} 
M^{ij} &=& {1 \over 2} \ddot Q^{ij} \;, \\
M^{ijk} &=& {1 \over 6} \ddot Q^{ijk} -{2 \over 3} \epsilon^{(i|ka}
\dot J^{a|j)} \;, \\
M^{ijkl} &=&  {1 \over 15} m^2 {\cal R} (2\delta^{i(k}\delta^{l)j} -{3 \over
2}\delta^{ij}\delta^{kl})  \nonumber \\
&&+{\rm term \; independent \; of \;} {\cal R} 
\label{appMijkl} \;,
\end{eqnarray}
\end{mathletters}
where we discard terms that fall off with increasing $\cal R$, but
retain all other terms.  
Although we never actually need the contribution from the moment
$M^{ijkl}$, we show the $\cal R$-dependent term to illustrate its
ultimate cancellation.

To evaluate $(W_{ij})_{{\cal C}-{\cal N}}$, we use the fact that, to
the required order, in the radiation zone, $\tau_{(1)}^{ij} =
(4\pi)^{-1}(m^2/{r^\prime}^4)(\hat n{^\prime}^{<ij>} - {1 \over
6}\delta^{ij} )$.  Using Eq. (\ref{houterzeta}), and remembering the
factor of 4 difference between $h^{ij}$ and $W_{ij}$, we obtain
\begin{eqnarray}
(W_{ij})_{{\cal C}-{\cal N}} = {1 \over 4} {m^2 \over r^2} \hat n^{ij}
-{1 \over 5} {m^2 \over r^3} {\cal R} \hat n^{<ij>} \;,
\label{appWouter}
\end{eqnarray}
where we again discard terms that fall off with $\cal R$.  It is easy
to see that  the $\cal R$-dependent term in Eq. (\ref{appWouter})
exactly cancels the corresponding term in $(W_{ij})_{\cal N}$ resulting
from Eq. (\ref{appMijkl}) and (\ref{appWexpand}).  
Combining the contributions to $W_{ij}$
through octupole order, and substituting them along with Eqs. (\ref{appVVi}) 
for $V$ and $V_i$ into Eqs. (\ref{h2iterated}) yields the
second-iterated radiation-zone fields $h^{\alpha\beta}$, Eqs.
(\ref{hforouter}).  It is interesting to note that the $(m^2/4r^2)\hat
n^{ij}$  term in $(W_{ij})_{{\cal C}-{\cal N}}$ is required in order
that the far-zone field correctly approximate the Schwarzschild
geometry in harmonic coordinates in the static limit, namely
\begin{mathletters}
\begin{eqnarray}
h^{00} &=& 4m/r + 7(m/r)^2 \;, \\
h^{0i} &=& 0 \;, \\
h^{ij} &=& (m/r)^2 \hat n^{ij} \;,
\end{eqnarray}
\end{mathletters}
[compare Eq. (\ref{hforouter})].  This contribution could not have
been found
using the EW approach without our new formulation of the
radiation-zone integrals.

We next consider field points within the near zone.   Expanding the
retardation about $t=u$ with $| {\bf x}-{\bf x}^\prime |$ as the small
parameter, we obtain Eqs. (\ref{VWapprox}) and (\ref{UP}).   The
compact contributions to $U$, $X$, $U_i$ and $P_{ij}$ can be evaluated
directly; the non-compact part of $P_{ij}$ is left unevaluated until
it is incorporated into an EW moment (see Appendix \ref{appcubic}).
It remains to evaluate the radiation-zone contribution
$(W_{ij})_{{\cal C}-{\cal N}}$ with a near-zone field point.  Using
the form of $\tau_{(1)}^{ij}$ above, and using the near-zone
field-point version of Eq. (\ref{houterzeta}), we find only
contributions proportional to $m^2 r^2/{\cal R}^4$ and $m^2/{\cal
R}^2$.  Thus we can discard such terms.

\section{Cubic Non-linearities in $I_{EW}^{\lowercase{ij}}$}
\label{appcubic}

At 2PN order, the non-linear field source $\Lambda^{00}$  Eq.
(\ref{lambda00inst}) contains
terms that are cubically nonlinear, {\it i.e.} that depend on effective
products of three
gravitational potentials.
The contribution of the 
final such term in Eq. (\ref{lambda00inst}), proportional to $U
U_{,k}U_{,k}$, to the integral $\int_{\cal M} \Lambda^{00}x^ix^jd^3x$
can be evaluated straightforwardly by integrating by parts.  However,
the two terms $2P_{,k}U_{,k} - P_{km}U_{,km}$ are more difficult
because $P_{ij}$ itself [Eq. (\ref{UP4})] is a potential, one of whose 
pieces is
produced by a non-linear source.  The contribution of the compact
source $\sigma_{ij}$ can be handled easily by the methods of Sec.
\ref{strategy}.  
Here we focus on the non-linear piece.  We define the non-linear
potential
\begin{eqnarray}
p_{ij}(u,{\bf x}) &\equiv& {1 \over {4\pi}}
\int { {d^3x^\prime} \over {|{\bf x}-{\bf x^\prime}|}}
( U_{,i}  U_{,j} - {1 \over 2} \delta_{ij}  U_{,k}  U_{,k} ) 
 (u,{\bf x^\prime})  \;,
\end{eqnarray}
We then need to evaluate the integral
\begin{eqnarray}
(1/\pi) \int_{\cal M} (2p_{,k}U_{,k}-p_{km}U_{,km}) x^ix^j d^3x \;.
\label{cubicterms}
\end{eqnarray}
We integrate the first term by parts, show that the surface terms fall
off with $\cal R$, and obtain $8\sum_A m_A p({\bf
x}_A)x_A^{ij} -(4/\pi)\int pU^{,(i}x^{j)} d^3x$.  The first of these
terms may
be evaluated using the non-linear pieces of Eq. (\ref{UP4new}).  The
second term may be written in the form 
\begin{eqnarray}
{1 \over {2\pi^2}} \int_{\cal M} |\nabla^\prime U^\prime |^2
d^3x^\prime \sum_A m_A \int_{\cal M} {1 \over {|{\bf x}-{\bf x^\prime}|}} 
{{(x-x_A)^{(i}x^{j)}} \over  {|{\bf x}-{\bf x}_A|^3}} d^3x \;.
\end{eqnarray}
In the $x$-integration, we change variables to ${\bf y}={\bf
x}-{\bf x}_A$ and integrate using the general method described in Sec.
\ref{strategy}.  The result is the integral $(1/2\pi)\sum_A m_A \int_{\cal M} 
|\nabla^\prime U^\prime |^2 d^3x^\prime ({x^\prime}^{ij} -x_A^{ij})/
|{\bf x}_A-{\bf x^\prime}|$, which can be easily evaluated by
integrating by parts.  

The second term in Eq. (\ref{cubicterms}) can be written
\begin{eqnarray}
&-&{1 \over {4\pi^2}} \int_{\cal M} \left ( U_{,k}^\prime U_{,m}^\prime - {1
\over 2} \delta_{km} |\nabla^\prime U^\prime |^2 \right ) d^3x^\prime
\nonumber \\
&& \quad \times \sum_A m_A \int_{\cal M} {1 \over {|{\bf x}-{\bf x^\prime}|}}
\left ( {{3(x-x_A)^{<km>}} \over  {|{\bf x}-{\bf x}_A|^5}} - {{4\pi}
\over 3} \delta^{km} \delta^3 ({\bf x}-{\bf x}_A) \right ) x^{ij} d^3x
\end{eqnarray}
Again we do the $x$-integration by changing variables to ${\bf y}={\bf
x}-{\bf x}_A$, and
using the method of Sec. \ref{strategy}.  Integration 
of the delta-function term is
straightforward.  The remaining $x^\prime$-integration takes the form
\begin{eqnarray}
{1 \over \pi} \int_{\cal M} && \left ( U_{,k}^\prime U_{,m}^\prime - 
{1 \over 2} \delta_{km} |\nabla^\prime U^\prime |^2 \right ) d^3x^\prime
\nonumber \\
&& \times \sum_A m_A \left ( {1 \over 6} \Phi_{,ijkm}^A 
 - \Psi_{,k(i}^A
\delta_{j)m} + {1 \over 2} \Psi_{,km(i}^A x_A^{j)} -2X_{,k}^A
\delta_{m(i} x_A^{j)} + {1 \over 2} X_{,km}^A x_A^{ij}  \right .
\nonumber \\
&& \left . - {1 \over 3}
{{\delta_{km}x_A^{ij}} \over {|{\bf x^\prime}-{\bf x}_A|}} + X^A
\delta_{k(i} \delta_{j)m} - {1 \over 5} {\cal R} \delta_{k(i}
\delta_{j)m} \right ) \;,
\label{PhiPsi}
\end{eqnarray}
where $\Phi^A \equiv |{\bf x^\prime}-{\bf x}_A|^5/15$, $\Psi^A \equiv
|{\bf x^\prime}-{\bf x}_A|^3/3$, and $X^A \equiv |{\bf x^\prime}-{\bf
x}_A|$.  The first five terms in Eq. (\ref{PhiPsi}) can be evaluated
simply by integrating by parts.  The sixth term is equivalent to the
cubically nonlinear term in $\Lambda^{00}$ 
proportional to $UU_{,k}U_{,k}$ [Eq. (\ref{lambda00inst})].  The final
term proportional to $\cal R$ is straightforward.

The seventh term requires extra work.  Dropping contributions with no
TT part, we find that the integral to be evaluated is $\pi^{-1} 
\int_{\cal M}
U_{,i}U_{,j} Xd^3x$.  Defining $U_A$ and $X_A$ to be the contribution to $U$
and $X$ from body $A$, respectively, we write
\begin{eqnarray}
\int_{\cal M} U_{,i}U_{,j} Xd^3x &=& \sum_A \int_{\cal M}
U_{A,i} U_{A,j}  X_A d^3x + \sum_{A \ne B} \int_{\cal M}
U_{A,i}U_{A,j} X_B d^3x \nonumber \\
&& + 2\sum_{A \ne B}\int_{\cal M} U_{A,(i}
U_{B,j)} X_Ad^3x + \sum_{A \ne B \ne C}\int_{\cal M} U_{A,(i}
U_{B,j)} X_C d^3x \;.
\label{splitup}
\end{eqnarray}
The first term has no TT part, while the second two terms can be
evaluated using the standard methods of Sec. \ref{strategy}, and lead
to the term $-3 \sum_{AB} m_A^2 m_B \hat n_{AB}^{ij}$ in Eq.
(\ref{Iijfinal}).  
We define the third term to be ${\cal G}_{(3)}^{ij}$, change
variables to ${\bf u}={\bf x}-{\bf x}_C$, ${\bf y}={\bf x}_A-{\bf
x}_C$, and ${\bf z}={\bf x}_B-{\bf x}_C$, verify that no surface
contributions at $\cal R$ are so generated, and show that $\cal G$ can
be written ${\cal G}_{(3)}^{ij} = \sum_{ABC} m_Am_Bm_C
\nabla_y^i \nabla_z^j F({\bf y},{\bf
z})$, where $F({\bf y},{\bf z}) \equiv \int_{\cal M} |{\bf u}-{\bf
y}|^{-1} |{\bf u}-{\bf z}|^{-1} ud^3u$.  The latter step involves
ensuring that the piece of $F$ that diverges with $\cal R$ contributes
no TT part to $\cal G$, so that the integration can effectively 
be commuted with
the $y$- and $z$-derivatives.  Note that $F$ has units of
(distance)$^2$, is symmetric on $y$ and $z$, is a function only of
$|{\bf y}|$, $|{\bf z}|$ and $w \equiv |{\bf y}-{\bf z}|$, and has the
property that $\nabla_y^2 F=-4\pi y/w$, $\nabla_z^2 F=-4\pi z/w$.  
It is then straightforward to show that the function with these
properties is given by $F({\bf
y},{\bf z})=-(2\pi /3)[(y+z)w - yz + (y^2+z^2-w^2)\ln (y+z+w)]$,
modulo terms that give no TT contribution to $\cal G$.  Thus the
solution for ${\cal G}_{(3)}^{ij}$ in Eq. (\ref{Iijfinal}) is 
\begin{mathletters}
\begin{eqnarray}
{\cal G}_{(3)}^{ij} &=& \sum_{A \ne B \ne C} 
m_Am_Bm_C \nabla_A^i \nabla_B^j
  F({\bf x}_{AC}, {\bf x}_{BC}) \;, \\
  F({\bf x}_{AC}, {\bf x}_{BC}) &=& -{2 \over 3} \left [
(r_{AC}+r_{BC})r_{AB} -r_{AC}r_{BC} +2{\bf x}_{AC} \cdot {\bf x}_{BC}
\ln (r_{AC}+r_{BC}+r_{AB}) \right ] \;. 
\end{eqnarray}
\end{mathletters}
Note that, because $ \nabla_A^i \nabla_B^j ({\bf x}_{AC} \cdot {\bf
x}_{BC})=\delta_{ij}$, no logarithmic dependence on source variables
actually survives in $h_{TT}^{ij}$.
{}For two-body systems, this term does not enter the formula for the EW
moment.

\section{STF-Multipole Decomposition}
\label{appSTFdecomp}
 
Although the Epstein-Wagoner multipoles arose very naturally 
in our retarded-time expansion of the relaxed Einstein equation, 
these are not the only multipoles
for displaying the answer.
An alternative set are the
symmetric tracefree (STF) multipoles, which arise naturally in angular
decompositions of the waveform (see, {\it e.g.} \cite{thorne80}), and
are multipoles of choice in the BDI framework.  Thus it is useful to
obtain a transformation between the EW multipoles and the STF
multipoles.

If the waveform is known then the STF multipoles can
be projected out.  
This is exactly analogous
to projecting the coefficients of spherical harmonics from a scalar
function.
The STF multipoles can be projected
from the TT waveform by integrating over the sphere [see
\cite{thorne80}, Eq. (4.11)]:
\begin{mathletters}
\begin{eqnarray}
{d^m \over du^m } {\cal I}_{STF}^{a_1 a_2 ... a_m} &=&
\left[ { m (m-1) (2m+1)!! \over 2 (m-1) (m+2) } {R \over 4 \pi }
        \int h_{TT}^{a_1 a_2} N^{a_3} ... N^{a_m} d \Omega 
\right] \;, \\
{d^m \over du^m } {\cal J}_{STF}^{a_1 a_2 ... a_m} &=&
\left[ { (m-1) (2m-1)!! \over 4 (m+2) } {R \over 4 \pi }
        \int \epsilon^{a_1 j k } N^j
        h_{TT}^{k a_2} N^{a_2} ... N^{a_m} d \Omega 
\right] \;,
\end{eqnarray}
\label{thorneformula}
\end{mathletters}
where $ {\cal I}_{STF}^{a_1 a_2 ... a_m}$ are called ``mass'' multipole
moments and ${\cal J}_{STF}^{a_1 a_2 ... a_m}$ are called ``current'' or
``spin'' multipole moments.   Substituting the expansion of
$h_{TT}^{a_1 a_2}$ in terms of EW moments, Eq. (\ref{EWseries}), and
adding the radiation-zone tail terms, Eq. (\ref{houterfinal}), we
obtain the transformations, correct to 2PN order:
\begin{mathletters}
\begin{eqnarray}
{\cal I}^{ij}_{STF} &=& \left [  I_{EW}^{ij} 
+ {1\over21}\left( 11 I_{EW}^{ijkk} 
 -12 I_{EW}^{k(ij)k} + 4 I_{EW}^{kkij} \right) \right .
\nonumber \\
&&\left . + {1 \over 63} \left( 23 I_{EW}^{ijaabb} - 32I_{EW}^{a(ij)abb}
                        +10 I_{EW}^{aaijbb} +  
2I_{EW}^{abijab} \right) \right ] _{STF} + {\cal I}_{TAIL}^{ij} \;, \\
\dot {\cal I}^{ijk}_{STF} &=& \left [  3 I_{EW}^{ijk} 
+ \left( 3 I_{EW}^{ijkaa}  
- 3 I_{EW}^{iaakj} + I_{EW}^{aaijk} 
\right) \right ] _{STF} + \dot {\cal I}_{TAIL}^{ijk} \;,\\
\ddot {\cal I}_{STF}^{ijkl} &=& \left [ 12 I_{EW}^{ijkl} + {72 \over 55} \left(
13 I_{EW}^{ijklmm} - 12 I_{EW}^{immjkl} +4 I_{EW}^{mmijkl} \right )
\right ]_{STF} \;,\\
 \stackrel{(3)\qquad}{{\cal I}_{STF}^{ijklm}} &=& \left 
[ 60 I_{EW}^{ijklm} \right
]_{STF} \;, \\
 \stackrel{(4)\qquad}{{\cal I}_{STF}^{ijklmn}} &=& \left 
[ 360 I_{EW}^{ijklmn} \right
]_{STF} \;, \\
{\cal J}^{ij}_{STF} &=& \left [ {1 \over 2} \epsilon_{ipq} I_{EW}^{jqp} +
{1 \over 28} \epsilon_{ipq} \left ( 9 I_{EW}^{jqpmm} -3 I_{EW}^{qmmjp} \right )
\right ] _{STF} 
+ {\cal J}_{TAIL}^{ij} \;, \\
\dot {\cal J}^{ijk}_{STF} &=& \left [ 2 \epsilon_{ipq} I_{EW}^{jqpk} +
{4 \over 15} \epsilon_{ipq} \left ( 7 I_{EW}^{jqpkmm} -2 I_{EW}^{qmmpjk} 
\right ) \right ] _{STF} \;, \\
\ddot {\cal J}^{ijkl}_{STF} &=& \left [ 9 \epsilon_{ipq} I_{EW}^{jqpkl} \right
]_{STF} \;, \\
 \stackrel{(3)\qquad}{{\cal J}^{ijklm}_{STF}} &=& \left [ 48 
\epsilon_{ipq} I_{EW}^{jqpklm} \right
]_{STF} \;, 
\end{eqnarray}
\label{STF-EW} 
\end{mathletters}
where the STF notation on the right-hand side means symmetrize and
remove all traces (note that the STF tensors are symmetric on all
indices, while the EW moments are symmetric only on selected pairs).  
These transformations can also be established
using Eqs. (5.23) and (5.24) of \cite{thorne80}. 

{}For two-body systems in general orbits, the resulting STF moments are
given by
\begin{mathletters}
\begin{eqnarray}
{\cal I}_{STF}^{ij} &=& \mu r^2 \biggl\{ \hat n^i \hat n^j + {1 \over 42} 
\biggl[ \hat n^i \hat n^j \left( 29(1-3\eta) v^2 - 6(5-8\eta) 
{m\over r} \right)
 \nonumber \\
&&
- 24(1-3\eta) \dot r \hat n^{(i} v^{j)} + 22(1-3\eta) v^i v^j \biggr]
\nonumber \\
&& + {1\over1512} \hat n^i \hat n^j \biggl[ 
  3(253 - 1835\eta + 3545\eta^2)v^4
 -6(355+1906\eta - 337\eta^2) \left( {m\over r} \right) ^2
\nonumber \\
&& +2(2021 - 5947 \eta - 4883\eta^2 ) {m \over r} v^2
 -2( 131 - 907 \eta + 1273\eta^2) {m \over r} \dot r^2 \biggr]
\nonumber \\ &&
+{1 \over 378} v^i v^j \biggl[ 2(742-335\eta -985\eta^2) {m \over r}
\nonumber \\ &&
+ 3 ( 41 - 337 \eta + 733 \eta^2 )v^2 + 30(1-5\eta+5\eta^2)\dot r^2 \biggr]
\nonumber \\ &&
-{1 \over 378} \hat n^{(i} v^{j)}\dot r 
\left[ (1085-4057\eta-1463\eta^2) {m \over r}
 + 12 (13 - 101\eta + 209 \eta^2 ) v^2 \right]
\biggr\}_{STF}
\nonumber \\ &&
+{\cal I}_{TAIL}^{ij} \;, \\
{\cal I}_{STF}^{ijk} &=& -\mu {{\delta m} \over m} r^3 \biggl\{ \hat n^i \hat
n^j \hat n^k \biggl[ 1+ {1 \over 6}(5-19\eta )v^2 -{1 \over
6}(5-13\eta ){m \over r} \biggr] \nonumber \\
&& +(1-2\eta )(v^iv^j \hat n^k - \dot r v^i \hat n^j \hat n^k )
\biggr\}_{STF} 
+{\cal I}_{TAIL}^{ijk} \;, \\
{\cal I}_{STF}^{ijkl} &=& \mu r^4 \biggl\{ \hat n^i \hat n^j \hat n^k \hat
n^l \biggl[ (1-3\eta) \nonumber \\ &&  
 +{1 \over 110} (103-735\eta+1395\eta^2)v^2 - {1 \over 11}(10-61\eta+105\eta^2)
{m \over r} \biggr] \nonumber \\
&&+ {6 \over 55} (1-5\eta+5\eta^2)(13v^iv^j \hat n^k \hat n^l -12 \dot
r v^i \hat n^j \hat n^k \hat n^l ) \biggr\}_{STF} \;, \\
{\cal I}_{STF}^{ijklm} &=& -\mu {{\delta m} \over m}r^5 \biggl\{  (1-2\eta)
 \hat n^i \hat n^j \hat n^k \hat n^l \hat n^m  \biggr\}_{STF} \;, \\
{\cal I}_{STF}^{ijklmn} &=& \mu r^6 \biggl\{ (1-5\eta+5\eta^2) 
 \hat n^i \hat n^j \hat n^k \hat n^l \hat n^m  \hat n^n \biggr\}_{STF} \;, \\
{\cal J}_{STF}^{ij} 
&=& -\mu {{\delta m} \over m} r \biggl\{ ({\bf x} \times {\bf v})^i
\biggl[ \hat n^j \biggl( 1+ {1 \over 28}(13-68\eta)v^2 + {3 \over
14}(9+10\eta){m \over r} \biggr) \nonumber \\
&&+ {5 \over 28}(1-2\eta) \dot r v^j \biggr] \biggr\}_{STF} +
{\cal J}_{TAIL}^{ij} \;, \\
{\cal J}_{STF}^{ijk} &=& \mu r^2  \biggl\{ ({\bf x} \times {\bf v})^i
\biggl[  \hat n^j \hat n^k \biggl ( 1-3\eta \nonumber \\
&& + {1 \over 90}(41-385\eta+925\eta^2)v^2 +{2 \over 9}(7-8\eta-43\eta^2){m
\over r} \biggr) \nonumber \\
&& +{1 \over 45}  (1-5\eta+5\eta^2)(10\dot r v^j \hat n^k + 7v^jv^k )
\biggr] \biggr\}_{STF} \;, \\
{\cal J}_{STF}^{ijkl} &=& -\mu {{\delta m} \over m} r^3 (1-2\eta) 
\biggl\{ ({\bf x} \times {\bf v})^i \hat n^j \hat n^k \hat n^l
\biggr\}_{STF} \;, \\
{\cal J}_{STF}^{ijklm} &=& \mu r^4 (1-5\eta+5\eta^2) 
\biggl\{ ({\bf x} \times {\bf v})^i \hat n^j \hat n^k \hat n^l \hat
n^m \biggr\}_{STF} \;, \\
\end{eqnarray}
\label{2bodySTF}
\end{mathletters}
where
the ``tail'' STF moments are given by 
\begin{mathletters}
\begin{eqnarray}
\ddot {\cal I}_{TAIL}^{ij}  &=& 2m \int_0^\infty ds
\stackrel{(4)}{Q^{ij}}(u-s) \left [ \ln \left ( {s \over
{2R+s}} \right ) + {11 \over 12} \right ]_{STF} \;, \\
\ddot {\cal I}_{TAIL}^{ijk}  &=& 2m  \int_0^\infty ds
\stackrel{(5)\quad}{Q^{ijk}}(u-s) \left [ \ln \left ( {s \over
{2R+s}} \right ) + {97 \over 60} \right ]_{STF} \;, \\
\ddot {\cal J}_{TAIL}^{ij}  &=& 2m \int_0^\infty ds
\stackrel{(4)\quad}{J^{ij}}(u-s) \left [ \ln \left ( {s \over
{2R+s}} \right ) + {7 \over 6} \right ]_{STF} \;. 
\end{eqnarray}
\label{tailstfmoments}
\end{mathletters}
Through 3/2PN order, these moments agree with \cite{magnum}, and in
the circular orbit limit, through 2PN order, they agree with BDI
\cite{bdi2pn}.

In terms of STF moments, the waveform and energy flux may be written
\cite{thorne80}
\begin{mathletters}
\begin{eqnarray}
h_{TT}^{ij} &=& {1 \over R} \sum_{l=2}^\infty \biggl[ {4 \over l!}
\stackrel{(l)\quad\quad\quad}{{\cal I}_{STF}
^{ija_1 \dots a_{l-2}}}(u) \hat N^{a_1
\dots a_{l-2}} + {8l \over {(l+1)!}} \epsilon_{pq(i}
\stackrel{(l)\quad \quad\quad}{{\cal J}_{STF}
^{j)pa_1 \dots a_{l-2}}}(u) \hat N^{qa_1
\dots a_{l-2}} \biggr]_{TT} \;,
\label{hSTF}\\
{dE \over dt} &=& \sum_{l=2}^\infty \biggl[ {{(l+1)(l+2)} \over
{l(l-1)l!(2l+1)!!}}
\stackrel{(l+1)\quad}{{\cal I}_{STF}^{a_1 \dots a_{l}}}(u)
\stackrel{(l+1)\quad}{{\cal I}_{STF}^{a_1 \dots a_{l}}}(u)
\nonumber \\ && + {{4l(l+2)} \over
{(l-1)(l+1)!(2l+1)!!}}
\stackrel{(l+1)\quad}{{\cal J}_{STF}^{a_1 \dots a_{l}}}(u)
\stackrel{(l+1)\quad}{{\cal J}_{STF}^{a_1 \dots a_{l}}}(u) \biggr] \;.
\label{EdotSTF}
\end{eqnarray}
\end{mathletters}
Substitution of Eqs. (\ref{2bodySTF}) into Eqs. (\ref{hSTF}) and
(\ref{EdotSTF}), using 2PN equations of motions in any
acceleration terms generated by time derivatives, and keeping terms
through 2PN order, yields Eqs. (\ref{hanswer}),
(\ref{hpieces}), (\ref{Edotanswer}) and (\ref{Edotpieces}).

\section{Spin Effects}
\label{appspin}

In this paper, we have used our augmented Epstein-Wagoner
formalism to give a complete description of the gravitational radiation
for inspiralling ``point-mass'' binaries
through
$O(\epsilon^2)$ beyond the lowest-order quadrupole contribution.
In this Appendix
we demonstrate that our formalism
is also adequate
for computing contributions to the radiation which arise from 
the finite spatial extent of the bodies.
Our primary goal will be to compute the contributions
to the radiation from the bodies' spin angular-momenta,
but in the process we will show how other extended-body
effects, 
such as those due to a body's intrinsic quadrupole moment,
could be computed with our formalism.
The results will be presented in such a way that the
spin contributions computed here can just be 
{\it added } to
results already presented here and elsewhere.
In particular we give the spin-orbit ($PQ^{ij}_{SO}$
and $P^{3/2}Q^{ij}_{SO}$)
and spin-spin ($P^{2}Q^{ij}_{SS}$)
contributions
to the waveform Eq. (\ref{hanswer}) for general orbits.  
We also give a restricted circular-orbit version of the results
which can be added to the ``ready-to-use'' waveforms in \cite{biww}.

In order to derive the spin corrections to the waveform,
we relax our ``point-mass'' assumption
and allow the bodies to have spatial extent {\it small} compared to
the interbody distances.
We further assume that the bodies are uniformly spinning fluid balls,
approximately spherical in harmonic coordinates.
(A full discussion of this ``fluid sphere'' formalism 
is given in Appendix A of \cite{magnum},
where it is used to derive the waveform produced by 
non-spinning bodies
through $O[\epsilon^{3/2}]$.)
Although formally, our PN approach restricts us to weak internal
gravity, we anticipate applying the results to neutron stars and black
holes, as in the non-spinning case, by 
relying upon the Strong Equivalence Principle (see Sec. II.B for
discussion of this point).  
It is now conventional, in treating spinning compact bodies, to view
the spin $\bf S$ of each body as a quantity measured in units of its
(mass)$^2$, as is the case for black holes.  Given that, formally,
$S \sim md \bar v$, where $d$ is the size of the body, and $\bar v$ is
its rotational velocity, our convention implies that
$S_{\rm Compact}/S_{\rm Formal} \sim m/d\bar v \sim \epsilon^{1/2}$, with the
result that spin effects are viewed as 1/2PN order smaller per factor
of spin than would
be the case formally (see \cite{kww,kidder} for further
discussion).

The leading-order spin corrections to the waveform arise solely from terms 
in the source [Eq. (\ref{effective})] directly dependent upon fluid velocities.
Since these terms have compact support,
they generate no contributions to the 
waveform from surface terms or from far-zone integrals,
at the order we are considering in this Appendix.
Thus the spin corrections can all be obtained from 
the compact support pieces of
the EW moments Eq. (\ref{EWmoments}).
We illustrate the procedure for computing the spin contributions
by examining the 
4-index EW multipole Eq. (\ref{EWmoments}c)
\begin{equation}
I_{EW}^{ijkl} = \int_{\cal M} \tau^{ij} x^k x^l d^3 x.
\end{equation}
Using Eq. (\ref{fluid})  and Eq. (\ref{effective})
we can write
\begin{equation}
I_{EW}^{ijkl} = \int_{\cal M} \left[ \rho v^i v^j + 
(\rm{terms \; independent \; of\;
velocity) \; }
+O(\rho \epsilon^2) \right] x^k x^l d^3 x \; .
\label{trunc4}
\end{equation}
Terms which are independent of the
fluid velocity will not contribute 
to the spin terms that we are computing here;
they give non-spin terms which we have already calculated.
Any spin terms that might result from the $O(\rho \epsilon^2)$
contributions will, in our convention, 
be at least $O(\epsilon^{1/2})$ smaller, beyond the
2PN order at which we are working.
We now write the source-point position and velocity as 
\begin{mathletters}
\begin{eqnarray}
x^i & \equiv & x_A^i + \bar x_A^i  \;,  \\
v^i & \equiv & v_A^i + \bar v_A^i  \; ,
\end{eqnarray}
\label{posvel}
\end{mathletters}
where $x_A^i$ is a suitably defined, PN-order, coordinate 
``center of mass'' of 
body $A$ and $\bar x_A^i$ is a coordinate displacement
vector from the center of mass
to the fluid element within the body.
Similarly $v_A^i = dx_A^i /dt$ is the coordinate
velocity of the center of mass.
(See {\it e.g.} \cite{tegp,kww,kidder} 
for the definition of the center of mass.)

Substituting Eq. (\ref{posvel}) into 
Eq. (\ref{trunc4}) and integrating we obtain
\begin{equation}
I_{EW}^{ijkl}
=  \sum_A m_A v_A^{ij} x_A^{kl}+ 
2 v^{(i}_A \epsilon^{j)m(k} x^{l)}_A S^m_A  \; ,
\label{I4spin}
\end{equation}
where we have defined the spin vector by the formula
\begin{equation}
\int_A \rho \bar x_A^i \bar v_A^j d^3x \equiv 
{1 \over 2} \epsilon^{kij}S_A^k \; ,
\label{spinvector}
\end{equation}
having assumed that $\int_A \rho {\bar x}_A^{(i}{\bar v}_A^{j)} d^3x =
(1/2) dI_A^{ij}/dt =0$, where $I_A^{ij}$ is the body's intrinsic
moment-of-inertia tensor.  
The first term in Eq. (\ref{I4spin}) is the leading-order
velocity-dependent term in Eq. (\ref{I4final}),
and the second term is the spin-orbit correction to this multipole, of
order $\epsilon^{1/2}$ smaller.
In obtaining Eq. (\ref{I4spin}) from Eq. (\ref{trunc4})
we have neglected a number of terms
because (1) they vanish because of our assumption of spherical symmetry,
(2) they have vanishing transverse-traceless projection, or
(3) they are higher order in the bodies' {\it small} dimension ($\sim
m$), and therefore effectively of higher PN order.
Such higher order moments can in principle be retained and
incorporated into the framework.

Keeping terms up to $O(\rho \epsilon)$ in the
source $\tau^{ij}$, and 
proceeding in precisely the same manner
we can compute the spin-orbit contributions to the other EW multipoles
\begin{eqnarray}
I_{EW}^{ij} &=& \sum_A \left[ m_A x^{ij}_A 
 +  x^{(i}_A ( {\bf v}_A \times {\bf S}_A )^{j)}
  \right] \; , 
\label{I2spinbeforessc} \\
{\tilde I}_{EW}^{ijk} &=& \sum_A ( m_A v_A^{i} x^{jk}_A  
+  x_A^{(j} \epsilon^{k)li} S_A^l  ) \; .
\label{I3spin}
\end{eqnarray}
Here again, the first terms are the leading-order non-spin contributions
to the multipoles, Eqs. (\ref{Iijfinal}) and (\ref{Iijkfinal}). The
spin-orbit correction terms are respectively of order $\epsilon^{3/2}$
and $\epsilon^{1/2}$ smaller than the leading terms.  

In generating the expressions for the multipoles
and waveforms,
we must include {\it spin}
corrections to the equations
of motion.
However, in the case of spinning
bodies there is a delicate point
to be considered in this procedure.
The center of mass of body-$A$, denoted by ${\bf x}_A$ used in our
derivation of the multipole expressions
turns out not to be
precisely the same as the definition 
of the body's position used in the derivation of the conventional spin-orbit
equations of motion,
as given, say, by Damour \cite{damour300},
or Eq. (\ref{spineom}) below.
The difference is related to the use of different 
so-called ``spin supplementary
conditions''  which fix the center of mass of spinning bodies
(see \cite{kww,kidder} for a thorough discussion).
We have previously shown \cite{kww} that, to bring our center-of-mass 
definition into accord with that used in the 
equations of motion we need to shift the position of
body-A  in the following manner
\begin{equation}
x^i_A \rightarrow  x^i_A + {1 \over 2 m_A}
( {\bf v_A} \times {\bf S_A} )^i \; .
\label{sscshift}
\end{equation}
Performing this transformation replaces Eq. (\ref{I2spinbeforessc}) with
\begin{equation}
I_{EW}^{ij} = \sum_A \left[ m_A x^{ij}_A  
 + 2  x^{(i}_A ( {\bf v}_A \times {\bf S}_A )^{j)}
  \right] \; .
\label{I2spinafterssc}
\end{equation}
Since we are working only to 3/2PN
order in the spin-orbit correction, the transformation
Eq. (\ref{sscshift}) has no effect on the other multipoles.
However if one were deriving the 2PN spin-orbit
correction to the waveform 
[{\it i.e.} $P^2Q^{ij}_{SO}$ in Eq. (\ref{hanswer})] it would be
necessary to use the tranformation on Eq. (\ref{I3spin}) as well.

The spin
pieces of Eqs. 
(\ref{I2spinafterssc}), (\ref{I3spin}) and (\ref{I4spin}) can just 
be added to their N-body point-mass counterparts in
Section (\ref{sec:ewmoments}), Eqs. (\ref{Iijfinal}).
(\ref{Iijkfinal}) and (\ref{I4final}), respectively.

We now wish to restrict our attention to the two-body case
and express our multipoles in terms of relative 
coordinates.
The reduction 
parallels the
2-body (non-spin) reduction given in Section VI.
We introduce the spin-corrections to the definition of the system
center of mass, Eq. (\ref{mQJ}c) (see \cite{kww,kidder}), 
find the relation between
the coordinates ${\bf x}_1$ and ${\bf x}_2$ and the relative
coordinate $\bf x$ corresponding to Eqs. (\ref{x1x2}), and substitute
into the two-body EW moments.
It is useful to define
two relative spin quantities
\begin{mathletters}
\begin{eqnarray}
 \mbox{\boldmath$\chi$}_s =  && {1 \over 2} \left ( { {\bf S_1} \over m_1^2 } 
+ { {\bf S_2} \over m_2^2 } \right ) \; , \\
 \mbox{\boldmath$\chi$}_a =  && {1 \over 2} \left ( { {\bf S_1} \over m_1^2 } 
- { {\bf S_2} \over m_2^2 } \right ) \; .
\end{eqnarray}
\label{spinquantities}
\end{mathletters}
With the spins normalized by the individual (masses)$^2$, 
these
vectors are essentially the vectorial sum and difference of
the dimensionless angular-momentum (Kerr) parameters
of the individual bodies.
{}For orbital systems composed of two Kerr black holes or neutron stars
these vectors will have a maximum magnitude of unity.
Stability studies of rotating neutron stars show that the dimensionless
angular momentum parameter is bounded above by  0.63 -- 0.74, 
\cite{nseos} depending on the equation of state.
Defining the vector spin quantities in this way also has the
advantage that they are comparable in maximum magnitude to the other
vectors that are used to form the terms in the waveform, 
namely ${\bf \hat n}$, ${\bf \hat N}$ and ${\bf v}$.
As one computes the 2-body multipoles, the waveform, the energy
flux, and the orbital phase evolution, the spins appear in many
combinations with the masses.  
With the spin-quantity definitions as above, the reduced mass
parameter $\eta$ never appears in any denominators, so that 
the extreme mass ratio limit ($\eta \to 0$) is always transparent
in all expressions below \cite{spinnotation}. 
This may seem like a minor aesthetic point, but it also
means that the equations in the form we present them
are suitable for stable numerical implementation
with mass parameters free to roam from the equal mass case
to the test mass case, and spin parameters free to roam 
independently of the mass choice from magnitude zero to unity.

The spin
corrections to the relation between ${\bf x}_1$, ${\bf x}_2$ and the
relative coordinate $\bf x$ \cite{kidder} take the form 
\begin{mathletters}
\begin{eqnarray}
{\bf x_1} =  && {m_2 \over m} {\bf x }
- m {\bf v} \times [ \mbox{\boldmath$\chi$}_s  (\delta m /m ) + 
\mbox{\boldmath$\chi$}_a ] \; ,\\
{\bf x_2} =  && - { m_1 \over m} {\bf x }
- m {\bf v} \times [ \mbox{\boldmath$\chi$}_s (\delta m /m ) + 
\mbox{\boldmath$\chi$}_a ] \; .
\end{eqnarray}
\end{mathletters}
Substituting
these transformations into the leading order term in 
Eq. (\ref{I2spinafterssc}), we find that 
these spin-orbit corrections cancel, to the required order
[compare Eq. (\ref{relquad})].
Substituting these definitions into the N-body multipoles 
gives the spin-orbit corrections to the two-body Epstein-Wagoner
multipoles
\begin{mathletters}
\begin{eqnarray}
I^{ij}_{EW(SO)}  = && 4 m^2 \eta^2 ( {\bf v} \times 
\mbox{\boldmath$\chi$}_s)^{(i} x^{j)} \;, \\
I^{ijk}_{EW(SO)}  = && 2 m^2 \eta x^{(i} \epsilon^{j)lk}
[ \,(\delta m /m) \mbox{\boldmath$\chi$}_s + 
\mbox{\boldmath$\chi$}_a ]^l \;, \\
I^{ijkl}_{EW(SO)}  = && 4 m^2 \eta^2 v^{(i} \epsilon^{j)m(k}
s^{l)} \chi_s^m  \; .
\end{eqnarray}
\label{EW2bodymoments}
\end{mathletters}
These corrections can be added to the 2-body multipoles 
given in Section VI.
STF multipoles can be projected 
from the EW multipoles using the formulae
given in 
Appendix \ref{appSTFdecomp}.  
The results are
\begin{mathletters}
\begin{eqnarray}
{\cal I}^{ij}_{STF (SO)} 
&=& {8 \over 3} m^2 \eta^2 \left[
 2 x^i ({\bf v} \times \mbox{\boldmath$\chi$}_s )^j
- v^i ({\bf x} \times \mbox{\boldmath$\chi$}_s )^j \right]_{STF}  \; , \\
{\cal J}^{ij}_{STF (SO)} 
&=& {3 \over 2} m^2 \eta \left[ \left ( (\delta m /m) \mbox{\boldmath$\chi$}_s
+ \mbox{\boldmath$\chi$}_a \right )^i  
x^j \right]_{STF} \; ,\\
{\cal J}^{ijk}_{STF (SO)} 
&=& 4 m^2 \eta^2 \left[ x^i x^j \chi_s^k \right]_{STF} \; .
\end{eqnarray}
\label{spinstf}
\end{mathletters}
Eqs. (\ref{spinstf}) are in agreement with \cite{kww,kidder}.
These spin-orbit contributions 
can be added to the STF multipoles given
in Appendix \ref{appSTFdecomp}.
It is interesting to note that the 4-index EW multipole
$I_{EW}^{ijkl}$ is needed to describe spin-dependence of the radiation,
but there is no spin contribution from the 4-Index STF-multipole
${\cal I}^{ijkl}_{STF}$. The multipole $I_{EW}^{ijab}$
does contribute to the multipole 
${\cal I}^{ij}_{STF}$ and ${\cal J}^{ijk}_{STF}$ 
through Eqs. (E2a) and (E2g).

In order to derive the spin contributions to the
waveform from the multipoles we must also augment
the equations of motion [Eq. (\ref{motion})]
with spin-orbit and spin-spin contributions.
These can be found in \cite{kww,kidder},
and in our notation are given by
\begin{mathletters}
\label{spineom}
\begin{eqnarray}
{\bf a}_{SO} =&& {m^2 \over r^3}
\biggl\{ 6 {\bf \hat n} ( {\bf \hat n \times v} ) \cdot
\left[ \mbox{\boldmath$\chi$}_s + (\delta m /m) 
\mbox{\boldmath$\chi$}_a \right] -2 
{\bf v} \times 
\left[ (2 -\eta) \mbox{\boldmath$\chi$}_s + 2 (\delta m /m ) 
\mbox{\boldmath$\chi$}_a \right] 
\nonumber \\ 
&& \;\; + 6 \dot r {\bf \hat n} \times 
\left[ (1-\eta) \mbox{\boldmath$\chi$}_s + (\delta m /m) 
\mbox{\boldmath$\chi$}_a \right]
\biggr\}  \; ,\\
{\bf a}_{SS} =&& - {m^3 \over r^4} 
\biggl\{ {\bf \hat n} 
\left[ \, |{\bf \chi_s}|^2 - |{\bf \chi_a}|^2 
-5({\bf \hat n \cdot }\mbox{\boldmath$\chi$}_s )^2 +5({\bf \hat n \cdot }
\mbox{\boldmath$\chi$}_a )^2 \right]
+2 \left[ \mbox{\boldmath$\chi$}_s ({\bf \hat n \cdot }
\mbox{\boldmath$\chi$}_s ) 
- \mbox{\boldmath$\chi$}_a ({\bf \hat n \cdot }\mbox{\boldmath$\chi$}_a) 
\right] \biggr \} \;.
\end{eqnarray}
\end{mathletters}
We now substitute our EW multipoles into 
Eq. (\ref{EWseries}) and use the equations of 
motion to eliminate acceleration terms
to obtain the final spin contributions to the waveform
\begin{mathletters}
\label{spinwaveform}
\begin{eqnarray}
PQ_{SO}^{ij} =&& 2 \biggl( {m \over r} \biggr)^2 \biggl\{ 
{\bf \hat N } \times 
\left[ (\delta m/m)\mbox{\boldmath$\chi$}_s + 
\mbox{\boldmath$\chi$}_a \right] \biggr\}^{(i} n^{j)} , \\
P^{3/2}Q_{SO}^{ij} =&& 4 \biggl({m \over r} \biggr)^2 
\biggl\{
3 ( {\bf \hat n \times v}) \cdot  
\left[ \mbox{\boldmath$\chi$}_s + (\delta m /m ) 
\mbox{\boldmath$\chi$}_a \right] n^i n^j \nonumber \\
&& \; \;
- \biggl[ {\bf v} \times [(2+\eta)\mbox{\boldmath$\chi$}_s
+ 2 (\delta m /m ) \mbox{\boldmath$\chi$}_a ] \biggr]^{(i}
n^{j)}
\nonumber \\
&& \; \; + 
3 \dot r \biggl[ {\bf \hat n} \times [ \mbox{\boldmath$\chi$}_s
+ (\delta m /m) \mbox{\boldmath$\chi$}_a ]
\biggr]^{(i} n^{j)} 
- 2 \eta  ({\bf \hat n \times} \mbox{\boldmath$\chi$}_s )^{(i} v^{j)}
\nonumber \\
&& \; \; + \eta
\left[ 2 ({\bf \hat N \cdot \hat n}) {\bf v} 
  + 2({\bf \hat N \cdot v})      {\bf \hat n} 
  - 3 \dot r ( {\bf \hat N \cdot n}){\bf \hat n} \right]^{(i}
      ({\bf \hat N \times } \mbox{\boldmath$\chi$}_s )^{j)}
\biggr\} \\
P^2Q_{SS}^{ij} =&& -6  \biggl( {m \over r} \biggr)^3 \eta \biggl\{
\left[ \, |{\bf \chi_s}|^2 - |{\bf \chi_a}|^2 
-5({\bf \hat n \cdot } \mbox{\boldmath$\chi$}_s )^2 
+5({\bf \hat n \cdot } \mbox{\boldmath$\chi$}_a )^2 \right] n^i n^j
\nonumber \\
&& \;\; 
+2 \left[ \mbox{\boldmath$\chi$}_s ({\bf \hat n \cdot }
\mbox{\boldmath$\chi$}_s ) -
\mbox{\boldmath$\chi$}_a ({\bf \hat n \cdot }
\mbox{\boldmath$\chi$}_a ) \right]^{(i} n^{j)}
\biggr \}  \; .
\end{eqnarray}
\end{mathletters}
Note that the spin-spin term comes entirely from the effects of the
equations of motion.
Thus we have computed the complete waveform, 
including leading-order spin effects, 
using our augmented EW formalism.
The formalism can be extended to compute
additional spin terms and other finite-size effects, such as the
2PN spin-orbit contribution to the waveform.

Either by a direct computation starting with the waveform or by 
using the STF-multipoles in  Eq. (\ref{EdotSTF}) we can 
compute the spin contributions to the rate of energy loss, Eq.
(\ref{Edotanswer}),
\begin{eqnarray}
\dot E_{SO} = && {8 \over 15} { m^3 \mu^2 \over r^5}
[{\bf \hat n \times v}] \cdot
\biggl\{ 
[ \mbox{\boldmath$\chi$}_s +(\delta m/m) \mbox{\boldmath$\chi$}_a ] 
\left( 27 \dot r^2 -37v^2 - 12 {m\over r} \right)
\nonumber \\
&& \;\;\; + 4 \eta 
\mbox{\boldmath$\chi$}_s \biggl( 12 \dot r^2 - 3 v^2 + 8 {m \over r} 
\biggr) \biggr\} \; , \\
\dot E_{SS} = && {8 \over 15} { m^4 \mu^2 \over r^6} \eta
\biggl\{ 
3 \left[  \, |{\bf \chi_s}|^2 - |{\bf \chi_a}|^2 \right] (47 v^2 - 55\dot r^2)
\nonumber \\
&& \;\; \; 
-3\left[ ({\bf \hat n \cdot } \mbox{\boldmath$\chi$}_s )^2 
+({\bf \hat n \cdot } \mbox{\boldmath$\chi$}_a )^2 \right] 
( 168v^2 - 269 \dot r^2 ) \nonumber \\
&& \;\; \; 
+ 71 \left[({\bf v \cdot} \mbox{\boldmath$\chi$}_s )^2 +({\bf v \cdot }
\mbox{\boldmath$\chi$}_a )^2 \right]
- 342 \dot r \left[ ({\bf v \cdot } \mbox{\boldmath$\chi$}_s )
({\bf \hat n \cdot } \mbox{\boldmath$\chi$}_s )
- ({\bf v \cdot } \mbox{\boldmath$\chi$}_a )({\bf \hat n \cdot }
\mbox{\boldmath$\chi$}_a ) \right]
\biggr\} \; .
\label{EdotSO}
\end{eqnarray}

Although they are not needed in our discussion,
for completeness
we include expressions 
for the precession of our spin vectors \cite{kww,kidder}
\begin{mathletters}
\label{spinprecess}
\begin{eqnarray}
m \dot {\mbox{\boldmath$\chi$}}_s = && {\bf \Pi_1} \times 
\mbox{\boldmath$\chi$}_s + {\bf \Pi_2} 
\times \mbox{\boldmath$\chi$}_a
- 2 (\delta m /m) \mbox{\boldmath$\chi$}_a \times 
\mbox{\boldmath$\chi$}_s   \;, \\
m \dot {\mbox{\boldmath$\chi$}}_a = && {\bf \Pi_2} \times 
\mbox{\boldmath$\chi$}_a + {\bf \Pi_1} 
\times \mbox{\boldmath$\chi$}_s
- 2 (1 - 2 \eta ) \mbox{\boldmath$\chi$}_s \times \mbox{\boldmath$\chi$}_a
\; .
\end{eqnarray}
\end{mathletters}
The precession vectors are given by
\begin{mathletters}
\label{spinprecess2}
\begin{eqnarray}
{\bf \Pi_1} = && {3 \over 4} \left ( {m \over r} \right )^2
\left [ ( 1 + 2 \eta / 3) ( {\bf \hat n} \times {\bf v} )
+ 2 {m\over r} [ (1-2\eta) 
{\bf \hat n \cdot } \mbox{\boldmath$\chi$}_s
+ (\delta m /m) {\bf \hat n \cdot }
\mbox{\boldmath$\chi$}_a ] {\bf \hat n} \right ] \; , \\
{\bf \Pi_2} = && - {3 \over 4} \left ( {m \over r} \right )^2
\left [ (\delta m /m) ( {\bf \hat n} \times {\bf v} )
+ 2 {m\over r} [ (\delta m /m ) 
{\bf \hat n \cdot } \mbox{\boldmath$\chi$}_s
+ (1-2\eta) {\bf \hat n \cdot } \mbox{\boldmath$\chi$}_a ] 
{\bf \hat n} \right ] \; .
\end{eqnarray}
\end{mathletters}

When spinning bodies are involved, the full gravitational-wave
signal can become quite complicated; the orbital plane and
the spin vectors of the individual bodies can precess, giving rise
to a complicated modulation of the 
signal \cite{kidder,apostolatos}.
However in the special case when the spins are
aligned (or anti-aligned) with the orbital angular momentum axis, the spin
vectors and the orbital angular momentum vector do not precess
[Eqs. (\ref{spineom}), (\ref{spinprecess}), (\ref{spinprecess2}) ].
In this special case there is a simple circular orbit solution
to the equation of motion and it is straightforward to 
compute the spin contributions to the phase evolution.
The spin contributions to orbital frequency
can obtained from Eq. (\ref{spineom}),
\begin{eqnarray}
\omega^2 = {m \over r^3} \biggl\{ 1 
- 2 \biggl( {m\over r}\biggr)^{3/2}
[ (1+\eta) \chi_s + (\delta m /m) \chi_a ]
- 3 \eta \biggl( {m\over r}\biggr)^{2} 
\left[  (\chi_s)^2 - (\chi_a)^2 \right]
\biggr\} \; ,
\label{spinfreq}
\end{eqnarray}
where $\chi_s$ and $\chi_a$ now represent the projections of 
$\mbox{\boldmath$\chi$}_s$ and $\mbox{\boldmath$\chi$}_a$ 
onto the angular momentum axis.
These quantities are positive when the spins are aligned in
the same direction 
as the angular momentum axis and negative when
they are anti-aligned.
The orbital energy and energy flux take the
simple form in the case of aligned spins
and circular motion,
\begin{mathletters}
\begin{eqnarray}
E  &=& - \eta {m^2 \over 2 r}
\biggl\{ 1 + 2 \biggl( {m\over r}\biggr)^{3/2}
\left[ (1-\eta) \chi_s + (\delta m / m) \chi_a \right]
+ \biggl({m\over r}\biggr)^{2}
\left[ (\chi_s)^2 - (\chi_a)^2 \right]
\biggl\} \; ,
\label{spinenergy} \\
\dot E &=& {32 \eta^2 \over 5} \biggl( {m\over r}\biggr)^{5}
\biggl\{ 1 - 
\biggl( {m\over r}\biggr)^{3/2}
\biggl[ \, {73 \over 12 } [ \chi_s + (\delta m /m)\chi_a] 
 - {\eta \chi_s \over 2}
\biggr] \nonumber \\
&& \;
- {71 \eta \over 8 } \biggl( {m\over r}\biggr)^{2}
\left[ (\chi_s)^2 - (\chi_a)^2 \right] 
\biggr\}\; . 
\label{spinenergyloss}
\end{eqnarray}
\end{mathletters}
These spin corrections can be added to 
the non-spin formulae
Eq. (\ref{energycirc}) and Eq. (\ref{edot}).
With these we can proceed as in 
Section VI to obtain the
orbital angular velocity and orbital phase
as explicit functions of time
\begin{mathletters}
\label{spinfreqphase}
\begin{eqnarray}
\omega(t) = && {1 \over 8 m} (T_c-T)^{-3/8} 
\biggl\{ 1 + 
\biggl[ {113 \over 160} [\chi_s + (\delta m/m)\chi_a] 
- {19 \over 40} \eta \chi_s \biggr]
(T_c -T)^{-3/8} \nonumber \\
 && \;\; - { 237 \over 512} \eta \left[ (\chi_s)^2 - (\chi_a)^2 \right] 
(T_c -T)^{-1/2}
\biggr\}  \;, \\
\phi(t) = && \phi_c - {1 \over \eta } (T_c-T)^{5/8} 
\biggl\{ 1 +
\biggl[ {113 \over 64} [\chi_s + (\delta m/m)\chi_a] 
- {19 \over 16} \eta \chi_s \biggr]
(T_c -T)^{-3/8} \nonumber \\
&& \;\;  
 - { 1185 \over 512} \eta \left[ (\chi_s)^2 - (\chi_a)^2 \right]
(T_c -T)^{-1/2}
\biggr\} \; .
\end{eqnarray}
\end{mathletters}
Again, the spin-contributions can be inserted directly
into Eqs. (\ref{freqphase}).
(The definition of the dimensionless time $T= \eta (u/5m)$
is unchanged.)
The explicit contributions 
to the $+$ and $\times$ polarizations for this
specialized circular orbit case 
can be obtained from Eq. (\ref{spinwaveform}).
In the notation of \cite{biww} they are given by
\begin{eqnarray}
h_{+,\times} = { 2 m \eta \over R} x
\biggl\{ H_{+,\times}^{0} + \dots 
+ x        H_{+,\times}^{(  1,SO)}
+ x^{3/2}  H_{+,\times}^{(3/2,SO)}
+ x^2      H_{+,\times}^{(  2,SS)}
\biggr\} \; ,
\label{spinhs}
\end{eqnarray}
where $x \equiv m \omega$ and
where the ``$\dots$'' represent the non-spin contributions
given in \cite{biww}.
In keeping with the notation used in \cite{biww}
the superscripts represent the post-Newtonian
order and the physical nature of each term.
The plus polarization spin-orbit and spin-spin contributions are 
\begin{mathletters}
\begin{eqnarray}
H_{+}^{(1,SO)} = && - \sin i  [(\delta m/m)\chi_s + \chi_a] \cos \phi \;,
\\
H_{+}^{(3/2,SO)} = && {4 \over 3} 
\left[ (1+\cos^2 i) [ \chi_s + (\delta m /m) \chi_a ] 
+ \eta (1-5\cos^2 i) \chi_s \right] \cos 2\phi \;, \\
H_{+}^{(  2,SS)} = && 
- 2 \eta (1+\cos^2 i ) [ (\chi_s)^2 - (\chi_a)^2 ] \cos 2\phi \;,
\end{eqnarray}
\label{spinhs+}
\end{mathletters}
and the cross polarization contributions are
\begin{mathletters}
\begin{eqnarray}
H_{\times}^{(1,SO)} = && - \sin i \cos i [(\delta m/m)\chi_s + \chi_a] 
\sin \phi \;, \\
H_{\times}^{(3/2,SO)} = && {4 \over 3}  \cos i 
\left[ 2[ \chi_s + (\delta m /m) \chi_a ] - \eta (1+3\cos^2 i) \chi_s \right] 
\sin 2\phi \;, \\
H_{\times}^{(  2,SS)} = &&
  4 \eta \cos i  [ (\chi_s)^2 - (\chi_a)^2 ] \sin 2\phi \; .
\end{eqnarray}
\label{spinhsx}
\end{mathletters}
We emphasize that these are only valid for quasi-circular orbits
in the case where the
the spins are aligned (or anti-aligned) with the
orbital angular momentum vector.
These restrictive assumptions about the configuration
of the system suppress many of the intricate features of
the waveform produced by spinning bodies \cite{kidder,apostolatos}.

%
{}Figure 9 shows an inspiral waveform for the same system
as in Figure 8
($10M_\odot$ black hole and a $1.4M_\odot$ neutron star
spiralling to coalescence), but
in this case the objects are spinning. The spins are
aligned with the orbital angular momentum axis.
The spin contributions to both the waveform Eq. (\ref{spinhs}) 
and the frequency evolution Eq. (\ref{spinfreqphase})
have been incorporated into the plot.
The black hole has been given a spin of $S_{BH}/m_{BH}^2 = 0.5$ 
and the neutron star has $S_{NS}/m_{NS}^2 = 0.1$ 
({\it i.e.} $\chi_s = 0.3$ and $\chi_a = 0.1$).
Notice the significant change in the frequency evolution;
the system only sweeps to about 130 Hz in the
same time it took for the non-spinning system to 
sweep to 180 Hz.
Consequently, the peaks are not as closely bunched
as they are in the non-spinning case.
This slower orbital decay  and frequency evolution 
is due to the dragging of inertial frames, which
is inherent in the equations of motion and thus in
our phase evolution equation Eq. (\ref{spinfreqphase}).
At the left side of Figures 8 and 9, the waveforms
are clearly in phase with each other,
but after a few cycles they are out of phase.
Since the phase evolution of the system is crucial
in analysing gravitational waves from an inspiral, 
it might seem that this sensitivity to spin in the phase evolution
could be exploited and the spins of the bodies be determined
with great accuracy.
However, by leaving the spins the
same but adjusting the masses slightly, we can recover the basic structure
of the non-spinning case almost exactly.
This is depicted in Fig. 10, in which   
the frequency sweep and the waveform itself
are virtually identical to the non-spinning waveform in 
{}Fig. 8.
This signal degeneracy in the spin and mass parameters
has been previously noted in \cite{cutlerflan94,poissonwill}.
It is also interesting to notice that the inclusion of the spins virtually
removes the jagged features from the troughs of the
waves.

\begin{figure}
\caption{
Past harmonic null cone $\cal C$ of the field point $(t,{\bf x})$
intersects the near zone $\cal D$ in the hypersurface $\cal N$.
}
\end{figure}

\begin{figure}
\caption{
Same as Fig. 1, for field point inside the near zone.
}
\end{figure}

\begin{figure}
\caption{
Taylor expansion of retarded time dependence on $\cal N$ results
in multipole moments integrated over the spatial hypersurface $\cal
M$.
}
\end{figure}

\begin{figure}
\caption{
Two-dimensional hypersurfaces $\cal F$ formed by intersection of
past null cone of field point with future null cones from the origin.
{}Field point is in radiation zone. For $u^\prime$ from $-\infty$ to
$u-2{\cal R}$, {\cal F} covers full $4\pi$ solid angle around the
origin.  From $u-2{\cal R}$ to $u$, {\cal F} terminates at boundary of
the near zone $\cal N$.
}
\end{figure}

\begin{figure}
\caption{
Same as Fig. 4, for field point in near zone.  Integral over
$u^\prime$ terminates at $u^\prime=u-2{\cal R}+2r$.
}
\end{figure}

\begin{figure}
\caption{
{}Fields contributing to $\Lambda^{\alpha\beta}$ at two representative
points $a$
and $b$ on {\cal F} have sources near same event $u^\prime$ at $r=0$.
Only orientation of near-zone source slice varies as angular integration moves
around {\cal F}.
}
\end{figure}

\begin{figure}
\caption{
Orientation of unit vectors defining $+$ and $\times$ waveform
polarizations.
Direction of detector is $\bf \hat N$; $\bf \hat p$ lies along line of
nodes and is the origin for orbital phase angle $\phi$.
}
\end{figure}
\begin{figure}
\caption{
(a) Orbital frequency
and (b) waveform
for a $1.4M_{\odot}$ neutron star spiralling into
a $10M_{\odot}$ black hole plotted vs. time in seconds. 
Orbit is viewed edge-on, therefore only ``$+$''-polarization
is present.
}
\end{figure}

\begin{figure}
\caption{
Same configuration as Fig 8, but bodies
are spinning.  Both spins are aligned with 
orbital angular momentum axis.
Angular momentum of black hole is 
$S_{bh} = 0.5 m_{bh}^2$ and of neutron star is $S_{ns} = 0.5 m_{ns}^2$.
Note frequency does not sweep as fast as non-spinning case
because of dragging of inertial frames.
}
\end{figure}

\begin{figure}
\caption{
Spins are same as in Fig. 9, but
heavier mass is now
$12M_{\odot}$.  Frequency evolution is same as 
non-spinning case.
Comparing this with Fig. 8 is an explicit
demonstration of
degeneracy in mass and spin parameters.
}
\end{figure}


\begin{references}
\bibitem{convention}
This convention holds sway, 
despite the fact that pure Newtonian gravity predicts no
gravitational radiation. It is often the source of confusion, since
the energy carried by the lowest-order ``Newtonian'' quadrupole
radiation manifests itself in a (post)$^{5/2}$-Newtonian, or $O[(v/c)^5]$
correction in the equation of motion.
\end{references}
\end{document}